\documentclass[twocolumn]{aastex62}
\usepackage{graphicx}
\usepackage{amssymb}
\usepackage{amsmath}
\usepackage{multirow}
\usepackage{hyperref}
\usepackage{natbib}
\usepackage{comment}
\usepackage{systeme}
\usepackage{footmisc}
\usepackage{booktabs}
\usepackage{lineno}
\newcommand{\rev}[1]{\textnormal #1}

\colorlet{myPurple}{blue!40!red}

\newcommand{\degree}{$^\circ$}
\newcommand{\minus}{$^{-3}$}
\newcommand{\efil}{\emph{E-fils~}}
\newcommand{\efilc}{\emph{E-fils}, }
\newcommand{\efilp}{\emph{E-fils}. }
\newcommand{\radmm}{rad~m$^{-2}$}
\newcommand{\ergss}{~ergs~s$^{-1}$}                 
\newcommand{\kms}{~km~s$^{-1}$}                                                        
\newcommand{\msun}{{M$_{\odot}$}}  
\shorttitle{Jet-magnetic filament encounter}
\shortauthors{L. Rudnick et al.}

\begin{document}
\title{Intracluster magnetic filaments and an encounter with a radio jet}% of the northern radio relic in the merging galaxy cluster CIZA\,J2242.8+5215}

\author{L. Rudnick}
\affil{Minnesota Institute for Astrophysics, University of Minnesota, 116 Church St. SE, Minneapolis, MN 55455, USA}
%[0000-0001-5636-7213]

%\author{\red{ORDER TBD}}
%\affil{Observatory, Country}
\author{M. Br\"uggen}
\affil{University of Hamburg,  Gojenbergsweg 112, 21029 Hamburg,Germany}
\author{G. Brunetti}
\affil{INAF - Istituto di Radioastronomia, Via P. Gobetti 101, 40129 Italy}
\author{W. D.  Cotton}
\affil{National Radio Astronomy Observatory, 520 Edgemont Road, Charlottesville, VA 22903 USA}
\author{W. Forman}
\affil{Center for Astrophysics $|$ Harvard \& Smithsonian, 60 Garden Street, Cambridge, MA 02138, USA}
\author{T. W. Jones}
\affil{Minnesota Institute for Astrophysics, University of Minnesota, 116 Church St. SE, Minneapolis, MN 55455, USA}
\author{C. Nolting}
\affil{Department of Physics and Astronomy, College of Charleston, 66 George Street, Charleston, SC 29424, USA}
\author{G. Schellenberger}
\affil{Center for Astrophysics, | Harvard \& Smithsonian, 60 Garden Street, Cambridge, MA 02138, USA}
\author{R. van Weeren}
\affil{Leiden Observatory, Leiden University, PO Box 9513, 2300 RA Leiden, The Netherlands}

\received{May 16, 2022}
\revised{June 25, 2022}
\accepted{June 28, 2022}

\begin{abstract}
    Thin synchrotron-emitting filaments are increasingly seen in the intracluster medium (ICM).  We present the first example of a direct interaction  between a magnetic filament,  a radio jet, and a dense ICM clump  in the poor cluster Abell~194. This enables the first exploration of the dynamics and possible histories of  magnetic fields and cosmic rays in such filaments.  Our observations are from the MeerKAT Galaxy Cluster Legacy Survey and the LOFAR Two Metre Sky Survey.   Prominent 220~kpc long filaments extend east of radio galaxy 3C40B, with very faint extensions to 300~kpc, and show signs of  interaction with its northern jet.  They curve around a bend in the jet and intersect the jet in Faraday depth space.  The X-ray surface brightness drops across the filaments; this suggests that the relativistic particles and fields contribute significantly to the pressure balance and evacuate the thermal plasma in a $\sim$35~kpc cylinder. We explore whether the relativistic electrons could have streamed along the filaments from 3C40B, and present a plausible alternative whereby magnetized filaments are a) generated by shear motions in the large-scale, post-merger ICM flow, b) stretched by interactions with the jet and flows in the ICM, amplifying the embedded magnetic fields, and c) perfused by re-energized relativistic electrons through betatron-type acceleration or diffusion of turbulently accelerated ICM cosmic ray electrons.  We use the Faraday depth measurements to reconstruct some of the 3D structures of the filaments and of 3C40A and B. \\
\end{abstract}

\section{Introduction}\label{sec:sintro}
The intracluster medium (ICM) is dynamic, driven by ongoing accretion along large-scale structure filaments and mergers with groups and clusters.  There are additional internal injections of energy from supernova explosions as the embedded galaxies evolve, and less frequent, but powerful injections of momentum, energy, cosmic rays and magnetic fields from the jets of active galactic nuclei \citep{2012ARA&A..50..455F}. Signatures of the dynamic ICM are visible in its X-ray emission, in the form of shocks, cold fronts, large-scale ripples \citep{2007PhR...443....1M, Perseus, 2019MNRAS.488.5259Z} and surface brightness and/or temperature fluctuations \citep{2012MNRAS.421.1123C,2014Natur.515...85Z,2016MNRAS.458.2902Z,2022arXiv220304977H}.

\subsection{ICM radio features - the emergence of filaments}\label{sec:emergence}
Radio emission provides a complementary window on ICM dynamics, including the distortions of tailed radio galaxy outflows, peripheral radio relics and cluster-wide halos primarily in merging clusters  and their smaller mini-halo counterparts, typically associated with BCGs in more relaxed clusters \citep[see, e.g., reviews by ][]{vWreview,stormy}.   These diffuse radio structures reveal the existence of a magnetized relativistic plasma, also seen through the Faraday rotation due to magnetic fields in the X-ray emitting thermal plasma \citep[e.g.,][]{Bonafede10}.  The diffuse radio structures typically require large-scale continuing inputs of energy to compensate for the radiative losses of the cosmic ray electrons \citep{BrunettiJones}.

Into this mix, a new radio phenomenon -- filamentary structure -- is emerging, as radio telescopes produce higher resolution and sensitivity images \citep{MGCLS,2022arXiv220104591B}. Their synchrotron emission, and its accompanying polarization when there is sufficient sensitivity, indicate that these are ordered magnetized structures \citep{2005A&A...430L...5G}.  Filaments, themselves, are not new; they have been known for a number of years to be prominent in radio galaxy lobes \citep{1989ApJ...347..713H} and in the dramatic  shock-related peripheral radio relic in Abell~2256 \citep{2014ApJ...794...24O,2022ApJ...927...80R}.  However, \emph{isolated} filaments are now also being observed in the ICM, not clearly associated with X-ray structures \citep[e.g.,][]{messA2255}.  In addition, deep, high resolution images of radio galaxies are revealing filaments in their surroundings; it is not clear if these are \emph{physically} connected with the radio galaxies themselves  or what they tell us about the underlying ICM magnetic fields (\citet{2020A&A...636L...1R}, A1314 in \citet{2021A&A...651A.115V}, and \citet{Condon21}). There are several known cases where ICM magnetic fields are interacting with radio galaxies; these include the Faraday rotation banding seen across diffuse lobes \citep{2011MNRAS.413.2525G,2012MNRAS.423.1335G} and the ram-pressure induced magnetic draping around fast-moving  radio and non-radio galaxies \citep{2021MNRAS.508.5326M,2021NatAs...5..159M}.

Filamentary structures present new opportunities to study the physical processes in the ICM, including their magnetic structures and the evolution of cosmic rays. In merger-driven flows, they can serve as probes of shear motions, which are otherwise invisible, and reveal the driving and dissipation scales of dynamic structures as the cluster evolves.  Their widths can provide information on the resistivity scales in the plasma.   Filamentary structures may also arise  when these  flows interact with bubbles of old emission from AGN \citep{Zuhone21}, although the filaments' appearance is heavily dependent on the modeling of the cosmic ray physics \cite{Zuhone21b}.   In this paper, we investigate a prominent pair of filamentary structures in Abell~194 that show clear signs of interactions with the northern jet from 3C40B, associated with the BCG NGC~547. 

Isolated filaments may provide yet another site for cosmic ray acceleration in clusters. Cosmic ray electron reacceleration to several GeV energies is already required for diffuse radio structures such as relics and halos to be visible at GHz frequencies.  For halos, turbulent reacceleration throughout the cluster appears to be the most promising, while peripheral radio relics are likely associated with (re-)acceleration at shocks produced from mergers (see review by \cite{BrunettiJones}.    \citet{Bell2019}  examined reacceleration in magnetic flux tubes in the backflows of radio galaxies, which would appear as filamentary features; they concluded that repeated encounters with weak shocks in such tubes could accelerate electrons to extremely high energies. While the backflow tubes likely differ from filamentary structures found in the external ICM, they illustrate the possibility for mechanisms associated with tubes to play an important role.

Ultimately, the seed electrons upon which the reacceleration operates likely come from current or past AGN activity.  Whether we can link specific filamentary structures to individual radio galaxies (as in cases where bars directly cross a tailed radio galaxy, \citet{Shapley,Lamee,messA2255}) or where they appear to be the remnants of a dissipating lobe \citep{Brienza}, or whether multiple AGN contribute to a mixed background seed population, or whether both play a role, are major open issues.

Filamentary structures are a natural consequence of turbulence in a ``high-$\beta$'' (high ratio of plasma to magnetic pressure) magnetised plasma.  Simulations show that extended bundles of magnetic fields, which would be observed as filaments,  are ubiquitous in turbulent MHD flows where their lengths reflect the local driving scales of the turbulence \citep{porter}.  These filamentary structures are formed from the tearing of thin current sheets; they are then stretched until the magnetic stresses approach the dynamical turbulent stresses.  Driven by ongoing merger activity, e.g., these filaments can reach cluster scales \citep{2018MNRAS.474.1672V}.

Except in the presence of shocks, the sound speeds in the ICM ($\sim$10$^3$~km/s) enforce pressure equilibrium.  When the dynamic pressure of AGN jets allows them to propagate in the ICM, the conversion of this directed energy to thermal energy can then evacuate cavities in the X-ray emitting plasma \citep{2000A&A...356..788C,2000ApJ...534L.135M}. One such cavity was previously reported in Abell~194, associated with the southern lobe \citep{Bogdan}. However, if the magnetic pressures in filamentary structures, along with the accompanying cosmic ray pressures,  can approach those in the surrounding medium, then they also may be sufficient to exclude the thermal plasma.  Thus, cavities in the X-ray ICM may appear even when there is no conversion from  dynamic to thermal pressures. This new physical process in the ICM is explored further below.

\begin{figure*}
\centering
    \includegraphics[width=0.8\textwidth]{ 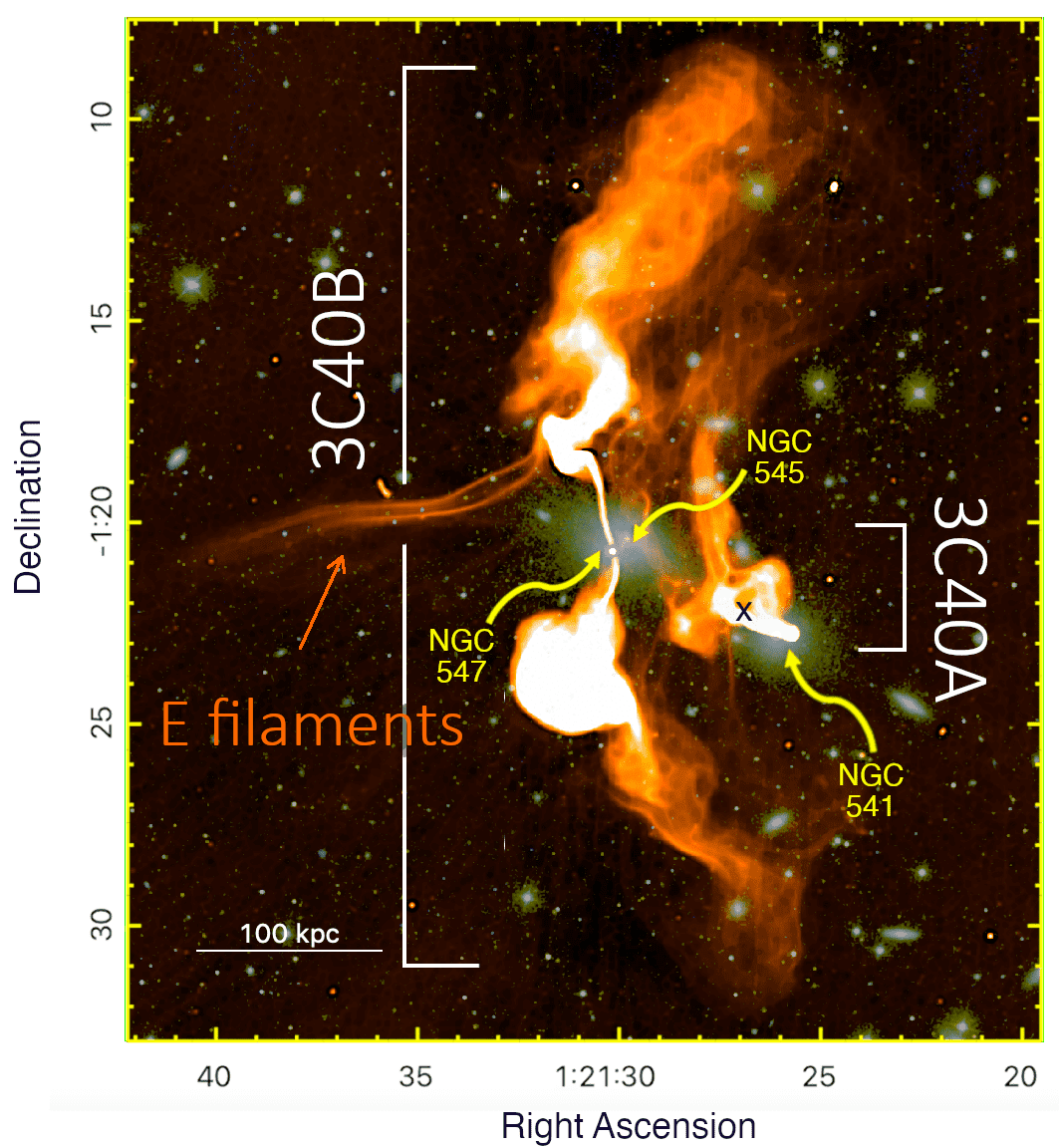}   
    \caption{Radio frequency map at 7.75\arcsec, identifying the main features under discussion. A small amount of edge enhancement has been added to emphasize the fine scale features; this image is not to be used quantitatively. Overlaid on optical SDSS \emph{gri} images \citep{2015ApJS..219...12A}.    ``\textbf{X}" marks the location of Minkowski's object.}
 \label{fig:intro}
\end{figure*}

\subsection{Abell~194}\label{sec:A194}

A194 is a richness class 0, BMII galaxy cluster \citep{Abell89}, at a redshift of 0.018 \citep{1999ApJS..125...35S}. \footnote{Using a flat $\Lambda$CDM with H=70~km/s/Mpc, $\Omega_m=0.286$, 1\arcsec~ = 0.366~kpc.}
As reported by \cite{Lovisari15} from XMM-Newton observations, A194 exhibits a low X-ray luminosity ($7.1\pm0.7\times10^{42}$~\ergss,    0.1 - 2.4 keV) and a low X-ray temperature ($kT=1.37\pm0.04$~keV), which is  characteristic of an X-ray bright group. The region of a superposed contaminating cluster at redshift 0.15 was removed from the analysis by \cite{Lovisari15} to derive the global parameters of A194. The hydrostatic mass yields
$M_{500} = 2.8\times 10^{13}$~\msun, which
%M_{500} = 2.8\times 10^{13}~\msun$ which 
corresponds to  $R_{500} = 460$ kpc ($21\arcmin$) at the distance of A194. Applying the scaling factor of 1.5 by \citet{Suhada11}, this yields 
$M_{200} = 4.2\times 10^{13}$~\msun based on the hydrostatic X-ray-derived cluster mass.  

% Due to superposed and contaminating sources, \citet{Lovisari} used the mass-temperature relation
%(M-T) to estimate a group/cluster mass  of $M_{500} = \sim6\times 10^{13}~\msun$ which corresponds to  $R_{500} = 500$ kpc ($3.5'$ at the distance of A194).
%Applying the scaling factor of 1.5 by \citet{suhada11}(Suhada et al (2011,http://dx.doi.org/10.1051/0004-6361/201116876 ) this yields $M_{200} = 9\times 10^{13}~\msun$ based on the X-ray %mass-temperature scaling relation.                                                                                                                          

Optically, \citet{Govoni2017} analyzed the spatial distribution and spectra for 143 cluster member galaxies. They found a 1-D velocity dispersion of $\sigma=425^{+34}_{-30}$~\kms~  and two  primary subgroups, both elongated in the NE-SW orientation (see Fig. 4  of \citet{Govoni2017}). They conclude there is no strong evidence of a major merger, but ongoing accretion of small groups along the NE-SW axis. \citet{Rines2003} performed a caustic analysis of A194.  The derived velocity dispersion within $R_{200}$, $\sigma=402^{+38}_{-29}$~km/s, is consistent with analysis of \citet{Govoni2017}.   The associated mass is $M_{200} = 1.1\times10^{14}$~\msun %(assuming $h=0.7$ as in Lovisari et al.) 
 which is significantly higher than the X-ray value of Lovisari et al. However, \citet{Govoni2017} could only trace the X-ray temperature profile to $\sim0.6 R_{500}$, so this higher derived cluster mass 
is based on extrapolation.

\subsubsection{Previous radio observations of A194}\label{sec:radioA194}
Abell~194 has been the subject of multiple radio studies \citep{Odea85,Jetha2006,Sakelliou2008, Bogdan,Govoni2017}. Most focused on the emission from the two prominent radio galaxies 3C40A and 3C40B.  3C40A contains the well-studied ``Minkowski's Object,'' a region of intense star formation which has likely been triggered by the impinging jet \citep{1985ApJ...293L..59B, 1985ApJ...293...83V,2006ApJ...647.1040C}.  The prominent Eastern Filaments in A~194 (hereinafter, \efil), which are the focus of the current paper, were observed and described as the double-stranded ``trail" by \citet{Sakelliou2008}, who also were able to measure the steepening of the spectrum as they extended eastward from their intersection point with the northern jet of 3C40B. \citet{Govoni2017} measured polarization structures in the Abell~194 system, combining archival 1.5~GHz VLA data with low resolution single-dish 6.6~GHz data from the Sardinia Radio Telescope.  At low resolution, the \efil are seen as a high polarization ``stub."  At higher resolution, \citet{Govoni2017} measured  low values of RM, and very high fractional polarization of the \efil (1.0$\pm$0.25) with a magnetic field aligned along the filaments, all features which are confirmed and shown in more detail here.  Most recently, the MeerKAT Galaxy Cluster Legacy Survey (MGCLS, \citet{MGCLS}) presented high resolution total intensity and spectral images of Abell~194, revealing the dense network of filamentary structures and the prominent \efil; these same data form the basis of the more detailed investigations in this paper.

The region under study in this paper extends out to only $\sim$330~kpc from the cluster center. For context, this is somewhat larger than R$_{2500}$=197$\pm$11~kpc \citep{Lovisari15}, while the caustic radius of A~194 extends beyond $\sim$8~Mpc  \citep{Rines2003}.

\subsection{Paper plan}\label{sec:plan}

Fig. \ref{fig:intro} introduces the objects under discussion in the paper, primarily the \efil and the large ($>$500~kpc) radio galaxy 3C40B.  % In this central region of Abell~194 we also see 3C40A and its host NG541, with the embedded Minkowski's object.
In Section \ref{sec:obs} we describe the observations, Section \ref{global} presents a global view of the cluster data products.   A more detailed look at the \efil and the region where they intersect the northern jet is given in Section~\ref{efilobs}. The underlying physical issues of the jet/filament interaction are explored in Section \ref{sec:discussion}, with concluding remarks and recommendations for further studies in Section \ref{sec:conclusion}.

\section{Observations and initial map production}\label{sec:obs}
\subsection{MeerKAT}\label{sec:MeerKAT}
\begin{figure}
    \centering
    \includegraphics[width=0.7\columnwidth]{ 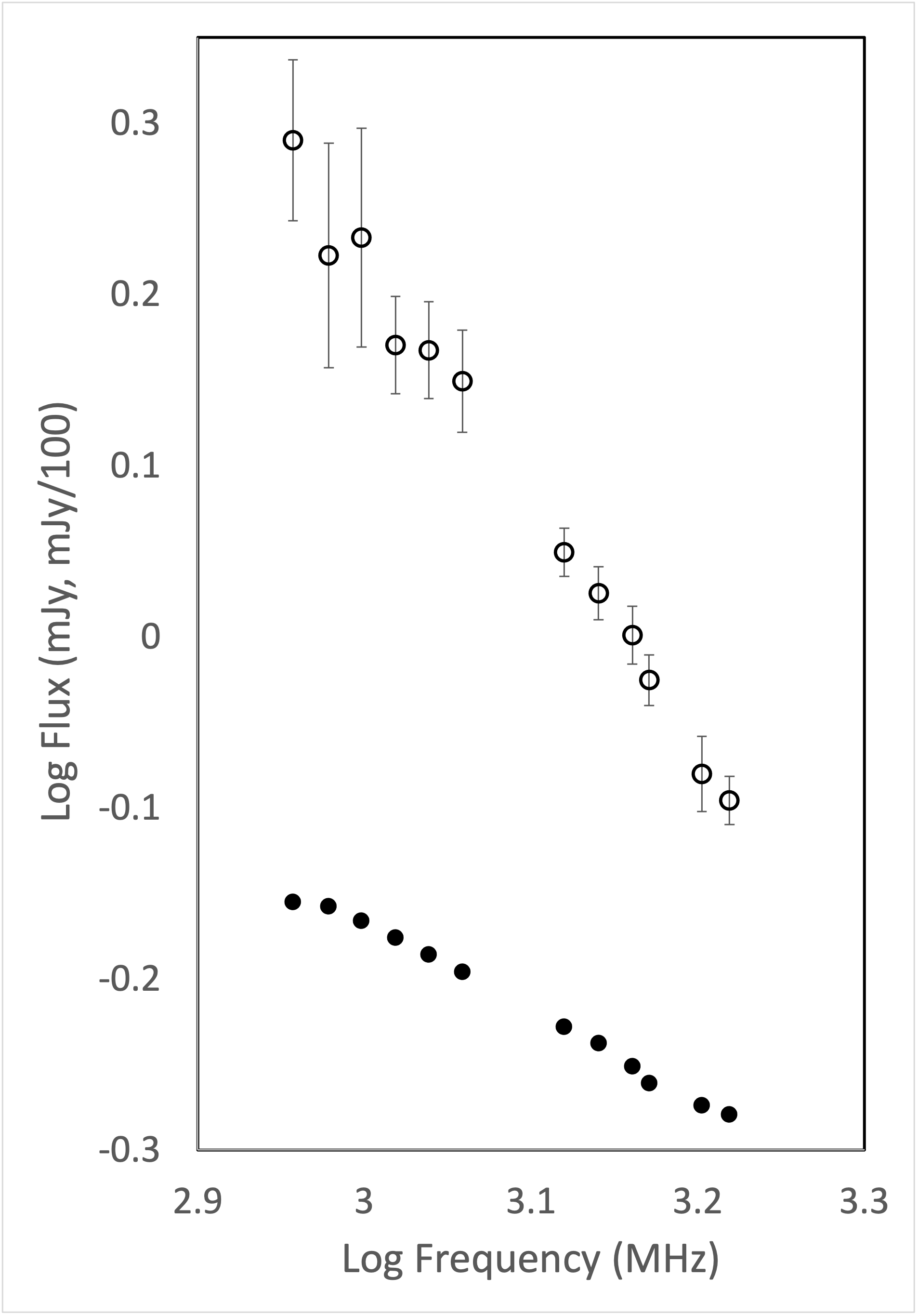}
    \caption{Plots of log(Intensity) vs. log(Frequency) for two illustrative positions to show the quality of the spectral information.  These are from maps convolved to 15\arcsec~ and represent the averages over small boxes on the filament (open circles) and at the first turn in the northern jet (point "A" in Fig. \ref{fig:RMsliceN}, filled circles) after dividing the fluxes by 100 for display purposes.  Statistical errors for the jet are smaller than the symbol size, although the non-linearities are probably due to residual bandpass calibration errors.}
\label{fig:specquality}    
\end{figure}

MeerKAT\footnote{Operated by the South African Radio Astronomy Observatory (SARAO).} observations were conducted in full polarization mode in the L band, 900--1670\,MHz, as described in \cite{MGCLS}. Telescope parameters are described in detail in \citet{Jonas2016} and \citet{Camilo2018}. The primary beam size at the band center of 1.28~GHz was 1.2$^\circ$; all of the observations presented here lie in the central 0.5$^\circ$. The  rms noise was 5.7$\mu$Jy/beam near the field center, with a nominal beam size of 7.69\arcsec$\times$7.55\arcsec~ at position angle 88\degree. 
After self-calibration, final images were produced and cleaned using the \textsc{Obit}\footnote{http://www.cv.nrao.edu/~bcotton/Obit.html} \citep{Cotton08} wide-band, wide-field imager \texttt{MFImage}, using robust weighting ($-1.5$), facets to correct for sky curvature, and a frequency dependent taper to maintain an approximately constant resolution over the $\sim$2:1 frequency range (see \cite{SourceSize}). The images were in the form of 14 frequency channel cubes, each with a 5\% fractional ($\Delta \nu/\nu$)
  bandpass. The quality of the data for calculating spectra are illustrated in Fig. \ref{fig:specquality}, showing one faint and one bright region. The spectral indices were calculated from a non-linear least-squares fit to $I(\nu)=I_0~\nu^{\alpha}$ for the 14 channels.  

\begin{figure}
    \centering

   \includegraphics[width=0.9\columnwidth]{ 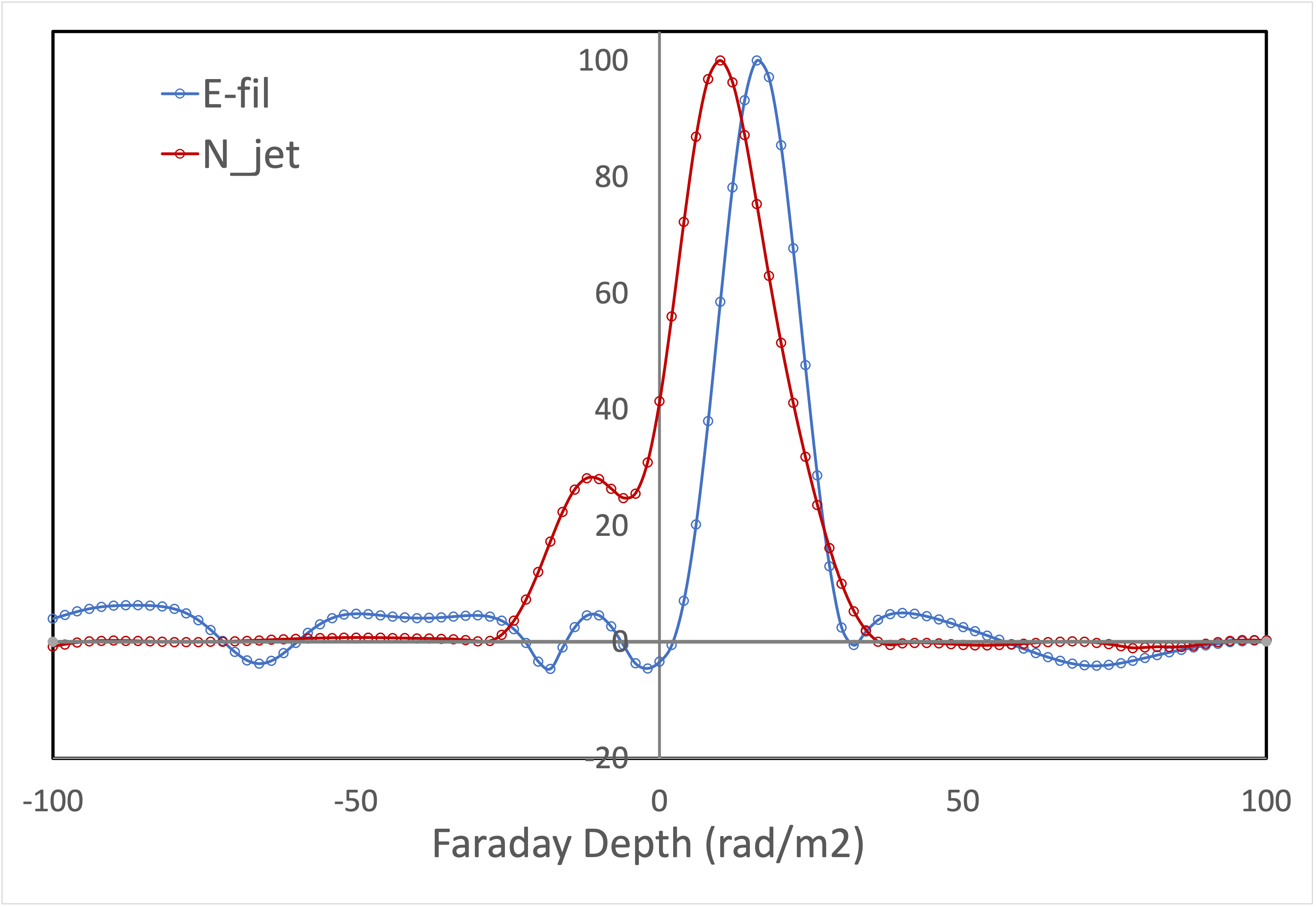}
    \caption{Faraday spectra, normalized to the same peak brightness, from the super-resolved Faraday cleaned spectrum for two illustrative locations, one on the southern filament of the \efil structure (01h26m13.76s, -01d13m37s, blue) and one at the location of an unresolved X-ray feature (see below) near the intersection of the northern jet with \efil (01h26m06.12s, -01d19m37s, red).  The filament is only marginally resolved spatially, with an observed width in Faraday space of 13.6 rad/m$^2$), while at the X-ray feature, there is Faraday structure over scales of $>$ 20 rad/m$^2$).}
    \label{fig:polquality}
\end{figure}

Polarization calibration and mapping was done as described in \cite{Condon21} and \cite{MGCLS}, using a series of steps to fix the polarization angle and rotation measure of 3C286 to  $-$33$^{\circ}$ and  zero, respectively \citep{2013ApJS..206...16P}. To better track the Faraday induced variations in Q and U, we divided the band into 68 frequency channels.  Examples of the  initial polarization results (polarized intensity, rotation measure and zero wavelength polarization angle) were derived from RM Synthesis (RMFit:Cube), using the peak value in the Faraday spectrum at each pixel. Unless noted otherwise, no normalization was done for the total intensity spectrum at each pixel; this does not affect the peak RM, as presented, but would broaden the observed Faraday spectrum.  Results are also presented from a new procedure (RMSyn) which achieves super-resolution in Faraday space by a factor of $\sim$3 \citep{Complex}.  The width of the restoring beam for the clean components is fixed by the real part of the dirty beam created using  $\lambda_0^2$=0 instead of the standard $\lambda_0^2 = <\lambda^2>$ averaged over the range of observations \citep{Brentjens}. 
Fig. \ref{fig:polquality} shows the Faraday spectra for two illustrative locations, restored with a Faraday beamwidth of 15.6~ \radmm; these are used to determine the peak rotation measure and the polarized intensity. 
We used these measurements to  characterize the fractional polarization, rotation measure and polarization position angle along the filaments, by sampling them at the same locations as the total intensity.  

\subsection{LOFAR}\label{sec:LOFAR}
3C\,40B was observed with LOFAR \citep{2013A&A...556A...2V} as part of the LOFAR Two-metre Sky Survey \citep[LoTSS][]{2017A&A...598A.104S,2019A&A...622A...1S,2022A&A...659A...1S} in pointing P020+01. Observations for this pointing were taken on 7 Jan and 16 May 2019, each observation having an integration time of 4 hrs and covering the 120-168 MHz range. The data reduction was performed in a standard way following the prefactor and Data Release 2 DDF-pipelines \citep{2021A&A...648A...1T}. These pipelines include flagging \citep{2012A&A...539A..95O}, calibration for direction independent \citep{2016MNRAS.460.2385W,2019A&A...622A...5D} and direction dependent effects \citep{2014A&A...566A.127T}, and applying the calibration solutions during imaging \citep{2018A&A...611A..87T}. The target visibilities were subsequently “extracted” by subtracting all sources in the field of view from the visibilities, apart from the 0.75 square degree region around 3C\,40B. The calibration for 3C\,40B was further optimized by carrying out a self-calibration procedure as described in \cite{2021A&A...651A.115V}, utilizing the WSClean imager \citep{2014MNRAS.444..606O} with multiscale deconvolution \citep{2017MNRAS.471..301O}.  The final LOFAR image is made with Briggs robust -0.5 weighting, (equivalent to $\approx -1.5$ in the AIPS/\textsc{Obit} implementations)  and has a r.m.s. noise of 0.42 mJy\,beam$^{-1}$, with a beam size $14.7\arcsec \times 6.5\arcsec$.

There is a large uncertainty in the LOFAR flux scales.  We thus used 11 compact sources in the field and measured their spectral indices $\alpha(144, 1283)$ between the LOFAR and MeerKAT maps at 15\arcsec~ resolution, and the $\alpha(900-1670)$ indices determined from fitting across the 15\arcsec~ MeerKAT frequency maps.  The compact sources were observed to scatter around the power law line, as expected for their spectra, with a correction of -0.085 to the low frequency spectral indices.   This represents a gain correction of $\sim$1.2 to the nominal LOFAR map values.  This correction was adopted for all further measurements. There were several other compact sources in the field with spectra flatter than -0.5; these were not used in the analysis because neither their likely true spectral shapes nor the effects of possible variability are known.

\subsection{X-rays}\label{sec:X-rays}
We employed archival X-ray data of Abell 194 from ESA's X-ray Multi-Mirror Mission (XMM-Newton, Obsid 0743700201, PI Farrell), taken in Jan 2015. This observation consisted of 5 exposures with the European Photon Imaging Camera (EPIC), one 130ks PN observation, and two MOS1 and two MOS2 observations with 87/50ks each. We processed the data with the XMM-Newton SAS software package version 19.0.0 and used the XMM-Newton-Extended Source Analysis Software \citep{2008A&A...478..615S} to produce exposure-corrected and background subtracted images. The lightcurve cleaning was performed with the mos-filter and pn-filter tools, resulting in a usable exposure time free of solar flares of 65ks for PN and 171ks for MOS. We followed the standard procedure described in the ESAS Cookbook\footnote{https://heasarc.gsfc.nasa.gov/docs/xmm/esas/cookbook/xmm-esas.html} and produced combined images in the 0.4-1.25~keV band, as well as in the 1.25-3~keV band. 
%\clearpage

\section{The Global View}\label{global}
We start by summarizing the key observational findings.  Readers with a primary interest in the physical interpretations may find it useful to skip to Section \ref{sec:discussion} and refer back here as needed.
\begin{itemize}
    \item {The \efil have narrow emission extending $\gtrsim$200~kpc and more diffuse emission extending $\sim$300~kpc from the 3C40B jet, with narrow components $\sim$3-8~kpc across and a broad component relatively constant at 12~kpc wide (FWHM). }
    \item {Minimum pressures in the \efil are $\sim$1-2 $\times$10$^{-12}$ (10$^{-11}$) erg~cm$^{-3}$, assuming proton/electron ratios of 1 (100). The local X-ray pressures are $\sim$10$^{-11}$ erg~cm\minus.}
     \item{The magnitude of the line-of-sight component of the ICM magnetic field, calculated using the Faraday structure,  is estimated to be 1.4$\mu$G near the bends in the northern jet, at a distance of $\sim$75~kpc from NGC~547; the  minimum pressure field in the jets themselves is 4 times higher.}
    \item{There is a $\sim$20\% deficit in X-ray surface brightness at the position of the \efilc consistent with an absence of X-ray emitting material in a $\sim$35~kpc cylinder encompassing the radio structure.}
    \item {The spectra of the \efil steepen mostly monotonically with distance from 3C40B, from about -1.3 to -2.5, likely reflecting the increasing dominance of the steep spectrum diffuse emission relative to the flatter spectrum narrow components.  The spectra show convex curvature, significantly smoother than exponentially cutoff spectra. The spectral shape of the \efil is indistinguishable, in the curved region, from the spectral shape of the emission from 3C40B.}
    \item {The \efil are $\sim$50\% polarized, with no detectable net RM with respect to the Galactic foreground, and only small rms variations ($9$~\radmm) along their length.  This enables a mapping between Faraday depth\footnote{\rev{A Faraday spectrum represents the brightness of the polarized emission at a given location as a function of ``Faraday depth" \citep{Brentjens}. In the simplest case, where emission is observed at only one Faraday depth, that depth is equivalent to the more commonly used term ``Rotation Measure (RM).}} and distance along the line-of-sight, revealing their 3D structure.}
    \item{The \efil are bent and wrap around the northern jet of 3C40B as it abruptly turns to the west and expands. The Faraday depths of the jet and \efil also coincide at this point,  indicating that this is a physical interaction in 3D, rather than a chance superposition along the line-of-sight.} 
    \item{There is an isolated unresolved X-ray feature of indeterminate origin at a sharp bend in the 3C40B jet, where it intersects with the \efilp}
\end{itemize}

\begin{figure*}[h]
    \centering
%    \includegraphics[width=0.33\textwidth]{ 3C40Bgreyscale.png}
%    \hspace{.35in}
%     \includegraphics[width=0.48\textwidth]{ A194_XR.png}
%  \includegraphics [width=0.49\textwidth]{ 3colorRM_nonCmpx_45,20,-5_enhance.png}%[width=\columnwidth]
     %\hspace{-0.45in}
%     \includegraphics[width=0.38\textwidth]{ 3C40spec_Jan22a.png}
%     \hspace{0.65in}
%     \includegraphics[width=0.3\textwidth]{ RMframescolorb.png}%[width=\columnwidth]
%     \hspace{.35in}
   \includegraphics[width=0.95\textwidth]{ 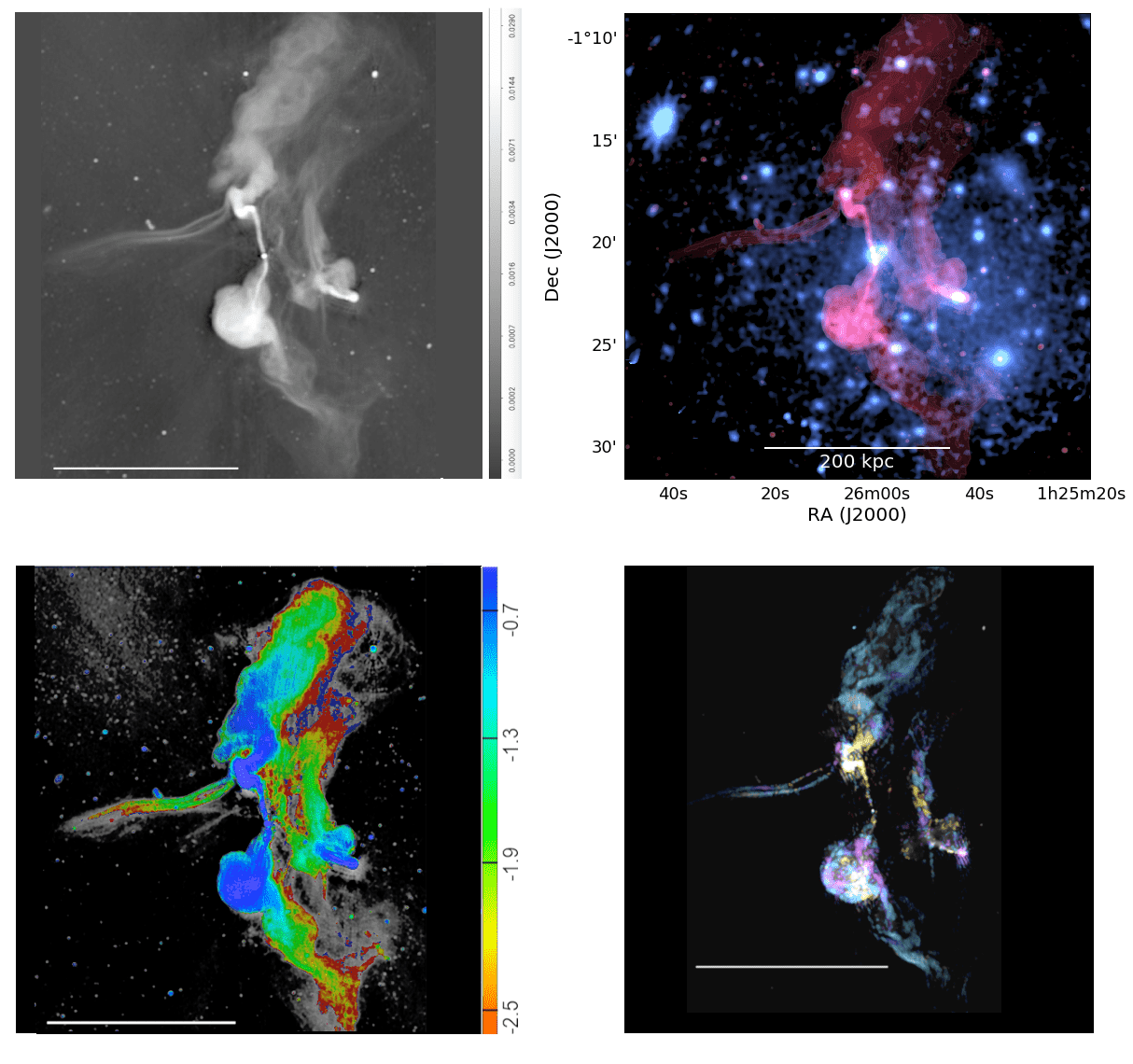}
    \caption{All radio images are at 7.75\arcsec~ resolution.  Top left: Total MeerKAT intensity image, showing the intensity scale.  Top Right: Overlay of MeerKAT total intensity image in rose  with XMM-Newton image in blue.  Bottom left: Spectral index image with regions in grey indicating the low signal:noise regions at this resolution, where no good spectral fits were achieved. Bottom right: Faraday depth composite, to illustrate the rich Faraday structure, color-coding the emission at a range of Faraday depths from  -14 to 50~ \radmm. Since emission from multiple Faraday depths contributes to each pixel, there is no simple translation from a single color to a single Faraday depth. Yellowish (blue, purple) colors tend to be dominated by emission at approximate depths of  -10 (10, 40 )~ \radmm. The images each have a bar indicating a length of 200~kpc (552\arcsec).}
     \label{fig:overview}
\end{figure*}

A global view of the data products used in this paper is given in Fig. \ref{fig:overview}.  The total intensity radio image from MeerKAT, (first presented in \citet{MGCLS}),  identifies the two central radio galaxies in Abell~194, 3C40B - the subject of this paper - and 3C40A.  These radio sources are  embedded in a rich network of filaments. Some of these, such as those going to the north from 3C40A, appear to be extensions of the radio galaxy;  others, especially between 3C40A and 3C40B are not clearly connected to either source.  The most prominent filaments are the \efilc seen extending over 200~kpc to the east from 3C40B, and studied in detail below.  An additional view of the filamentary structures, including ones found inside 3C40B's lobes, is presented in Appendix \ref{sec:Afilaments}.

The brighter central regions of the X-ray emission from XMM-Newton are also seen in Fig. \ref{fig:overview}, overlaid with the MeerKAT image. The bright patch at the southwest end of the diffuse X-ray emission is from a background cluster at z=0.15 \citep{Mahdavi}.  A pointed ROSAT PSPC image \citep{Bogdan}, shows that the X-ray emission extends for $\sim 37.5\arcmin$ ($\sim$800~kpc) along a NE-SW axis, far beyond that visible in this image.  \citet{Bogdan} also detected an X-ray cavity associated with the southern lobe of 3C40B.

Fig. \ref{fig:overview} also shows the in-band spectral indices of the radio galaxies.  The bright regions of 3C40B north have a spectral index of $\sim$-0.5, and show no characteristic steepening with increasing distance from the core until there is a significant drop in brightness at a distance of $\sim$100~kpc. In the south, the -0.5 spectral regions extend for a similar distance from the core, although they are embedded in slightly steeper emission in the southern lobe.  The spectra continue to steepen further out, although the S/N at the 7.75\arcsec~ resolution shown here is too low to produce reliable indices for the fainter filamentary structures.  A more detailed discussion of spectral trends in the \efil is presented below.

\begin{figure}[h]
    \centering
    \includegraphics[width=0.7\columnwidth]{ 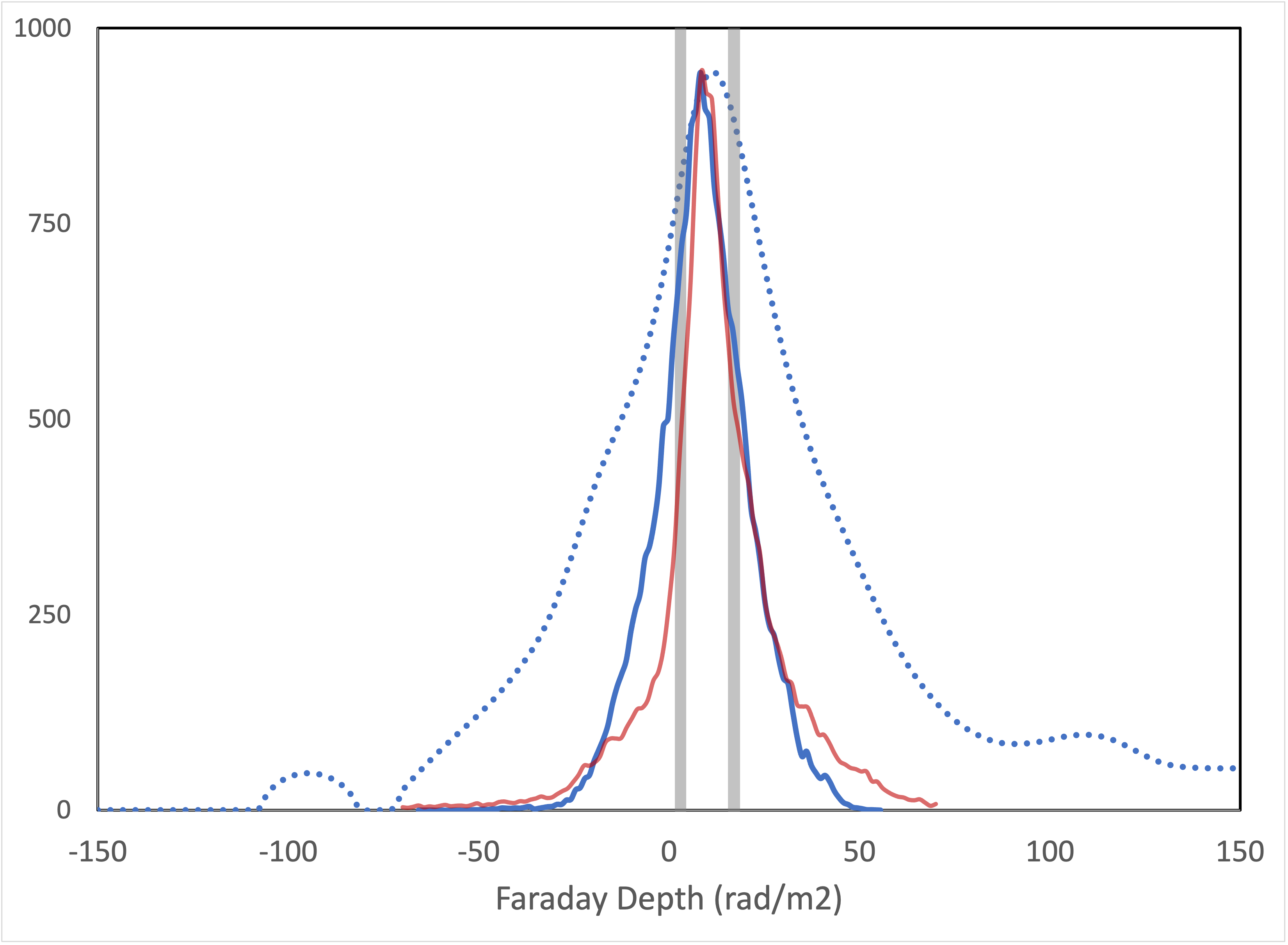}  
    \caption{Distribution of Faraday depths in the central regions of Abell~194, encompassing 3C40B, 3C40A, and the surrounding filamentary features.  Red -  peak RMs from the standard Faraday synthesis. Blue -  the first moment in Faraday depth space of the amplitude of the super-resolved Faraday spectrum in each spatial pixel.  Both of these used only polarized intensities $\gtrsim$150$\mu$Jy/beam. Blue dashed -  sum of power in each Faraday depth channel, with a noise contribution subtracted in quadrature. This includes all the polarized emission, on and off the peak in the spectrum. The approximate range of Faraday depths produced by the foreground Milky Way is shown in grey. } 
 \label{fig:RMhist} 
\end{figure}

The final data product previewed in Fig. \ref{fig:overview} illustrates the rich Faraday structure available for this system.  Unlike other images in the literature, this is \emph{not} a Faraday rotation measure (RM) image, although it is closely related.  Instead, Faraday synthesis often reveals a distribution of Faraday depths, as opposed to a single RM, along individual lines of sight.\footnote{Where the Faraday spectrum is dominated by a component at a single Faraday depth, this depth is designated as the ``Rotation Measure" (RM).}   The colors here result from various combinations of emission at different Faraday depths, and simply illustrate the complex structures. Fig. \ref{fig:RMhist} shows the distribution of the peak RMs (which have a spread of $\sim$9~\radmm), and the fuller picture of the Faraday depth distribution (with a spread of $\sim$34~\radmm) from summing the power in the image at each depth.  % In the following sections, we will present the Faraday structure more quantitatively, including illustrations of the dominant depth (RM) as appropriate.

%\clearpage

%When Faraday synthesis can be performed, a fuller picture of how much emission is present at each Faraday depth can also be obtained by summing the power in each Faraday depth channel.  For Abell~194, this is also shown in Fig, \ref{fig:RMhist}  and has an equivalent rms scatter of 34~rad/m$^2$.
\subsection{The Diffuse Emission}\label{sec:diffuse}
\begin{figure}[h]
\centering
\includegraphics[width=0.9\columnwidth]{ 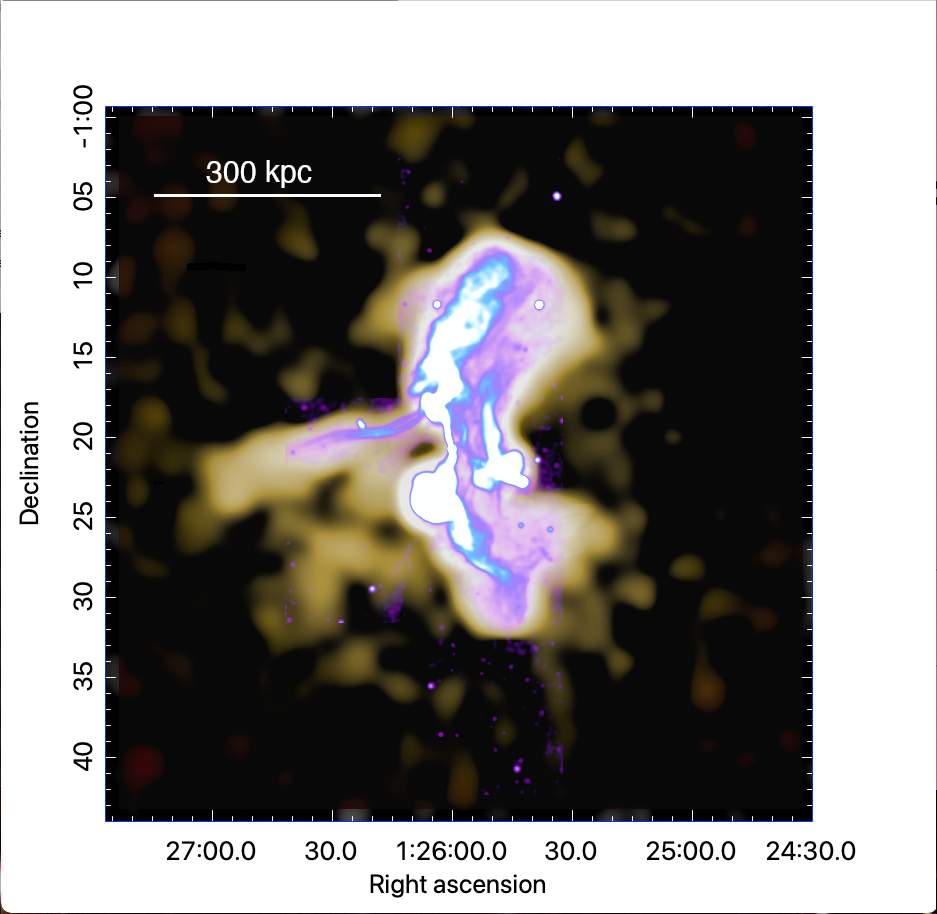}
\caption{Overlay of 15\arcsec~ MeerKAT image (purple) onto diffuse LOFAR emission (yellow) at 75\arcsec~ resolution, as described in the text, showing the 300~kpc extent of the diffuse component of the \efilp}
\label{fig:diffuse}
\end{figure}

In order to probe the lowest brightness emission in Abell~194, we filtered the LOFAR image using the multiresolution filtering technique of \cite{Rudnick02}, with a box size of 76.5\arcsec, and then convolved it with a 75\arcsec~ beam.  The results are shown in Fig. \ref{fig:diffuse}.  Note that this diffuse emission represents only emission on scales $\gtrsim$75\arcsec; it is \emph{not} a low resolution version of the total intensity image, in which the convolved finer scale emission would completely dominate the brightness.  The diffuse component of the \efil is seen to extend to $\sim$850\arcsec~ (300~kpc) from 3C40B's jet.  Near the end its brightness is $\sim$6.7$\mu$Jy/(\arcsec)$^2$. The bright portions of the radio galaxies 3C40A and B and their surrounding filamentary structures are all embedded in a low brightness region of $\sim$60$\mu$Jy/(\arcsec)$^2$. %This structure is clearest to the north of the host galaxies, and is bordered on the east by emission surrounding 3C40B's northern jet, and on the west by emission surrounding the filaments that extend north from 3C40A. 
Faint emission covering a region $\sim$200~kpc across can also be seen south of the \efilc at a level of $\sim$1.6$\mu$Jy/(\arcsec)$^2$, with no obvious association to other radio structures.

\section{The East Filaments}\label{efilobs}

\subsection{Structure}\label{sec:structure} The \efil consist of a pair of narrow, curved, mostly parallel structures embedded in a more diffuse background which traces  a similar locus. The filaments' total extent is $\sim$600\arcsec~ (220~kpc), along their curved path, with a very diffuse extension to 300~kpc shown above.  Their total flux at 1283~MHz is $\sim80$~mJy, corresponding to a monochromatic luminosity of 10$^{19}$ W/Hz.  Additional low surface brightness regions and faint intersecting filaments at different angles are also present in this region; these are fainter than seen in the maps presented here and are not obviously connected to the \efilp The brightnesses, and thus emissivities of the \efil are %comparable to those in the faint parts of 3C40B's lobes, and 
a factor of 10$^4$ below those of the 3C40B jets.%; in many places, the jets and lobe filaments are only slightly resolved in width.

\begin{figure}[h]
    \centering
    \includegraphics[width=0.9\columnwidth]{ 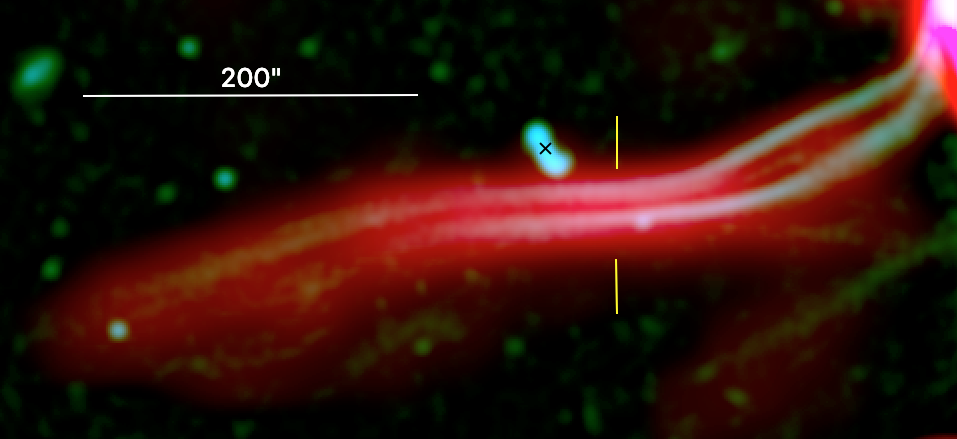} 
    \includegraphics[width=0.9\columnwidth]{ 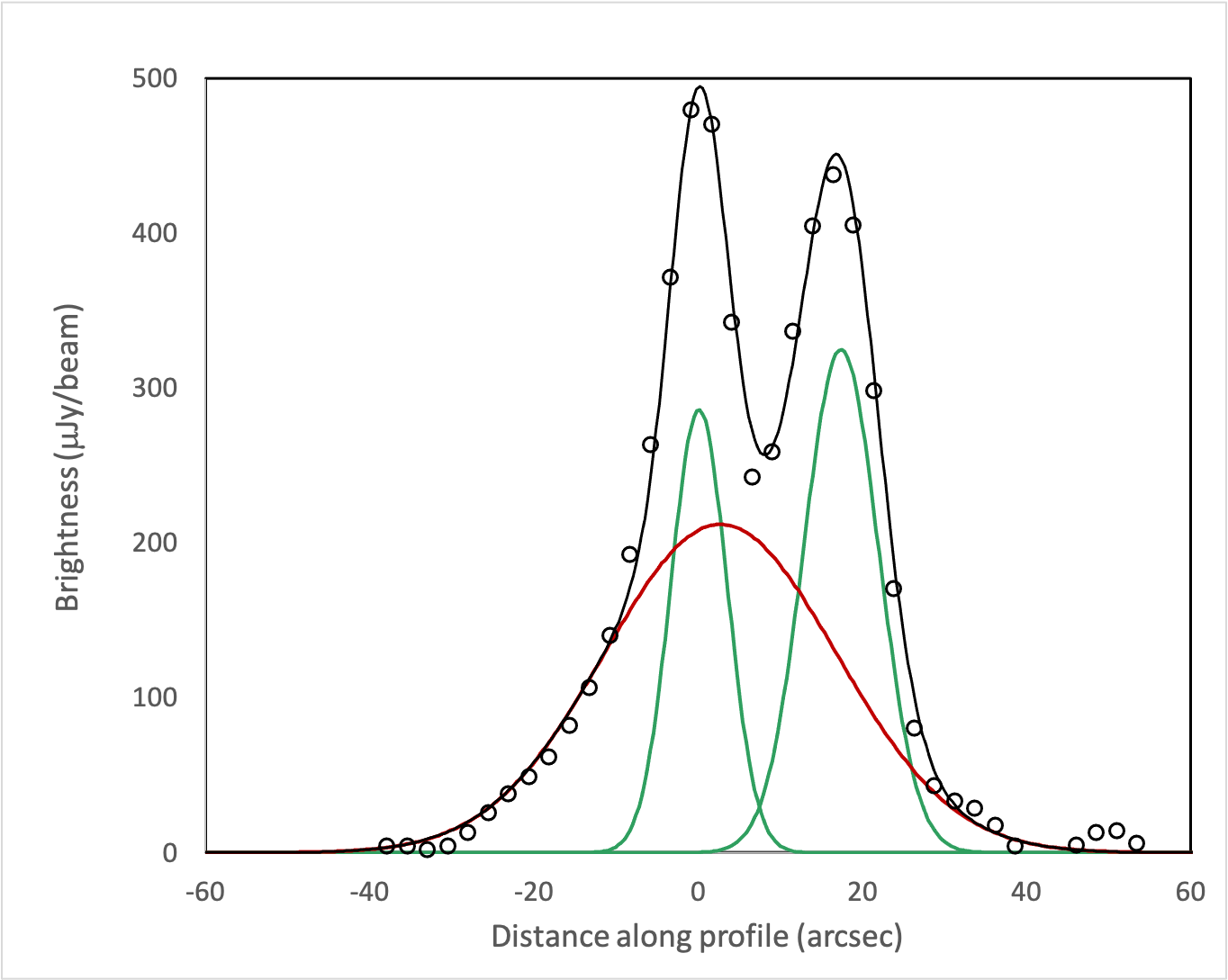}
    \caption{Top: The eastern filaments separated by color into narrow and broad components, as described in the text. The black X marks the location of an unrelated background radio double, at RA 01:26:23.41, Dec -1:19:10. Bottom: Profile across the eastern filaments as indicated by the broken vertical yellow line. The data are shown as open circles with errors on the same order as the symbol size.  The solid curves show a three Gaussian fit, with two green narrow components, one red broad component, and the total fit shown in black. }
  \label{fig:structure1} 
\end{figure}

The \efil are at least slightly resolved in most places in our 7.75\arcsec~ maps and are a combination of narrow and broad features which can be seen in Fig. \ref{fig:structure1}. For this figure,  the separation between narrow and broad features was done using the multiresolution filtering method of \cite{Rudnick02} using a box size of 30\arcsec. The broader features are enhanced in the color figure, to increase their visibility. To illustrate the actual typical relative brightness contributions of these components, Fig. \ref{fig:structure1} shows a north-south profile cut near the midpoint of the filament structure, along with a three-gaussian fit.  The deconvolved narrow component widths are 0.85$\pm$0.2~kpc (south) and 2.8$\pm$0.1~kpc (north) while the broad component width is 12$\pm$0.2~kpc.  Across this strip, the brightnesses of the narrow and broad components are comparable, as well as their total fluxes of 750$\mu$Jy and  910$\mu$Jy, respectively, from integrating along the 7.75\arcsec~ strip.  The relative contributions of narrow and broad features changes along the filaments; closer to 3C40B the narrow components dominate, while the broad component dominates further to the east (Fig. \ref{fig:Broad_Narrow}). Trends of brightness and other parameters along the \efil are presented below.

\begin{figure}
    \centering
    \includegraphics[width=0.9\columnwidth]{ 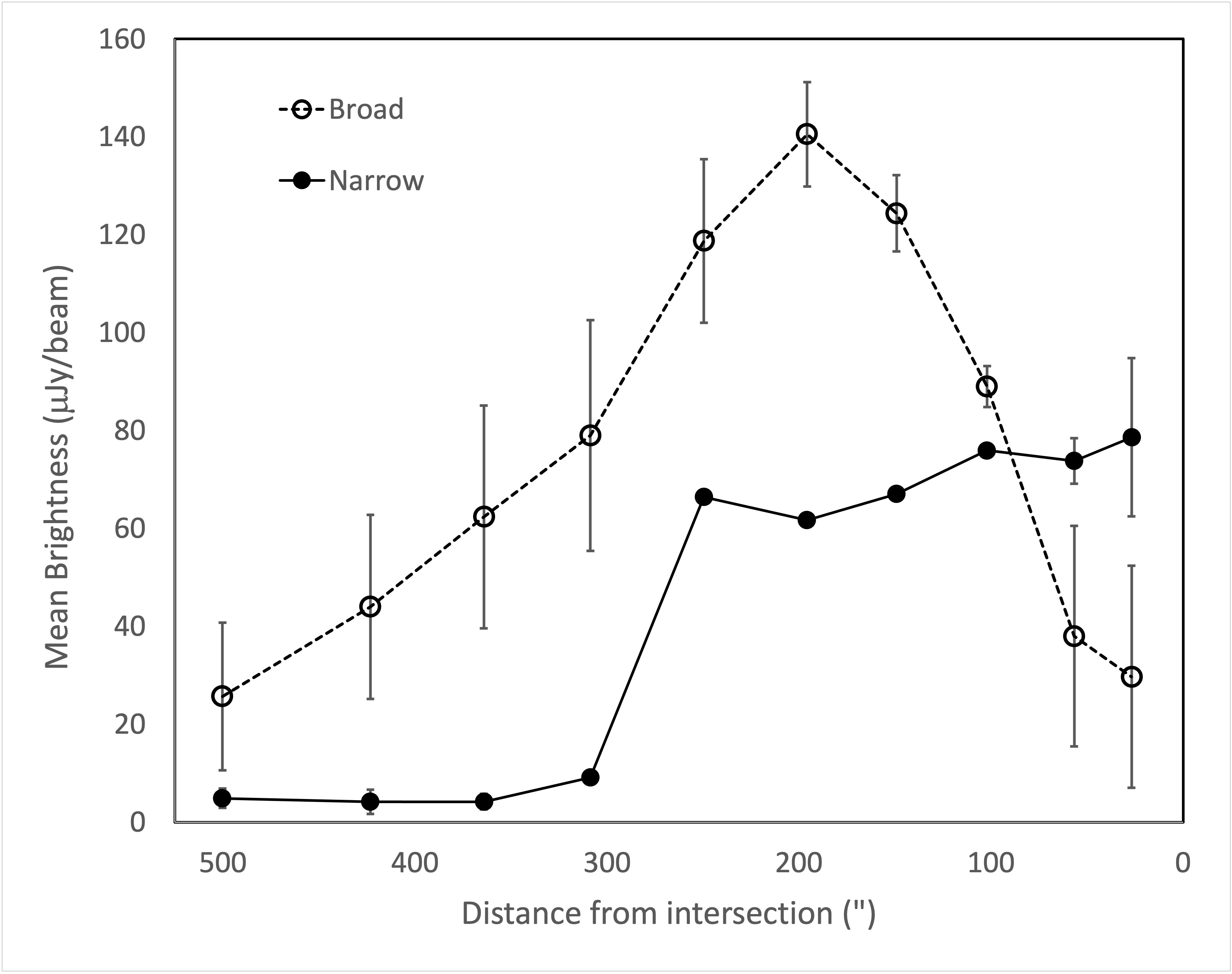} 
    \caption{Average brightness in boxes of the broad and narrow emission as a function of position along \efilp The 7.75\arcsec~ maps were decomposed into broad and narrow components using the multiresolution filtering technique of \citet{Rudnick02} with a (somewhat arbitrary) box size of 30\arcsec~(11~kpc). The details, but not the overall trend, would change with a different choice of box size.}
\label{fig:Broad_Narrow}   
\end{figure}

\subsection{Spectra}\label{sec:spectra}

To characterize the intensity, spectral and polarization behavior along the \efilc we calculated the mean of the respective quantities in small boxes (1-2 beam areas) positioned along the ridge line of the N and S \efil in the 7.75\arcsec~ maps.   

\begin{figure}
    \centering
  \includegraphics[width=0.85\columnwidth]{ 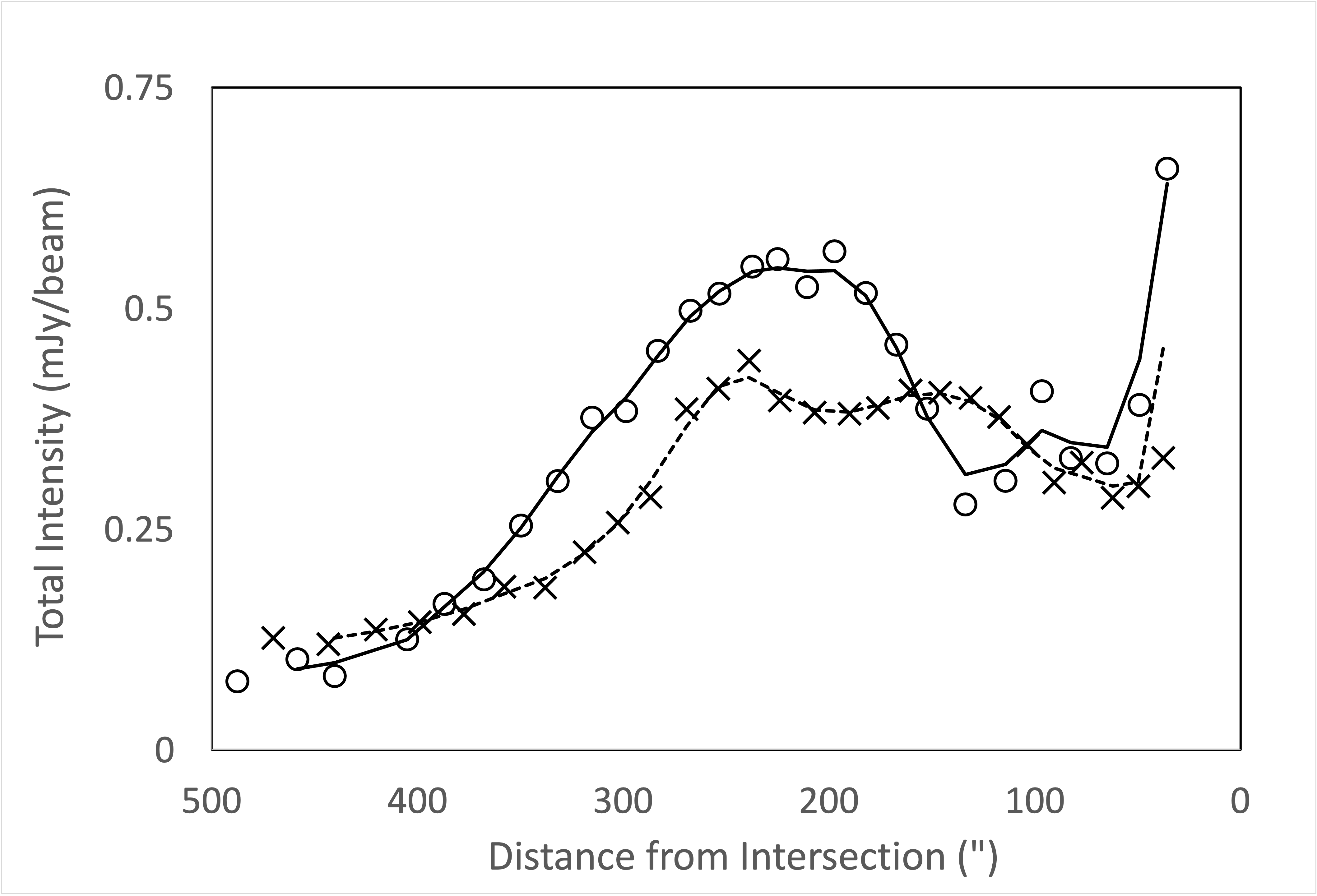}
  \includegraphics[width=0.85\columnwidth]{ 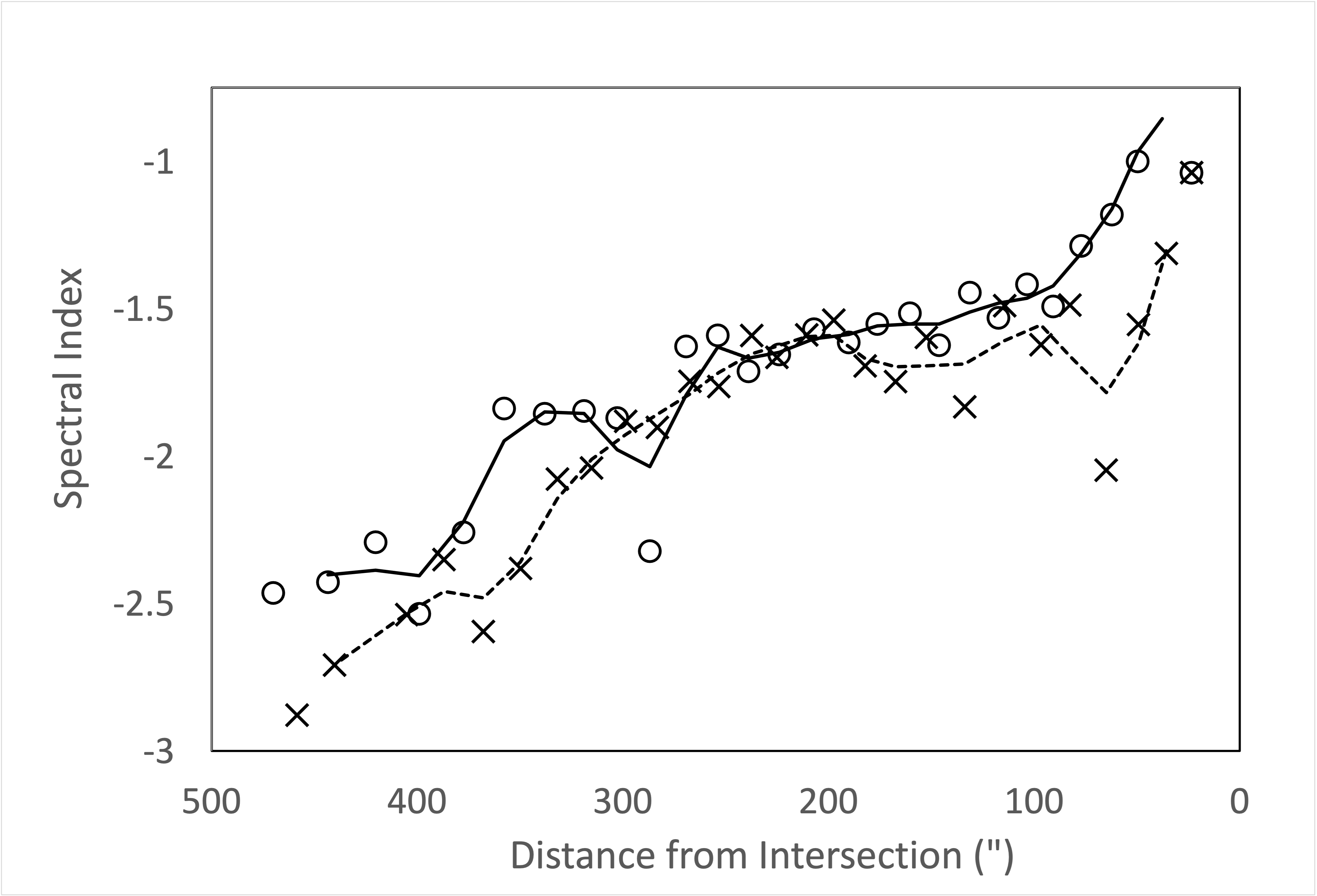}
    \caption{Intensity (top) and MGCLS in-band spectral index (bottom) as a function of distance from the intersection of \efil with 3C40B, highlighting the near-monotonic spectral steepening with distance.  Values are from the means in small boxes along the ridge lines of the north (circles, solid line) and south filaments (Xs, dotted line).  Lines are hanning smoothed.}
    \label{fig:filI}  
\end{figure}

The peak brightnesses of the \efil occur in a broad region $\sim$200\arcsec~ from where they intersect with 3C40B (Fig. \ref{fig:filI}) while the spectra steepen mostly monotonically with distance from 3C40B.  The steepening spectra could result from four different causes:  a) A decrease in particle energies due to radiative and/or adiabatic losses as a function of distance from 3C40B (e.g., if the electrons were streaming from 3C40B); b) A decrease in the magnetic field strength away from 3C40B, so that for a fixed electron energy distribution falling off at high energies (assuming no particle acceleration), higher energy electrons (and thus steeper spectra) would be sampled at larger distances; and c) A decrease in the relative strength of the narrow and broad components of \efilc as discussed below; and d)  Changes in the local relativistic particle acceleration processes as a function of distance away from 3C40B.%; this could then lead to steeper power laws (for first order processes), or more curved spectra.    

\begin{figure}
    \centering
    \includegraphics[width=0.5\columnwidth]{ 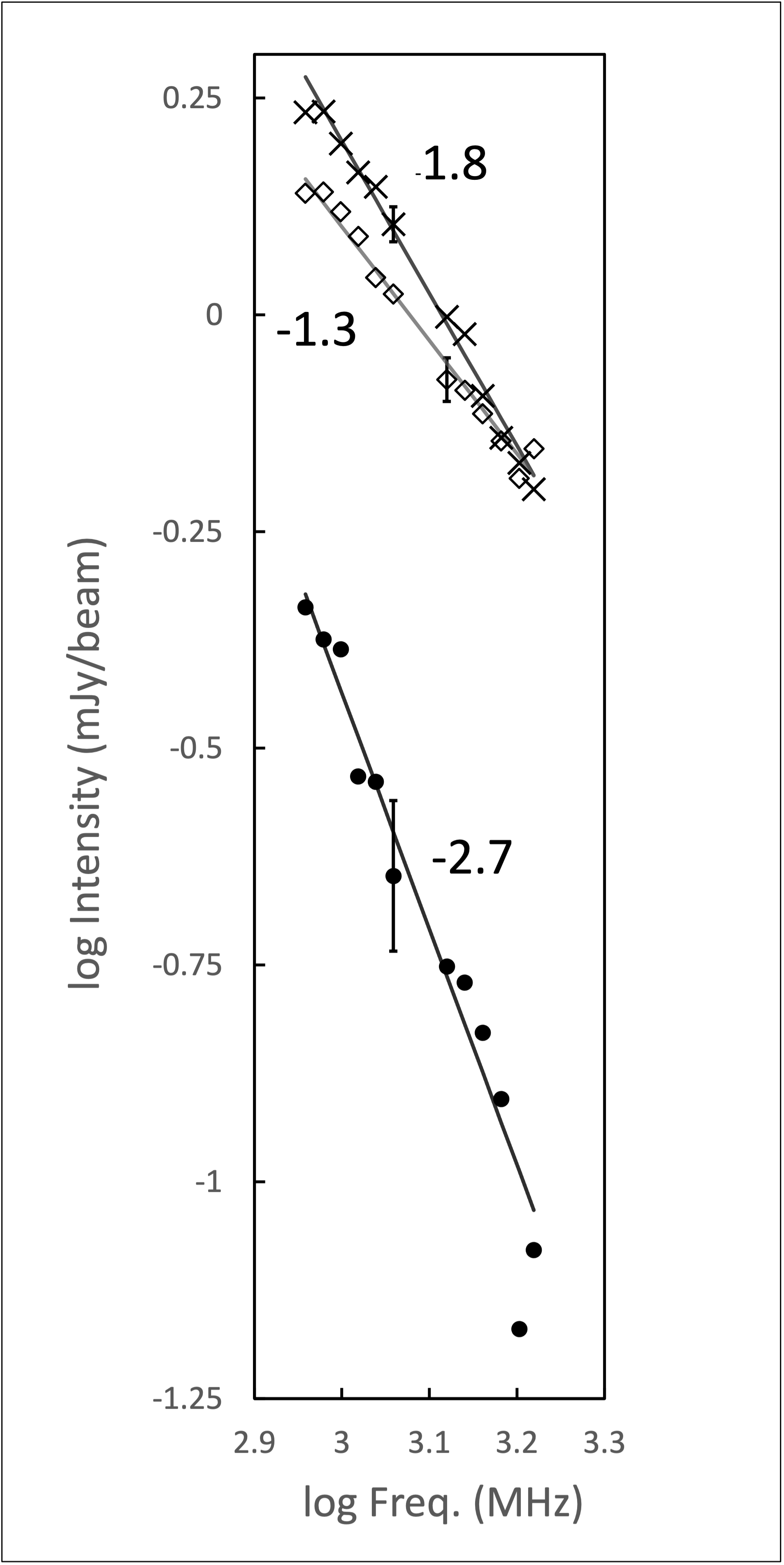} 
    \caption{Spectra of the north (X's), south (diamonds) and broad (filled circle) components from a transverse cut across \efil after convolving each frequency channel by 35\arcsec~  along the structure. Typical errors are shown. }
  \label{fig:specnarrowbroad} 
\end{figure}

 It would be very useful to separate the spectral indices as a function of position separately for the narrow and diffuse components of \efilp   However, at full resolution, there was insufficient signal:noise to measure the low brightness emission, and at lower resolution, there is too much blending between the components. To increase the signal:noise at full resolution, we convolved the 15\arcsec~ maps by 35\arcsec~ along \efilc and made transverse profiles at each frequency. At one position both the narrow and broad emission were strong enough to determine their spectra simultaneously, and the results are shown in Fig. \ref{fig:specnarrowbroad}. The narrow filaments had spectra of -1.3$\pm$0.07 (north), -1.8$\pm$0.08 (south) and the broad component -2.7$\pm$0.2.   These values are very similar to those shown in Fig. \ref{fig:filI} where each of these components dominate the emission.  It is plausible, and even likely, that the spectral steepening as a function of position along the jet is due to the changing relative dominance of the diffuse and fine-scale emission, as shown in Fig. \ref{fig:Broad_Narrow}
 
\begin{figure}
    \centering
    \includegraphics[width=0.9\columnwidth]{ 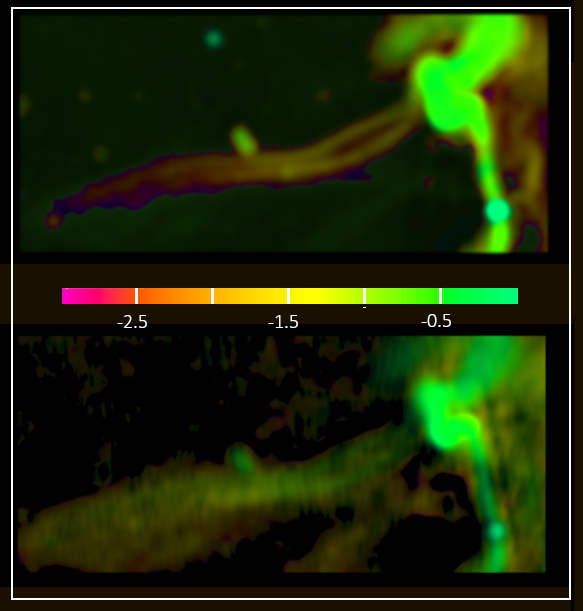}   
    \caption{Total intensity maps at 15\arcsec~ resolution, color coded by spectral index.  Top: MeerKAT image with $\alpha_{900-1670}$. Bottom: LOFAR image with $\alpha_{144,1283}$.  The color scales are the same, so the low frequency indices are flatter, as expected for convex spectral shapes.}
 \label{fig:alphas15}
\end{figure}

%Some constraints on the gain and loss processes can be obtained by studying the shape of the relativistic electron population in the filaments. 
To characterize the shape of the relativistic electron population, we used measurements at 15\arcsec~ resolution of the spectral indices between  960~MHz and 1670~MHz ($\alpha_{900-1670}$\footnote{Values of $\alpha_{900-1670}$ at 15\arcsec~ resolution are  available as a MGCLS data product, derived from fitting the 15\arcsec~ total intensity images as a function of frequency across the MeerKAT GCLS band, as described in \citet{MGCLS}.}) and between 1283~MHz (the central MGCLS frequency) and the LOFAR map at 144~MHz, convolved to 15\arcsec. ($\alpha_{144,1283}$).   Both of these spectral index maps are shown in Fig. \ref{fig:alphas15} for the region containing the \efil and the bright northern region of 3C40B.

\begin{figure}
    \centering
        \includegraphics[width=0.75\columnwidth]{ 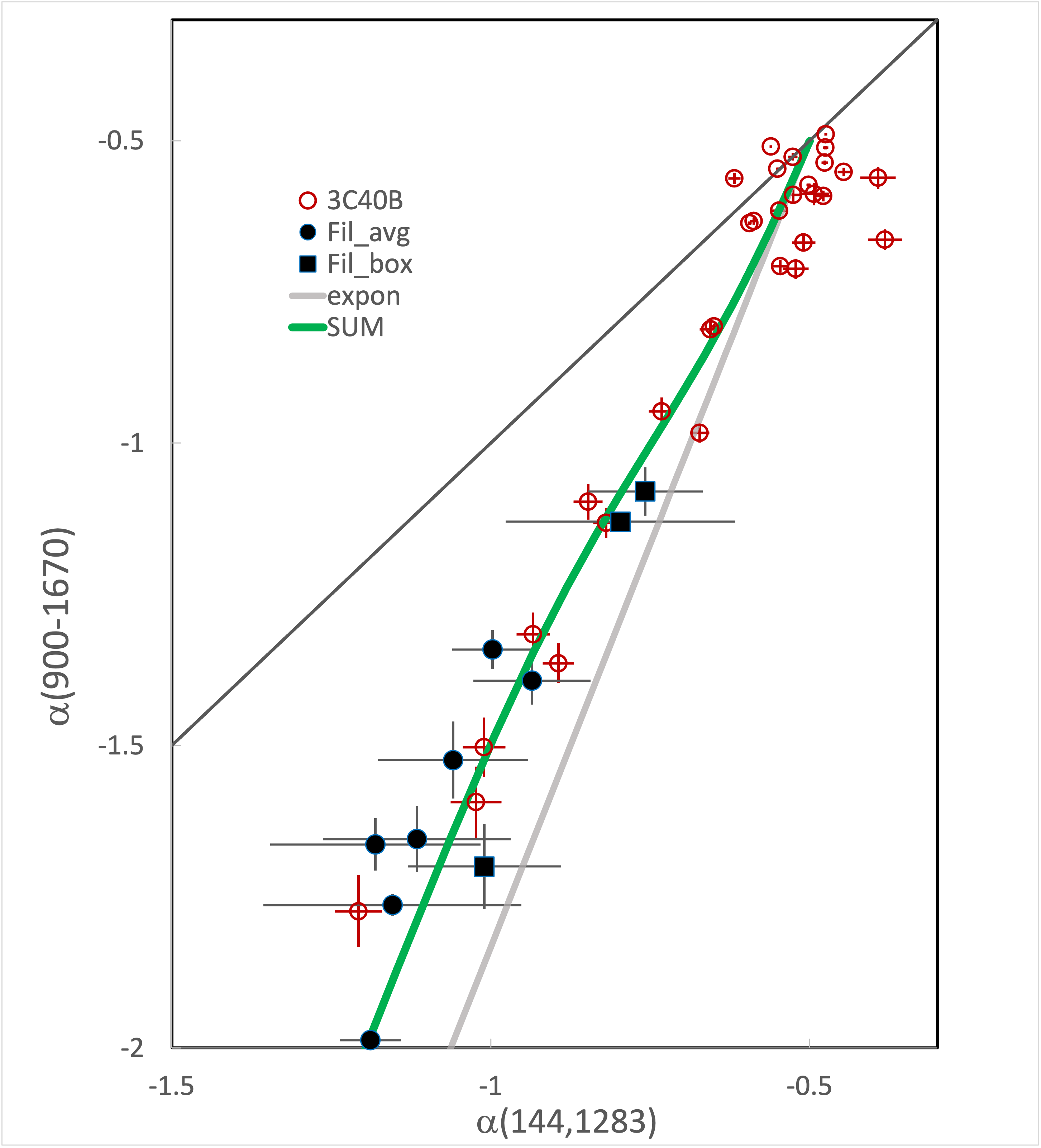} 
    \caption{Color-color diagram for 3C40B (red) and \efil (black symbols, see description in text). For comparison purposes we show the line corresponding to power laws with different slopes (black), an exponentially cut-off spectrum with low frequency index of -0.5 (grey) and \rev{an illustrative sum of two cut-off spectra with different cutoff frequencies in solid green}, as described in the text. See also Fig. \ref{fig:expmods}.  
      }
 \label{fig:cc}  
\end{figure}

We used the color-color diagram in Fig. \ref{fig:cc} \citep{color-color}, plotting  $\alpha_{900-1670}$ vs. $\alpha_{144,1283}$, to examine the shape of the underlying spectra, using the correction to the LOFAR flux scale described earlier.  If the shape of the underlying population of electrons is the same throughout a source, then there is a unique mapping between the locus of points from different positions in the source and the shape of the spectrum in $log(I) vs. log(\nu)$ space.   Changes in local magnetic field strength shift the observed spectrum in $log(I) vs. log(\nu)$ space, but preserve the locus in color-color space.  The same is true for adiabatic changes to the electron energy distribution. The locus is unchanged even in the presence of radiative losses for idealized energy distributions \citep{color-color}.  The locus of points occupied in the color-color diagram thus can be used to determine the shape of the underlying electron population. However, this can become difficult to interpret when each line-of-sight contains regions with different magnetic field strengths, or  different particle gain/loss processes \citep[e.g.,][]{sausagespec}.  

The results of the color-color analysis are shown in Fig. \ref{fig:cc}, for both 3C40B and the eastern filaments. For 3C40B, the spectra were averaged over boxes with sizes which increased at low brightness levels to increase the signal:noise.  For the relatively faint filaments, we calculated the spectra over large areas ($\sim$50 beams) in two different ways, to check their robustness. Filled circles represent noise-weighted spectra averaged over locally brighter regions;  filled squares denote spectra re-calculated by summing the fluxes in the individual MGCLS frequency channels and the LOFAR image. Their results are consistent within the errors.

For 3C40B, the flattest points scatter in a narrow region near the power-law line, suggesting a low frequency spectral index of  $\approx-0.5$.   In regions where the spectra are curved, the shape of the spectrum from 3C40B is indistinguishable from that of the \efilp

\begin{figure}
    \centering
        \includegraphics[width=0.48\columnwidth]{ 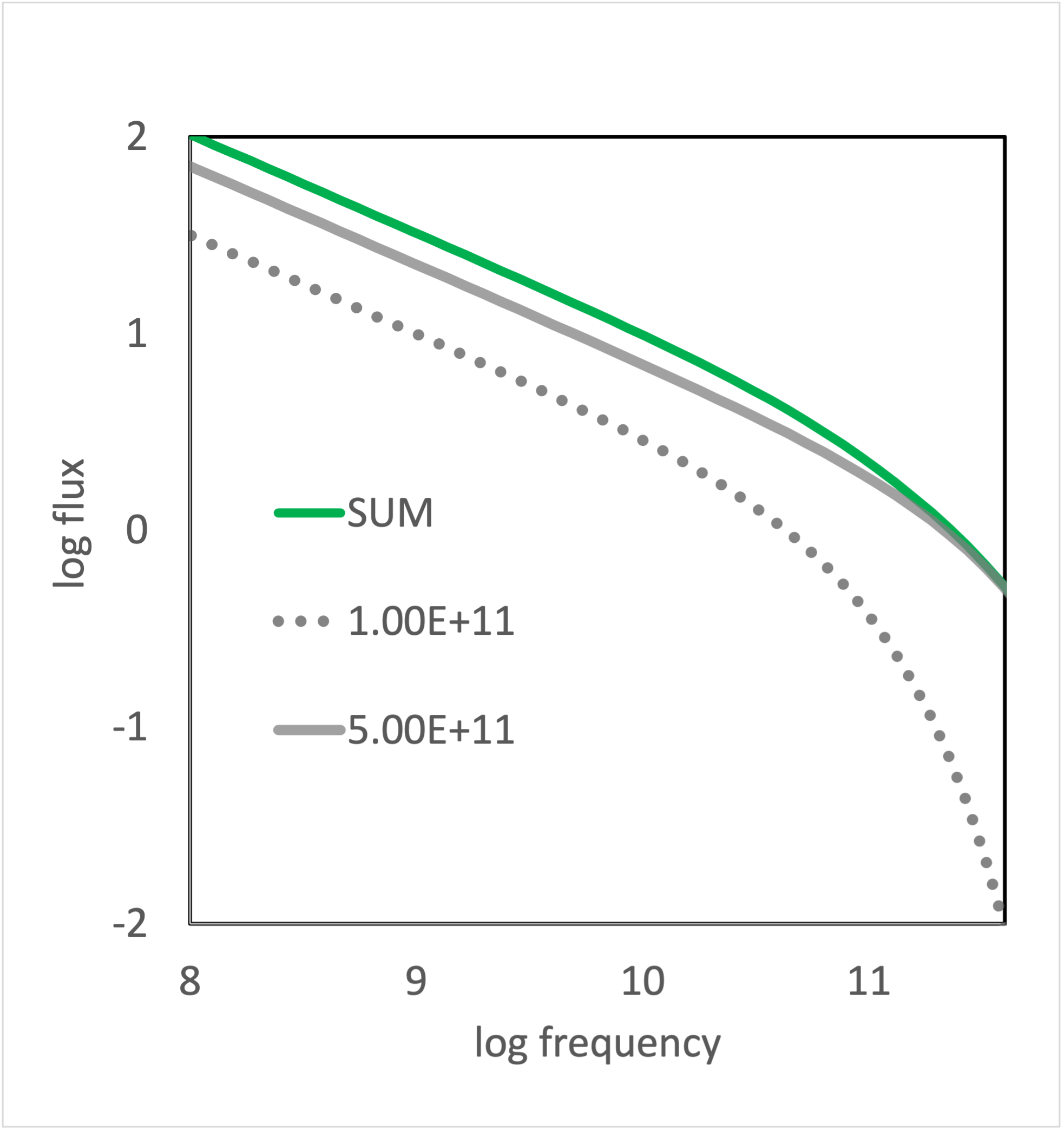}
    \includegraphics[width=0.48\columnwidth]{ 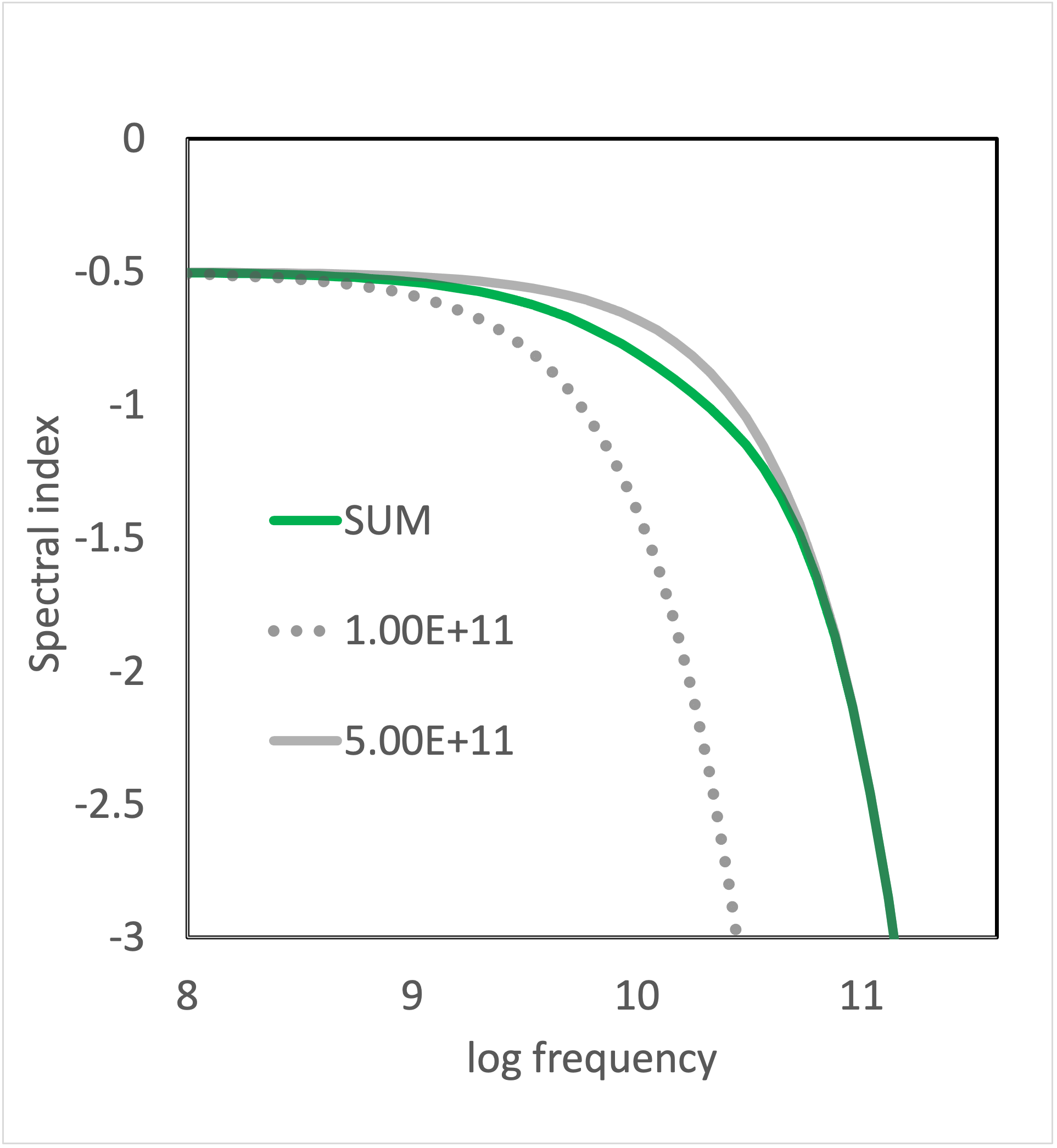}
    \caption{Spectra of two exponentially cut-off spectra in grey (each labelled with their cutoff frequency) and their sum, shown as green line here and in Fig. \ref{fig:cc}.  Left: log flux \emph{vs} log frequency.  Right: Spectral index as a function of log frequency.  Note that each of the two grey curves maps to the \emph{same} gray line in the color-color diagram, while their average maps to a different locus (green line in both plots).}
  \label{fig:expmods}  
\end{figure}
For comparison, we show three illustrative idealized spectral shapes, each with a low frequency index of $-0.5$. The spectra are shown in Fig. \ref{fig:expmods}, while their color-color mapping is shown in Fig. \ref{fig:cc}.   The grey line represents an exponentially cutoff spectrum $S(\nu)=S_0~\left ( \frac{\nu}{\nu_c}\right )^{\alpha}\times e^{-\left (\frac{\nu}{\nu_c}\right )}$, an approximation to the idealized spectrum of \cite{1973A&A....26..423J}. This would result if the cutoff frequency were changing as a function of position in the source due to magnetic field changes, adiabatic changes, or radiative losses, but where the underlying spectral shape (relativistic electron distribution) remains the same.  This idealized homogeneous spectrum falls off more rapidly than the data.   In addition, we can rule out different power laws at different locations in the filaments, which would trace out the line of all power laws. Different power laws would have been  expected in some scenarios where particle acceleration varied with distance from 3C40B.

There are many ways to broaden a spectrum, to move the locus in color-color space into the region between exponentially cut-off spectra and power laws, as is observed. We know, e.g., that there is more than one single spectral component along each line-of-sight, from the different spectra of the narrow and broad components (Fig. \ref{fig:specnarrowbroad}.  The green line in Figures \ref{fig:cc} and \ref{fig:expmods} represents the sum, at each position in the sky, of two  exponentially cutoff spectra with cutoff frequencies that differ by a factor of 5. Then, at a fixed observing frequency, if the magnetic field were changing, e.g., the observed spectra would trace out the solid green locus in Fig. \ref{fig:cc}.  Alternatively, if the spectra do not shift with position along \efilc but the spectrum with the high cutoff dominates the sum near the 3C40B jet, and the spectrum with the low cutoff dominates further away, then this would follow a similar locus.

The green line accurately represents the spectral shape of the data. However, there are other ways to create this same shape, as well. Instead of two exponentially cutoff populations with different cutoff frequencies along the line-of-sight, e.g., there could be a single broader convex distribution resulting from particle acceleration. There could also be  electron populations with different shapes at different positions in the source, embedded in their own respective magnetic field strength regions; these can still create what appears to be a simple locus in color-color space.  In the discussion below, we do not attempt to identify a unique electron energy distribution for the jet and \efil; we instead present physical scenarios that are compatible with the observations.   
\subsection{Pressure Estimation}\label{sec:pressure}
Using the above values for the brightness and diameters of the filaments, we can obtain a rough estimate of their emissivities; combining these with the spectral index, we obtain an estimate of the minimum combined pressures of the cosmic rays and magnetic field \citep{1997A&A...325..898B}.
  To calculate a first order, fiducial number, we assumed that the structures have a uniform emissivity (filling factor=1), that the electron energies extend down to a somewhat arbitrary value $\sim$300~MeV,  and that the ratio of relativistic protons to electrons is 1 at a given energy.  Using a position midway along the filaments, we made estimates for both the total emission and the individual filaments, used a spectral index of -1.6, and derived minimum pressures (=1/3 minimum energy density) of 1-2$\times 10^{-12}$~erg~cm$^{-3}$, corresponding to magnetic field values of $\sim$7~$\mu$G. The actual pressures are likely to be higher, even by an order of magnitude or more, if the ratio of relativistic proton to electron ratio is 100, if the filling factor is smaller, if the electron distribution extends down to, e.g., 100~MeV, or if the ratio of magnetic field to cosmic ray energy densities doesn't match these minimum conditions.   These pressure estimates for the fields and relativistic particles in \efil will be used below to compare with those in the ambient medium, calculated from the X-rays.

\subsection{Polarization}\label{sec:polar}
\begin{figure}
    \centering
    \includegraphics[width=0.95\columnwidth]{ 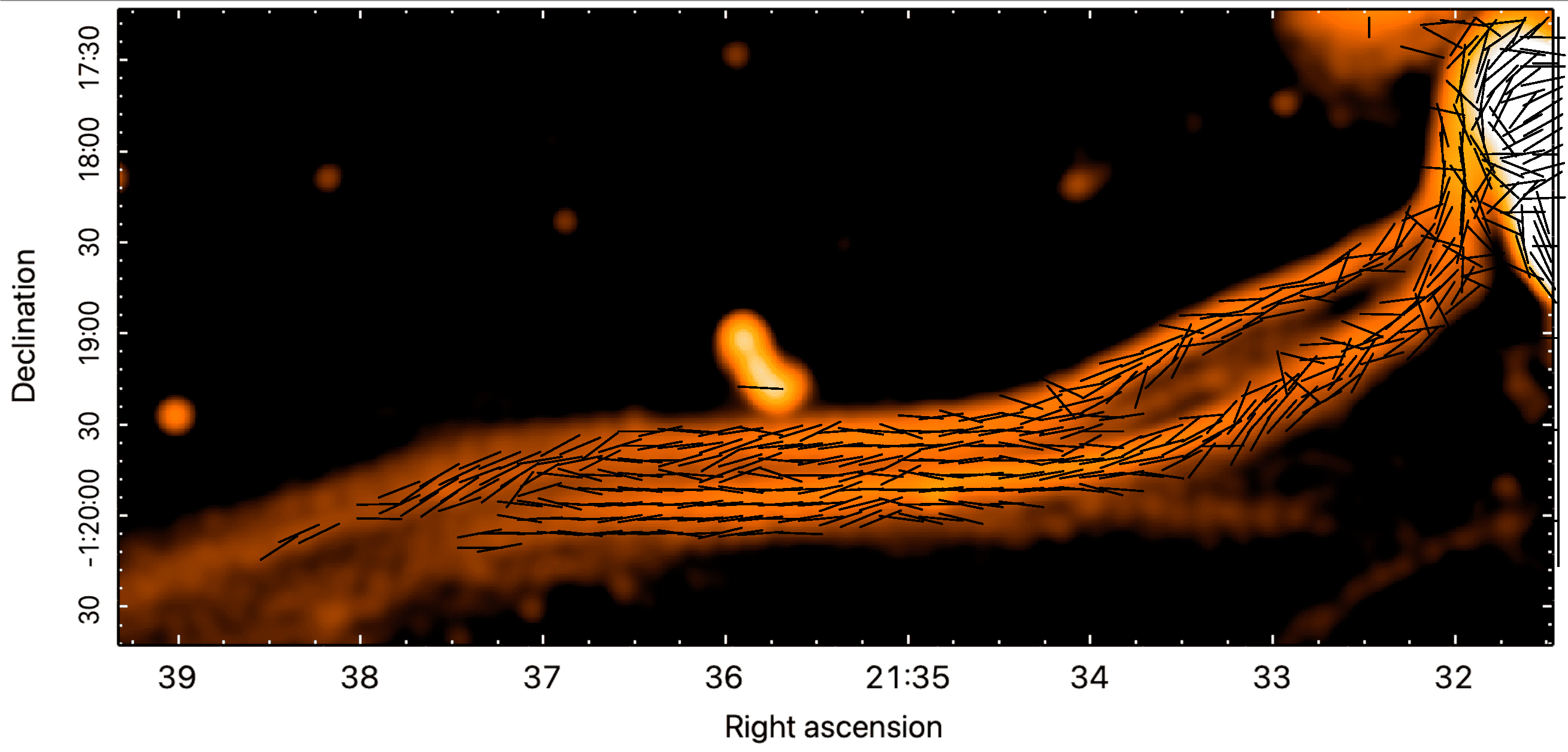}
    \caption{Inferred magnetic field vectors after correction for Faraday rotation, superposed on the total intensity image. The lengths of the vectors are fixed;  actual fractional polarizations in the northern lobe of 3C40B are $\sim$10\%, while the filaments are typically at least 50\% polarized.}
 \label{fig:my_polvec}   
\end{figure}

%We start with results from  the  68 channel, Faraday spectrum analysis, using the standard products of the peak RM, amplitude and angle.
The peak amplitude in the cleaned Faraday spectrum at each pixel, along with its corresponding Faraday depth (RM) and phase, are used to map the polarized emission.
A map of the \efil and the adjacent northern section of 3C40B is shown in Fig. \ref{fig:my_polvec}.  Where the signal:noise is sufficient, polarized emission is detected everywhere, except in isolated regions where a rapid change in angle causes local depolarization.

\begin{figure}[h]
    \centering
 \includegraphics[width=0.8\columnwidth]{ 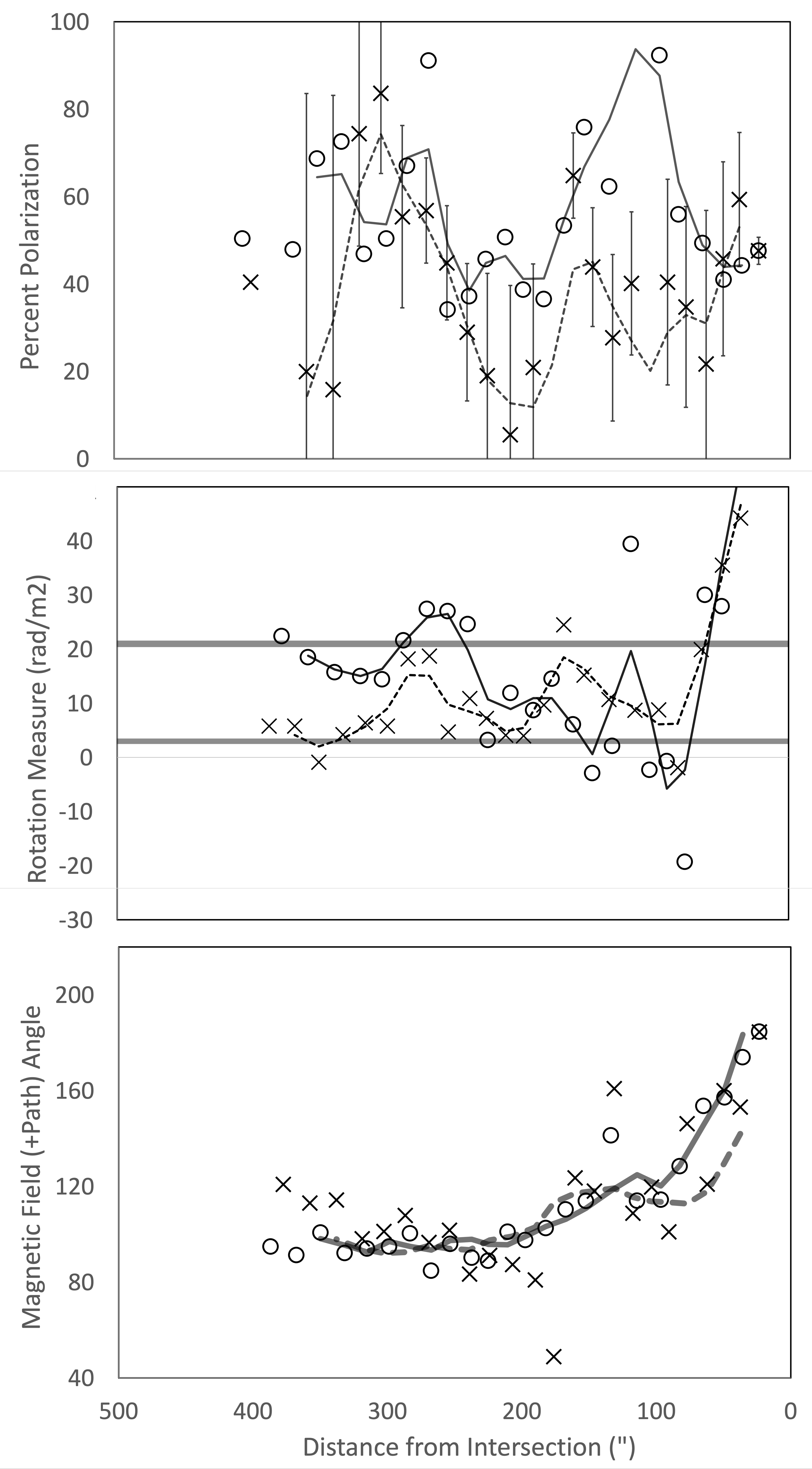}
       \caption{Polarization properties of the north (circles, solid line) and south (Xs, dashed line) filaments of \efil as a function of distance from 3C40B.  Top: Percent polarization after correction for bias (errors shown only for \rev{south}.  Middle: Peak RM. %from peak in 68 channel Faraday spectra. 
      Lines are hanning smoothed.  Horizontal lines show the range of contributions from the Milky Way foreground.  Bottom: Inferred magnetic field orientation after RM correction.    On the bottom panel, the lines (solid, north; dashed, south) are \emph{not} derived from the polarization data;  they indicate the \emph{local orientation} of the filament.}%, calculated from the orientation of the two neighboring sampled points. }
\label{fig:filP}    
\end{figure}

At the same positions as used in Fig. \ref{fig:filI}, we sampled the 7.75\arcsec~ maps in polarized flux, RM, and electric vector position angle corrected for the local RM.   The resulting behavior of the polarization as a function of distance from 3C40B, is shown in Fig. \ref{fig:filP}. The fractional polarizations were calculated after correcting for the polarized flux noise bias.

There are several important findings from these measurements.  First, the calculated percentage polarizations are high, with median values of 40\% (50\%) in the northern (southern) filament.  
The correction for the noise bias resulted in only 2 points out of 46 yielding  negative percent polarizations, supporting the high median percentages, even in the presence of large uncertainties.

Second, the rotation measures are found to lie within the narrow band corresponding to the local foreground RM from the Milky Way, so there is no detectable \emph{net} RM from the cluster itself. They have an rms variation of only 9~\radmm~ along the filaments, similar to the variations across all the structures in the central region of Abell~194 (see Fig. \ref{fig:RMhist}); these variations are much smaller than seen in richer clusters.  A considerably more detailed investigation of the Faraday structure is presented below.

Finally, the magnetic field direction is closely aligned, on average, with the local orientation of the filaments.  The Faraday corrected magnetic field angles (based on the observed single dominant RM contribution), when smoothed, trace the orientation of the filaments as they range over $\sim70$\degree, mostly vertical at the intersection of \efil with 3C40B, to nearly horizontal at large distances to the east (Fig. \ref{fig:filP}).

To examine the Faraday structure of the filaments in more detail, we used a new implementation of Faraday synthesis, (see Section \ref{sec:obs} and  \citet{Complex}), which employs ``super-resolution" to achieve a Faraday resolution of 15.6~ \radmm. %A source with a single Faraday depth $\phi$ will produce an unresolved peak in the Faraday spectrum once the spectral dependence is removed; this could be done, e.g., by using Q($\nu$)/I($\nu$) and U($\nu$)/I($\nu$) in the calculation of the Fourier transform.  To do this successfully, however, I($\nu$) must be free of dynamic range problems, RFI, cleaning systematics, etc.,  and becomes unstable at low signal:noise values in the individual frequency channels.  For our initial exploration, we  instead 
We used $\nu \times$Q($\nu$) and $\nu \times$U($\nu$), to approximate the removal of the spectral dependence. %; this would be appropriate if the spectral index were -1.  As seen in Fig. \ref{fig:filI}, the spectral index of the filaments actually ranges from -1 to -2.5 .  
The effect of not fully correcting for spectral index is that the peaks in Faraday space are broadened, while the mean Faraday depths remain unchanged. % We restrict ourselves here to the variations in the distribution of peak Faraday depths.

\begin{figure}
    \centering
    \includegraphics[width=0.93\columnwidth]{ 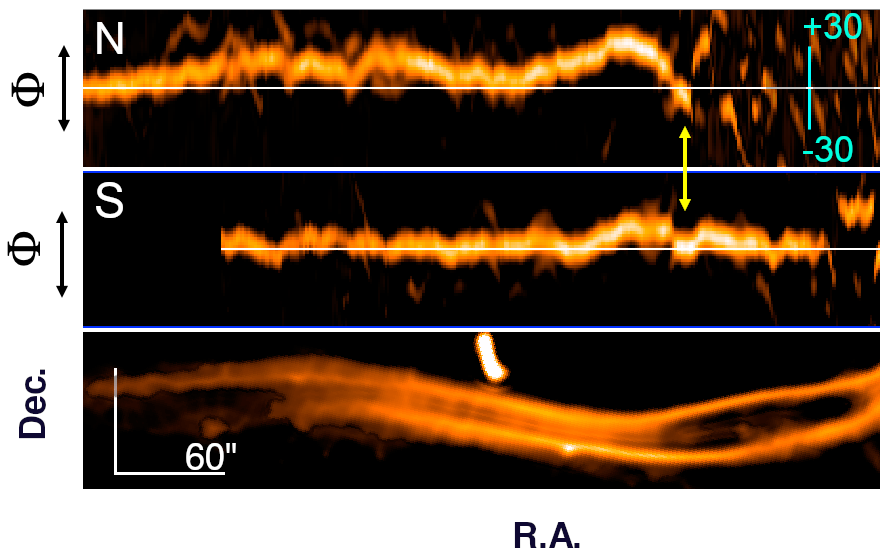}
    \caption{Top two frames: 2D images of F($\phi$) in $\phi$ vs. RA space, i.e, the Faraday spectrum for the north and south filaments separately.  The integrated flux in the Faraday spectrum is set to a constant at each RA. Bottom: The \efil region in total intensity at 7.75\arcsec~ resolution, aligned in RA with the top images.  The yellow arrows indicate a feature discussed in the text.  }
\label{fig:rmpatfil}    
\end{figure}

The cleaned Faraday spectrum F($\phi$) at each pixel forms a cube, with axes corresponding to RA, Dec, and Faraday depth $\phi$.  From this cube, one can vary the position on the sky and extract a $F(\phi$) vs. position plane. %We have performed the equivalent of that procedure, except that the filaments are not at a fixed Dec. 
To produce the F($\phi$) image in the $\phi$ vs. RA plane, shown in Fig. \ref{fig:rmpatfil}, we extracted the cleaned Faraday spectrum at each pixel in RA, at the value of Dec corresponding to the ridge line of the filament.  For each pixel in RA, we then normalized the spectrum -- for display purposes --  so that the integrated polarized flux between Faraday depths of -40~ \radmm~ and +65~ \radmm~ was set to a constant value.  %The resulting two dimensional images in $\phi$ vs. RA are .  

We briefly explore the type of information available in these RA, Dec, $\phi$ cubes.   The common interpretation of Faraday structures is that they result from variations in the foreground plasmas that are not directly related to the source.  The clearest exceptions to this are cases where the synchrotron structure and the Faraday structure are clearly related, e.g., cases of banding or draping \citep{2012MNRAS.423.1335G,2011MNRAS.413.2525G,2021NatAs...5..159M,2021MNRAS.508.5326M}.  \cite{RudMag} suggested that local effects might actually be a major contaminant to estimates of cluster magnetic fields which are based on the assumption that they are unrelated foreground variations.  

In  Abell~194, we are in the unusual position where the scatter in RM is quite low \citep[see Fig. 15 in ][]{Govoni2017}, in contrast with rich clusters \citep[e.g.,][]{2004A&A...424..429M}.  A more thorough discussion of Abell~194's special nature is presented in Appendix \ref{sec:3D}.  With minimal confusion from foreground fluctuations, we suggest that the changes in Faraday depth, $\phi$ from pixel to pixel can be mapped to distance along the line of sight. The 2D F($\phi$), $\phi$ vs. RA images are then equivalent to a ``top view'' of the filaments, illuminating their 3D structure.  This is perhaps best illustrated by its exception, where a foreground patch of ionized plasma likely creates a sharp jump in the Faraday depth in both filaments (the yellow line in Fig. \ref{fig:rmpatfil}), with no corresponding structure in the plane of the sky.  By contrast, most observed changes in Faraday depth appear associated with structures in the total intensity images.

Further discussions of isolating local Faraday effects from those of an unrelated foreground are found in Sec. \ref{sec:interact} and in  Appendix \ref{sec:3D}.  At this point, however, we  note that neither the magnitude nor the sign of the scaling between Faraday depth ($\phi$) and distance along the line-of-sight $l$ are known.  Since $\phi~=~n_e~\vec{B}~\cdot~\vec{l}$, changing the direction of the magnetic field reverses the sign of the inferred change along the line-of-sight, while its magnitude scales with the electron density n$_e$ and the projected magnetic field.

\subsection{X-rays}\label{sec:xraydip}
The X-ray surface brightness in the region of the \efil was fit with a $\beta$-model plus an additional component for NGC~547 (which contributes very little at these distances).  Pressures were calculated assuming a temperature of 2~keV, based on the kT map in Fig. 6 of \citet{Bogdan} and normalizations based on a 4\arcmin~ circle.  At 2.9\arcmin~ from NGC~547, at the western end of the \efilc the density was 1.9$\times$10\minus~cm\minus~  with a corresponding pressure of 1.15$\times$10$^{-11}$ erg~cm\minus.  At the eastern end, 9.6\arcmin~ from NGC~547, the corresponding values were 0.98$\times$10\minus~cm\minus~  and 0.6$\times$10$^{-11}$ erg~cm\minus.  

\begin{figure}
    \centering
    \includegraphics[width=0.9\columnwidth]{ 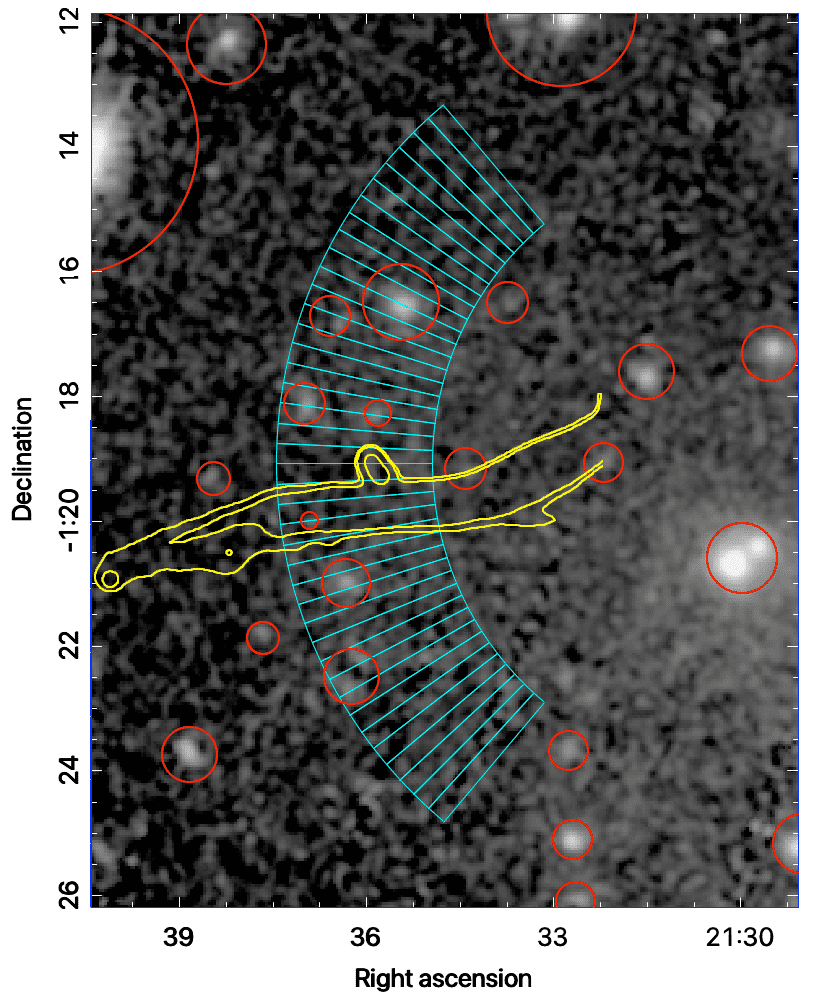} 
    \caption{XMM-Newton image in a soft energy band (0.4-1.25 keV) with pixel size 
$2.5"$, and smoothed with a 3 pixel Gaussian. The cyan sectors are used to extract an azimuthal profile crossing the location of the \efilc  (shown with yellow contours from the 15\arcsec~ MeerKAT image). The X-ray sources circled in red are excluded from the analysis. The resulting surface brightness distribution is shown in Fig.\ref{fig:sbrxmm}.}
  \label{fig:sbrimg} 
\end{figure}
\begin{figure}[h]
    \centering
    \includegraphics[width=0.9\columnwidth]{ 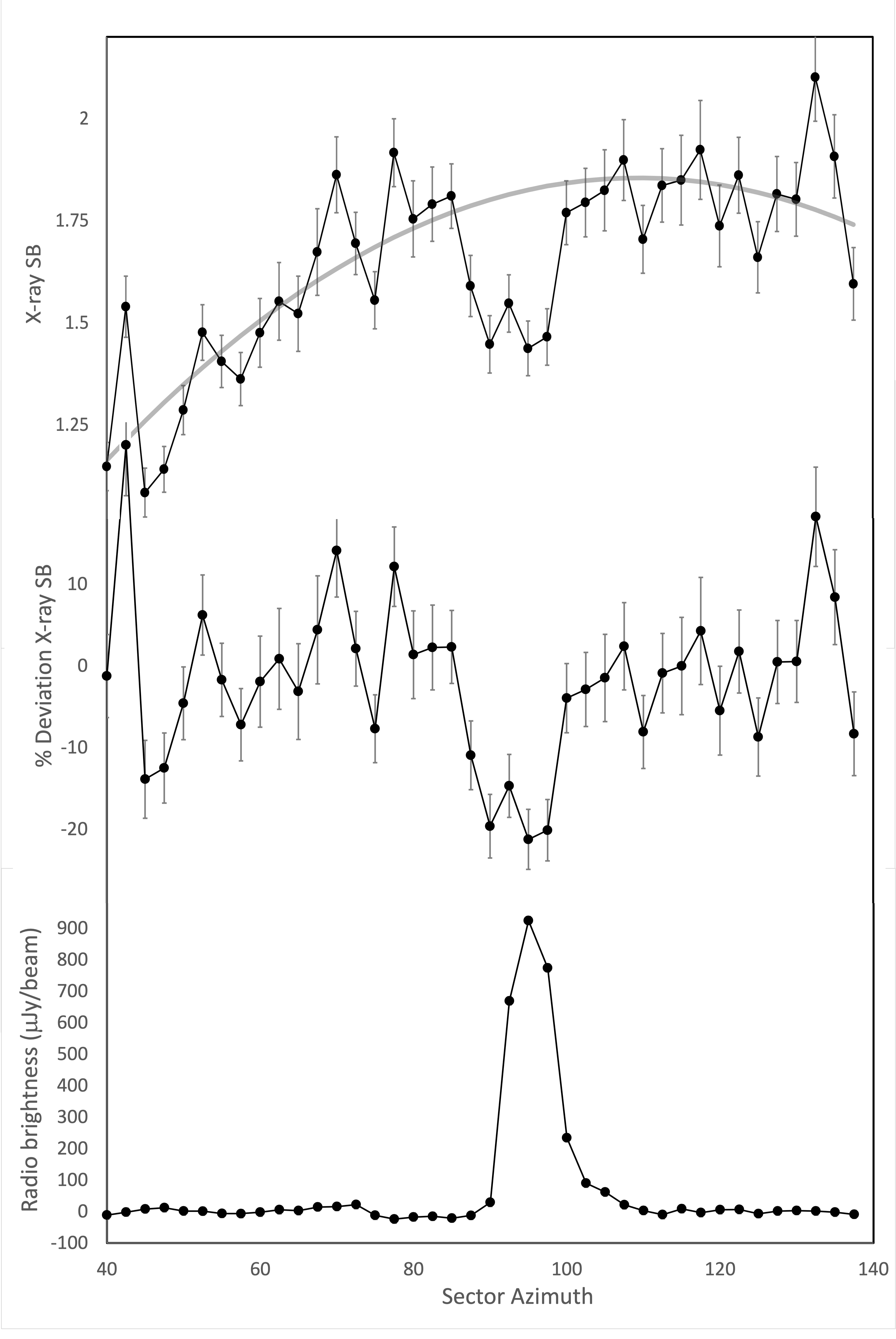} 
    \caption{Top: X-ray surface brightness profile using the sectors in Fig. \ref{fig:sbrimg} \rev{in units of counts/(ksec~arcmin$^2$)}. 
    The profile is made using the unsmoothed data that are shown (smoothed) in Fig. \ref{fig:sbrimg}. \rev{A second order baseline is fit to the profile, ignoring the region of the dip, and is shown as a grey line.  Middle: Baseline-subtracted X-ray brightness, in terms of percent deviation from the baseline.}  \rev{Bottom: Radio profile from the 15\arcsec~ MeerKAT image with a linear baseline subtracted.} The emission from the compact background double just north of the filaments has been excluded. A dip of 19$\pm$2\% \rev{in the X-ray surface brightness is coincident with the peak in the radio.}  }
 \label{fig:sbrxmm}  
\end{figure}

After removing detected point sources in the area we extract a surface brightness profile along sectors of an annulus which extends from 5-7.5\arcmin~ from a center slightly north of the X-ray peak (R.A. 1:26:00,  Decl. -1:19:06); The sectors align with and cross the trajectory of \efil (Fig. \ref{fig:sbrimg}). Surface brightness values are taken counter-clockwise for each of the two bands separately, indicating a dip of the surface brightness in the location of \efilp 

By  adding the two subbands to construct a combined surface brightness profile from 0.4-3keV and subtracting out a large-scale parabolic baseline, we identify a significant dip of  19$\pm$2\%, spanning five sectors or $\sim$1.25\arcmin~(27~kpc) at the location of \efilp    The X-ray surface brightness dip extends somewhat beyond the radio emission, so for the purposes of calculation we assumed an empty 35~kpc empty cylinder.  With the nominal beta model, this would produce a dip of $<$10\%.  So if an empty cylinder is appropriate, then the line of sight distance through the cluster would have to be only 190~kpc, instead of the inferred 966~kpc (2 $\times R_{500}$).  It is therefore possible that the X-ray emission has a flattened distribution along the line of sight;  if this were true, then our estimates of the densities and pressures at the position of the \efil would go up by $\sim\sqrt{~2}$. The brightness structure in the plane of the sky also indicates that the ICM is non-spherical, so a flattened distribution would not be surprising.  In any case, having at least a 35~kpc region that is completely empty of X-ray emitting material is difficult to avoid.

\subsection{Closeup - the \efilc north jet intersection} \label{sec:closeup}
 %If the \efil and 3C40B were simply unrelated structures, then we would expect the \efil to maintain their east-west orientation and perhaps reappear on the west wide of 3C40B.  Instead, t
 The narrow north and south filaments appear to merge and turn towards the north as they approach 3C40B, until they blend in with the brighter jet emission. This suggests that the two structures could be interacting. In this section we examine the apparent interaction region.

We start by separating the different spectral components using the spectral tomography technique of \cite{tomog}. Since the \efil have steeper spectra than the 3C40B jet, an image dominated by steep spectrum material isolates them from the flatter jet-related emission.\footnote{This involves taking two maps at different frequencies, and subtracting the higher frequency map from the lower frequency one to yield $\Delta$S$_t$  with a scaling equivalent to an arbitrary spectral index $\alpha_t$. In such a difference map, all structures with $\alpha = \alpha_t$ disappear, leaving positive residuals for regions with steeper $\alpha$ and negative residuals for flatter ones. To create the spectrally separated maps, we used the broadband 1283~MHz map (S$_{1283}$) and the corresponding spectral index map, to create two ``pseudo-maps" at the arbitrary frequencies of 850~MHz and 2000~MHz.}

\begin{figure}
    \centering
     \includegraphics[width=0.8\columnwidth]{ 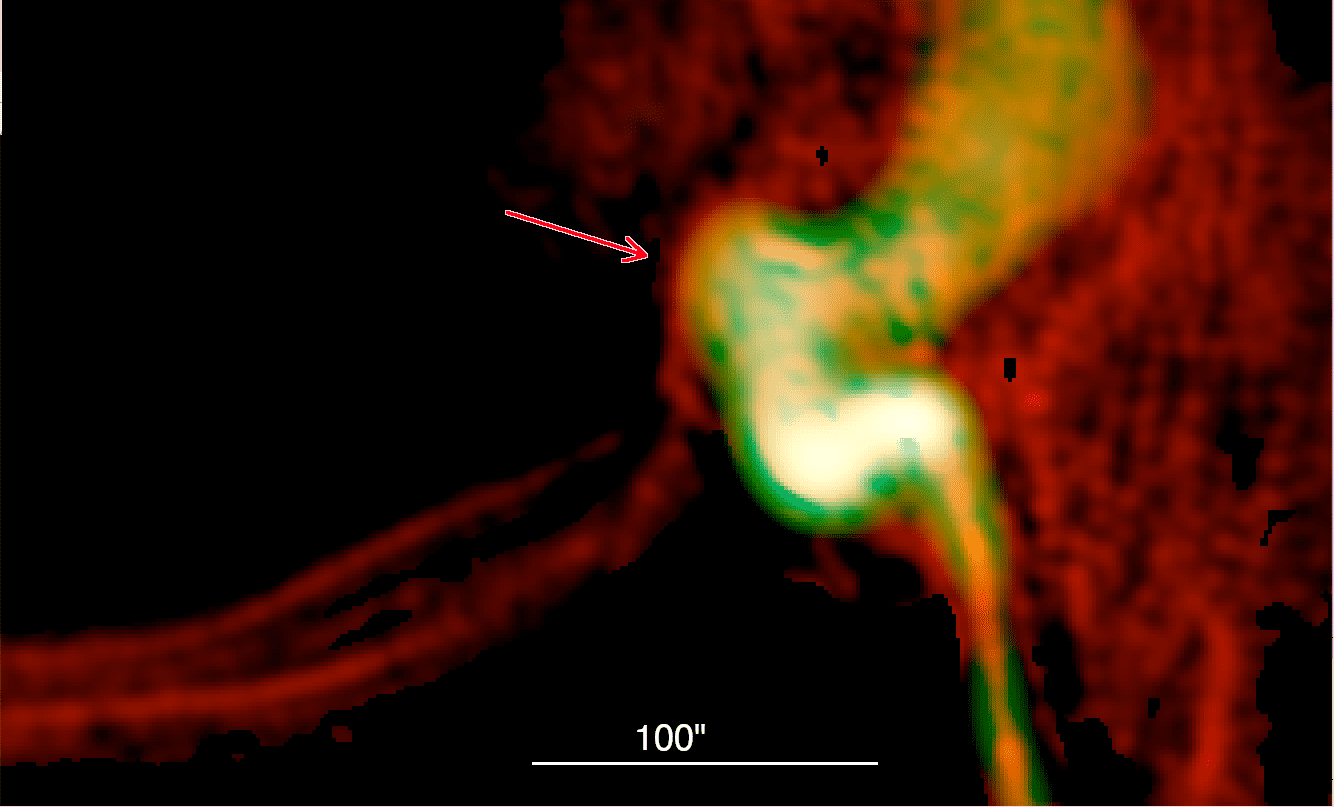}
     \includegraphics[width=0.8\columnwidth]{ 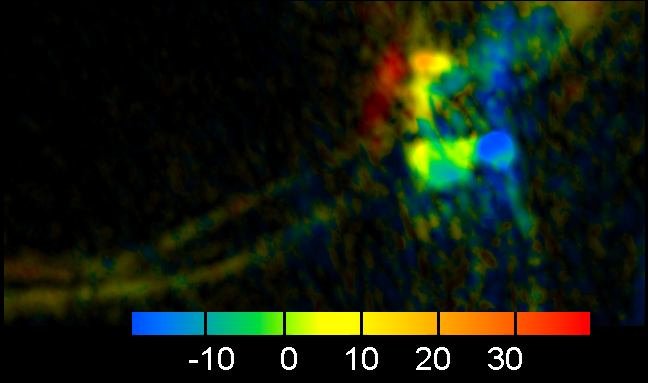} 
    \caption{Top: Separation of steep spectrum (red) and flat spectrum (green) structures around the northern jet of 3C40B, constructed as described in text. $\Delta$S$_{0.5}$ in red, where most of the contribution from the flatter spectrum source 3C40B disappears, overlaid with the original 1283~MHz map in green.  The red arrow indicates the apparent extension of the filaments to wrap around the top of a bend in the jet.  Bottom: Polarized intensity of the jet/filament interaction region color coded by RM, as shown in the color scale. }
 \label{fig:wrapfil}   
\end{figure}

The results are shown in Fig. \ref{fig:wrapfil}. The key finding is that the steep spectrum \efil not only turn to the north as they approach 3C40B, but appear to wrap around the north end of this portion of its jet, as the jet abruptly bends to the west. In Section \ref{sec:discussion} we explore what can be learned about jet and filament physics from this apparent interaction. 

There is also a dramatic shift in the Faraday structure at this point, as shown in  Fig. \ref{fig:wrapfil}. %This map was created from the peaks in the Faraday spectrum, as described earlier. 
Where the \efil and 3C40B intersect, the peak RM jumps by $\sim$30~ \radmm; adjacent to this is a depolarized region of 3C40B and some hint that higher RMs are also found in the jet at this point.  

\begin{figure}
    \centering
    \includegraphics[width=0.9\columnwidth]{ 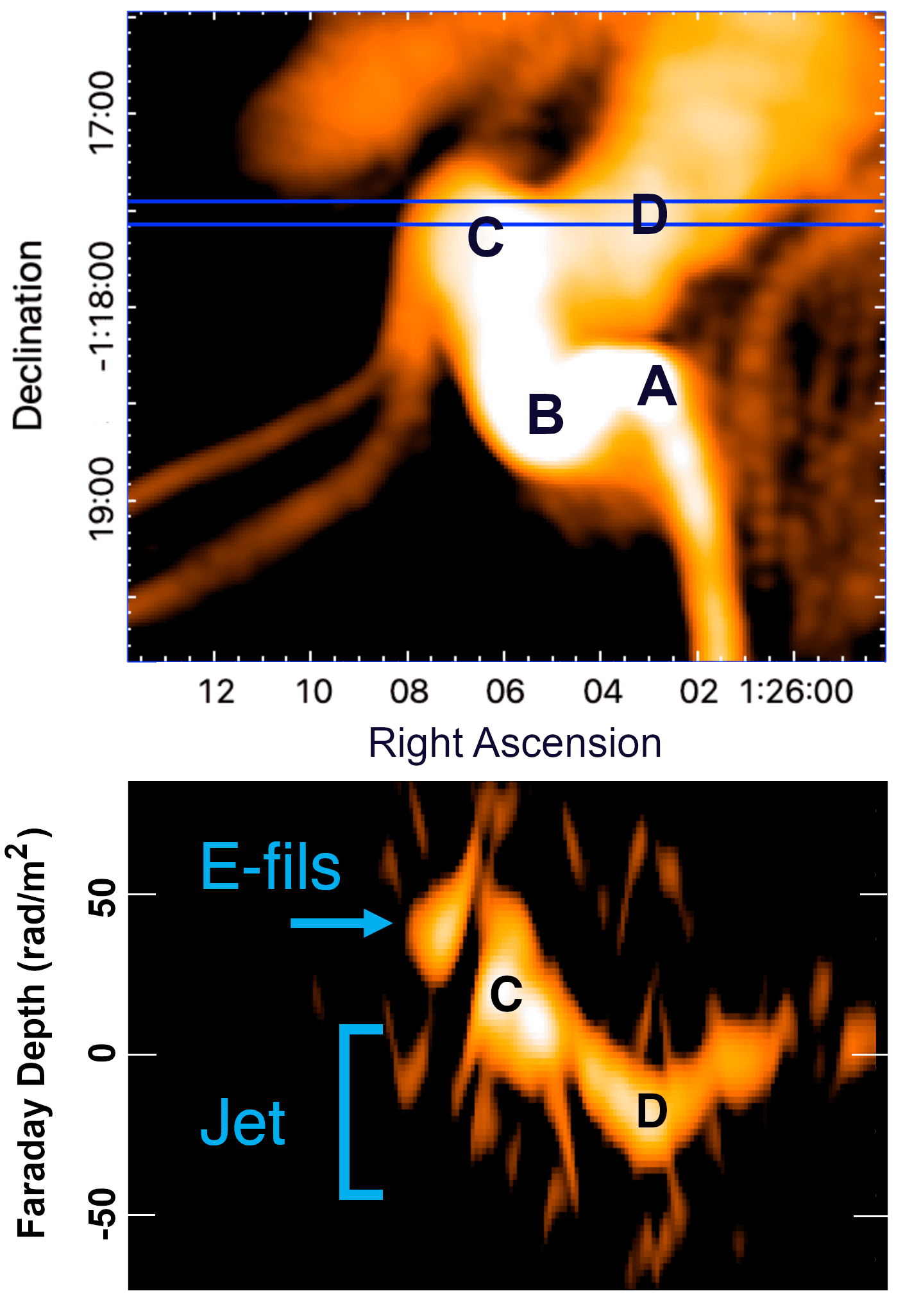} 
    \caption{Top: MeerKAT total intensity of 3C40B in the region of the interaction with the \efilp Letters mark the locations of bends in the jet. Bottom: Polarized intensity produced by the super-resolved  cleaned Faraday spectrum.  Plotted in the Faraday depth vs. RA plane, at the fixed  declination marked by dark blue lines in the top panel.  The cyan arrow shows the polarized emission from the \efilc while the rest of the emission is from the jet. }
 \label{fig:RMsliceN}  
\end{figure}

To examine this transition more carefully, we again turn to the super-resolved version of the Faraday spectrum, this time correcting for an effective spectral index of -0.5, since we are primarily interested in the flat spectrum northern jet of 3C40B.  Fig. \ref{fig:RMsliceN} shows the Faraday depth \emph{vs.} RA plane at fixed declination;  as in Fig. \ref{fig:rmpatfil}, this is equivalent to a ``top-view" if we interpret the Faraday depth as a distance along the line-of-sight.

The most important conclusion from this display is that the northern segment of the jet (between points \textbf{C} and \textbf{D}) has a smooth gradient in Faraday depth which connects directly to the Faraday depth of the \efil --  there is \emph{not} a sudden jump at the \efil as one might conclude from a superficial look at the peak RM distribution shown in Fig. \ref{fig:wrapfil}.  In retrospect, one can see the same gradient in Fig. \ref{fig:wrapfil}, while Fig.\ref{fig:RMsliceN} makes it much more obvious.  If we now interpret the Faraday depth changes as line-of-sight changes, then the jet from locations \textbf{C} to \textbf{D} meets up with \efil along the line-of-sight, just as they meet up in the plane of the sky.  This confirms that the jet and \efil are indeed interacting, and not simply superposed in the plane of the sky.

Note also that there is a change in the 3C40B jet at point \textbf{D}, where it bends by 45\degree~ to head NW into the much more diffuse faint northern lobe.  At this same position, the trend in Faraday depth reverses, showing that this is a change in 3D, not simply in the plane of the sky.  The correspondence between the changes in trajectory in the plane and in Faraday depth would be an (unlikely) coincidence if it were due to an unrelated foreground screen;  this provides additional support for the line-of-sight interpretation of Faraday depth in this system.

The connection between Faraday depth and distance provides us a new tool to estimate the magnetic field strength in the ICM.  If we make the reasonable assumption that the orientation of the jet between \textbf{C} and \textbf{D} is at 45\degree~ to the line-of-sight, then the linear distance of 18~kpc between \textbf{C} and \textbf{D} corresponds to a change in Faraday depth of $\sim$40~ \radmm.  Using the density of 1.9$\times$10$^{-3}$cm\minus~ from the X-rays yields a magnitude of 1.4$\mu$G for the component of the magnetic field along the line-of-sight.  

This is the first estimate of magnetic field strength from a \emph{single region} of Faraday rotating material in a cluster ICM. Previous estimates are based on the statistical properties of RM fluctuations, scale sizes and variations in field strength and densities over the cluster.  Our single region estimate is, of course, uncertain by factors of order unity because of unknown projection effects for both the jet and the magnetic field orientation.  The value of 1.4$\mu$G for the surrounding ICM is a factor of 4 below the minimum pressure magnetic field of 5.8$\mu$G calculated for the portion of the jet between points \textbf{C} and \textbf{D}, using the observed spectral index of -0.52 and the same assumptions as used for the filaments, above.

\subsection{X-rays in the northern jet}\label{sec:xraypatch}

\begin{figure}
    \centering
    \includegraphics[width=0.9\columnwidth]{ 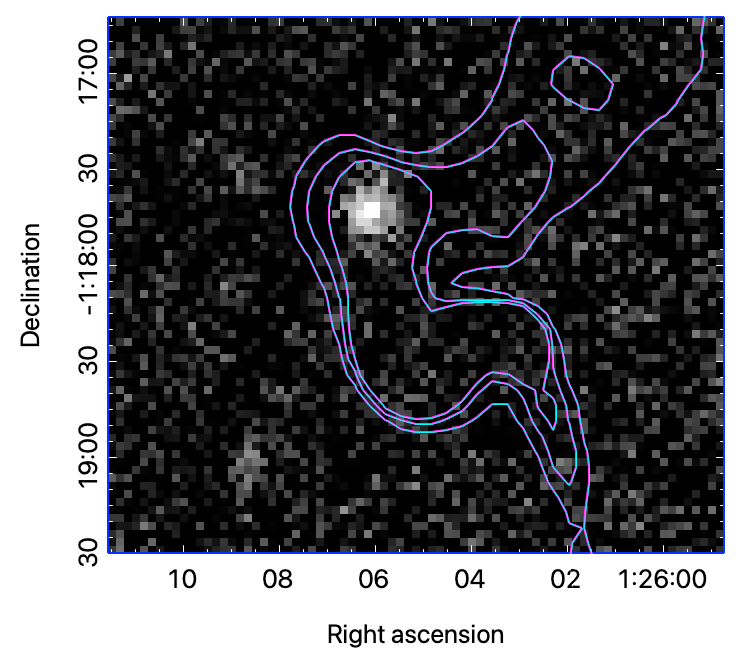}
   \caption{X-ray XMM EPIC image overlaid with total intensity contours of the northern jet of 3C40B.} %Right: Normalized profile cuts of total and polarized intensity and X-ray brightness on a line south (negative) to north through the X-ray feature.  The percent polarization scale is on the vertical axis.}
 \label{fig:xray}   
\end{figure}

The XMM-Newton observation shows an X-ray source at the location of the intersection of the northern jet with E-fils (Fig. \ref{fig:xray}). There is no hint of any extent in the XMM-Newton images. This X-ray source is also seen in the Chandra observation (OBSID 7823,~ 65ks) at the edge of the ACIS S2 (ccd\_id=6) during an observation of A194 centered on ACIS-S3. %Hence, the source is far from the optical axis and at the very edge of the FOV.  
Using the Chandra PSF calculator for the off-axis location, at the very edge of the FOV, the source is also consistent with a point source having FWHM $=2.5"$.  Hence, we refer to this X-ray emission as an unresolved X-ray feature (UXF). It is possible that the X-ray and radio superposition is random; it is difficult to estimate this probability \emph{a posteriori}.  Still, the X-ray source density is low and a random alignment on such a unique location of the radio structure seems small.
%The XMM-Newton observation shows an X-ray source at the location of the intersection of the northern jet with \efil (01 26 06.06, -1 17 43.3).  This  X-ray source is also seen in the Chandra observation (OBSID 7823,65ks) at the edge of the ACIS S2 (ccd\_id=6) during an observation of A194 centered on ACIS-S3.  Hence, the source is far from the optical axis and at the very edge of the FOV.  Using the Chandra PSF calculator for the off-axis location, the source is consistent with a point source having FWHM $=2.5\arcsec$.  Nor is there any hint of extent in the XMM-Newton images.  Hence, we refer to this X-ray emission as an unresolved X-ray feature (UXF).

The UXF is listed in the compilation of X-ray sources in the field of A~194 by \citet{Hudaverdi06}. There are a number of other X-ray sources coincident with faint, compact radio sources in the field. \citet{Hudaverdi06} note a large excess of compact X-ray sources in this field, which they attribute to the fueling of AGN during cluster infall.  However, aside from the UXF and those X-ray sources associated with NGC~541, 545 and 547, none have any apparent connection to the radio structures 3C40B and 3C40A, or the accompanying filamentary features. 

The XMM-Newton spectrum of the UXF is consistent with both a power-law (spectral index $2.15\pm 0.12$), and a thermal model with a temperature of $1.84\pm 0.28$\,keV, and a low abundance of $Z=0.17 \pm 0.12 Z_\odot$. The model predicted (unabsorbed) flux is $2.8\times 10^{-14}~ {\rm erg\,s^{-1}\,cm^{-2}}$ in the 0.5 to 2\,keV band.  Assuming the X-ray emission is associated with a cloud of thermal gas with the size of the Chandra PSF ($\sim$1~kpc), the gas mass is  about $5\times 10^6\,M_\odot$. 

We found no evidence for a counterpart to UXF in SDSS u,g,i,r, or z images, or the near- or far-UV images from GALEX.  There is a very faint apparent source at this location in the unWISE W1 co-adds, although the background density at this brightness means the probability is high for a spurious detection.

\section{Discussion}\label{sec:discussion}
The following section addresses  the origin of the \efilc the ways in which they interact with the radio jet/ICM, and the implications for relativistic particle (re-)acceleration. We briefly summarize below the key points from this section to provide a framework for the more detailed discussions.  In our concluding remarks, we  emphasize the more general and important implications of this work for the relativistic particles and fields and thermal plasmas in groups and clusters of galaxies. 

\subsection{Key points}\label{sec:key}
\begin{itemize}
    \item {Magnetized filaments, composed of bundles of smaller ``fibers," arise naturally in turbulent magnetic plasmas, and will reflect the eddy turnover lengths and times of the driving forces.}
    \item{Filaments will be be further stretched, and their magnetic fields amplified, by bulk motions in the ICM and by encounters with radio jets.}
    \item{Magnetic pressures in filaments can reach significant fractions of the ambient thermal pressures and together with accompanying cosmic rays, can plausibly expel the local thermal plasma.}
    \item{A scenario in which a dense moving cloud encounters the northern jet of 3C40B, bending it and the \efilc appears plausible.}
    \item{Either streaming of cosmic ray electrons from a radio galaxy, or, more likely, the (re-)acceleration of seed electrons from the ICM through betatron or other mechanisms, is required to illuminate the filaments.} 
\end{itemize}

\subsection{The ubiquity of synchrotron filaments}\label{sec:ubiquity}
Long, thin synchrotron-emitting filaments are becoming ubiquitous in radio maps of sufficient resolution and sensitivity.  They are found in peripheral radio relics, e.g. Abell~2256 and Abell~3376 \citep{2014ApJ...794...24O, 2022A&A...659A.146D}, and as polarized features e.g., in Abell~2255 and MACS~J0717.5+3745, which are likely peripheral but projected onto the cluster interior \citep{2011A&A...525A.104P,2022A&A...657A...2R}. They are found in the neighborhood of radio galaxies \citep{2020A&A...636L...1R,2022arXiv220104591B}, and even inside radio lobes, e.g., Centaurus~A \citep{2014MNRAS.442.2867W} or M87 \citep{1989ApJ...347..713H}.  Sometimes these filaments connect  ``old’’ tails of radio galaxies to diffuse radio  emission \citep{2017SciA....3E1634D, 2018MNRAS.476.3415W, 2020ApJ...897...93B, 2020A&A...634A...4M}. Possibly analogous complex and extended synchrotron structures include those in the wake of radio galaxy NGC1272 in the Perseus cluster \citep{2021ApJ...911...56G}. Synchrotron filaments are also prominent in the Galactic Center region \citep{2021MNRAS.500.3142Y}, and in supernova remnants \citep{1995MNRAS.277.1435M}.   Given the wide variety of physical conditions in these environments, it becomes clear that filament formation is a natural result of dynamic magnetized astrophysical plasmas. %There is, however, a small, but growing population of other clusters exhibiting ensembles of lengthy radio filaments in apparent association with radio galaxies. 

In Abell~194, we see a rich network of filaments, shown in more detail in Appendix \ref{sec:Afilaments}.  Many of these appear associated with the luminous radio galaxies 3C40A and 3C40B.  The most prominent filaments are the \efilc which are the subject of this paper. Consideration of the observational findings listed above, and the filamentary magnetic structures seen in MHD simulations lead us to a preferred origin scenario. In that scenario,  the \efil are initially generated by shear flows in the Abell~194 cluster, followed by interactions with the 3C40N jet and a dense ICM clump, which stretch and amplify the filaments and lead to the reacceleration of the embedded cosmic ray electrons.

\subsection{Shear and the generation of filamentary magnetic ``bundles"}\label{sec:shear}
Simulations show that extended bundles of magnetic fields are ubiquitous in turbulent MHD flows in ``high-$\beta$'' environments, where transverse magnetic field pressure gradients can be balanced by ambient pressure gradients \citep[e.g.,][]{porter}. The strongest fields tend to be substantially stronger than the global mean field and are separated into distinct bundles of fibers with lengths reflecting the driving scales of the local turbulent flow \citep{porter, vazza18}. The individual thicknesses of the fibers are set by the limiting \textbf{transverse resistivity scales}, so are likely less than a kpc \citep{Zhura}. % \red{Why does this follow?}
The apparent $\sim$3~kpc filament thickness of the narrow component radio filaments is then not the physical width of an individual magnetic fibre, but more likely the width of a bundle of illuminated, aligned fibers, held together by the tension along the field lines. More broadly distributed ensembles of weaker fiber structures could then be responsible for the even broader component of \efilp  

MHD simulations have  demonstrated that magnetic folds that occur in a turbulent dynamo succumb to tearing at high Reynolds numbers leading to increasingly longer field structures. In general such studies show that when the seed magnetic field for a turbulent dynamo is weak, then the amplified magnetic energy initially peaks at the small, resistivity-based scales.  The proceeds until the developing magnetic field becomes strong enough by stretching that the back-reaction on the small-scale ambient turbulent dynamical stresses becomes significant \citep[see, e.g., ][for a review]{2007mhet.book...85S}, and \citet{2022arXiv220107757G} %\red{(see, e.g., Schekochihin \& Cowley, 2007 for a review; also, \cite{galishnikova2022tearing}).} 

Subsequently, thin current sheets are torn into fibre bundles that extend to increasingly larger scales via stretching, until at ``saturation," the magnetic stresses become comparable to the fluid dynamical stresses on turbulence driving scales.  Then, the longest filamentary magnetic structures approach a substantial fraction of that driving scale. When such turbulence is driven by highly non-isotropic dynamics, such as we may expect in merging clusters, these filament lengths would reflect the scales associated with the active drivers, which can extend to cluster scales \citep[e.g.,][]{2018MNRAS.474.1672V}.

If the magnetic fields have indeed been amplified by shear stretching, as described above, we expect the field strengths to scale as  $\rho v^2/\ell$, where $\rho$ is the density of the turbulent plasma, with a characteristic velocity $v$ on the driving scale $\ell$.  For turbulent Mach number of order 1/2, as commonly assumed for ICMs, the maximum magnetic pressures would then be several times smaller than the ICM pressure \citep[e.g., ][]{porter,2015Natur.523...59M}.  This is what the observations suggest for the \efilp

In the low-mass system Abell~194, the field stretching is likely dominated by  the large-scale bulk flows from continuing infall suggested by the motions of \rev{NGC~545} and NGC~541 with respect to the X-ray medium centered on \rev{NGC~547} \citep{Bogdan}. \rev{These authors use inferred pressures in the galaxies to show that their velocities in the plane of the sky are of order 800~km/s, \rev{for NGC~541,5} while their radial velocities are $<$100~km/s with respect to the cluster. The infall of these galaxies, and any accompanying gas, would thus be primarily from the West.}  The $>$200~kpc long gently curved \efil structure, with its high fractional polarization and aligned magnetic field are consistent with this picture \rev{of flow-induced stretching}.  The lack of strong, much smaller-scale (10s kpc) RM variations, seen by contrast in rich merging clusters again indicate the dominant  role of  stretching by large-scale bulk motions in Abell~194, instead of by smaller-scale turbulence.  

Whether or not this picture is viable depends on the lifetimes of the filaments.  The \emph{dynamical lifetime} of a filament of length $L$ , formed in a sheared flow pattern with characteristic velocity $v$, is set roughly by  $$\frac{L}{v} \sim \frac{L}{(M\cdot c_s)} \sim \frac{200~ kpc}{(M\cdot1000~ km/s)} \sim \frac{200}{M}~  Myr.$$
$M<1$ is the turbulent Mach number, and $c_s$ the sound speed.
This is likely comparable to the synchrotron lifetimes in the absence of a resupply of cosmic ray electrons into the high field regions (see Sec. \ref{sec:cosmic}).  Therefore, this overall scheme appears plausible to explain observed filaments.  It also implies that additional weaker magnetic field bundles are likely to intermittently thread the entire cluster volume  and become ``lit up'' above current detection limits once a sufficient population of high energy cosmic-rays is locally accelerated. %\red{this reference or another? %(see recent simulations of radio galaxies by Vazza et al. 2021)}.
%$L/c_s \sim 200 kpc/1000km/s \sim 200 Myr$.  c_s$ the sound speed.
%The \textbf{dynamical lifetime} of a filament formed this way in a turbulent flow is set roughly  by the eddy turnover time of the turbulent flow on the given length scale, i.e., $\gtrsim L/c_s \sim 200\, {\rm kpc}/1000\, {\rm km/s} \gtrsim 200$ Myr, where $L$ is the length of the filament and $c_s$ the sound speed.  
 
The discussion above does not explain the changes in structure or spectra of the filaments as a function of distance from 3C40B.  For that, we need to examine the region where the \efil and 3C40B's northern jet appear to interact, as we do in Sec. \ref{sec:interact}.

\newcommand{\crn}[1]{\textcolor{blue}{CRN: #1}}

\subsection{Jet/Filament/ICM interactions}\label{sec:interact}

The northern jet of 3C40B goes through a series of sharp bends, both as projected onto the plane of the sky and along the line of sight (Fig. \ref{fig:RMsliceN} and \ref{fig:RMcubes}).  These bends are not mirrored in the southern jet, and are thus unlikely to be the result of jet precession \citep[e.g.,][]{precess} or displacement due to orbital motions \citep[e.g.,][]{orbit}.  The remaining alternative is that the jet interacts with a clump of higher density ICM material, causing one or more of the observed bends.  

\begin{figure}[ht]
   \centering
  \includegraphics[width=0.75\columnwidth]{ 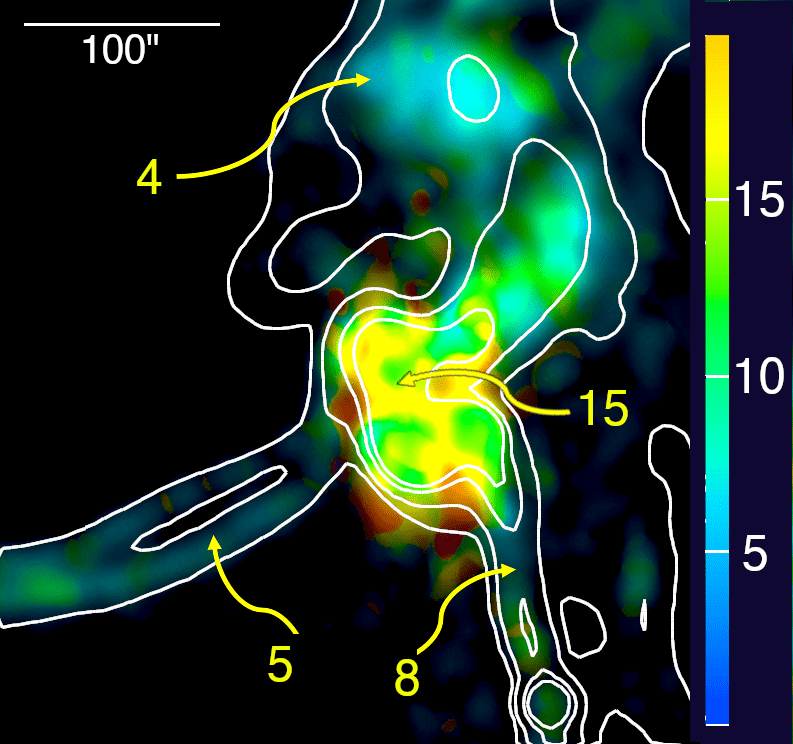}
   \caption{Map of rms width of the super-resolution Faraday spectrum at 15\arcsec~ resolution, with an assumed spectral index of -0.5 .  The colorbar and numbers for different regions are in units of~ \radmm.  The rms width is an indicator of multiple Faraday depths along each respective line-of-sight; any value below six is essentially unresolved in Faraday space. The intensity and contours are from the total intensity image.}   
   \label{fig:moment2}
\end{figure}

Fig. \ref{fig:moment2} provides evidence from Faraday rotation that the thermal plasma in the region of the bends is indeed denser than elsewhere along the jet.  The dramatic increase in the width of the Faraday spectrum near the bends suggests a denser environment which is likely mixed, on either microscopic or macroscopic scales, with the synchrotron emitting plasma.  The UXF at the top bend in the jet (Fig. \ref{fig:xray}) may be another indication of local dense thermal material, though its origin remains unclear. 

Turning now to the \efilc  their largely east-west structure appears to curve north and drape over the final sharp bend in the jet (Fig. \ref{fig:wrapfil}). Along the \efil the narrow filaments are stronger relative to the broad component as they approach the jet  (Fig. \ref{fig:Broad_Narrow}),  accompanied by an overall flattening of the spectral index (Fig. \ref{fig:filI}).  These systematic changes would not arise naturally from only the shear motions discussed above, which do not depend on the jet. We thus need to investigate how the jet, \efilc and a high density clump in the ICM could mutually interact.

\subsubsection{Filament stretching and magnetic field amplification}\label{sec:amplify}
A filament embedded in a shear flow will be stretched, as long as the tension stress from the field (roughly the magnetic pressure divided by the length over which it is being stretched) is less than the flow-based, dynamical stretching force. The field strength then increases in proportion to the increased filament length.   The strongest filament fields built from ICM-based flows are limited,  since within subsonic ICM flows the local plasma $\beta$ values will remain significantly greater than unity, while flow speeds are generally subsonic. In strong flows with major transitions (such as the flow around an advancing jet), 
the lengths and field intensities can increase by factors of a few.

 We observe gentle bends in the \efil ($\sim$15\degree~ and $\sim$22\degree~ at 100~kpc and 50~kpc from the jet, respectively).
A much sharper bend of $\sim$60\degree~ to the north is seen at $\sim$7~kpc ; this strongly indicates an interaction with the jet. Using simple geometry, we can estimate the amount of stretching associated with these bends --  for the gentle bends it is $\leq16\%$ , but for the large bend near point \emph{C} (Fig. \ref{fig:RMsliceN}), the \efil have likely been stretched locally by a factor of $\sim$3, from $\sim$16-46~kpc. This is a likely explanation for the relatively brighter narrow components of \efil closer to the jet (Fig. \ref{fig:Broad_Narrow}).  However, since the narrow components appear enhanced out to  $\sim$250\arcsec~ from the jet, an additional source of stretching from the ICM is required in the direction away from the jet and cluster center.  

The above discussion assumes for simplicity that the stretching happens only in the plane of the sky. The \efil likely also stretch along the line of sight, increasing the  total amplification of the magnetic field.

\subsubsection{Interaction Scenarios}\label{sec:scenarios}

\begin{figure}
%    \centering
    \includegraphics[angle=-90, origin=c, width=0.9\columnwidth]{ 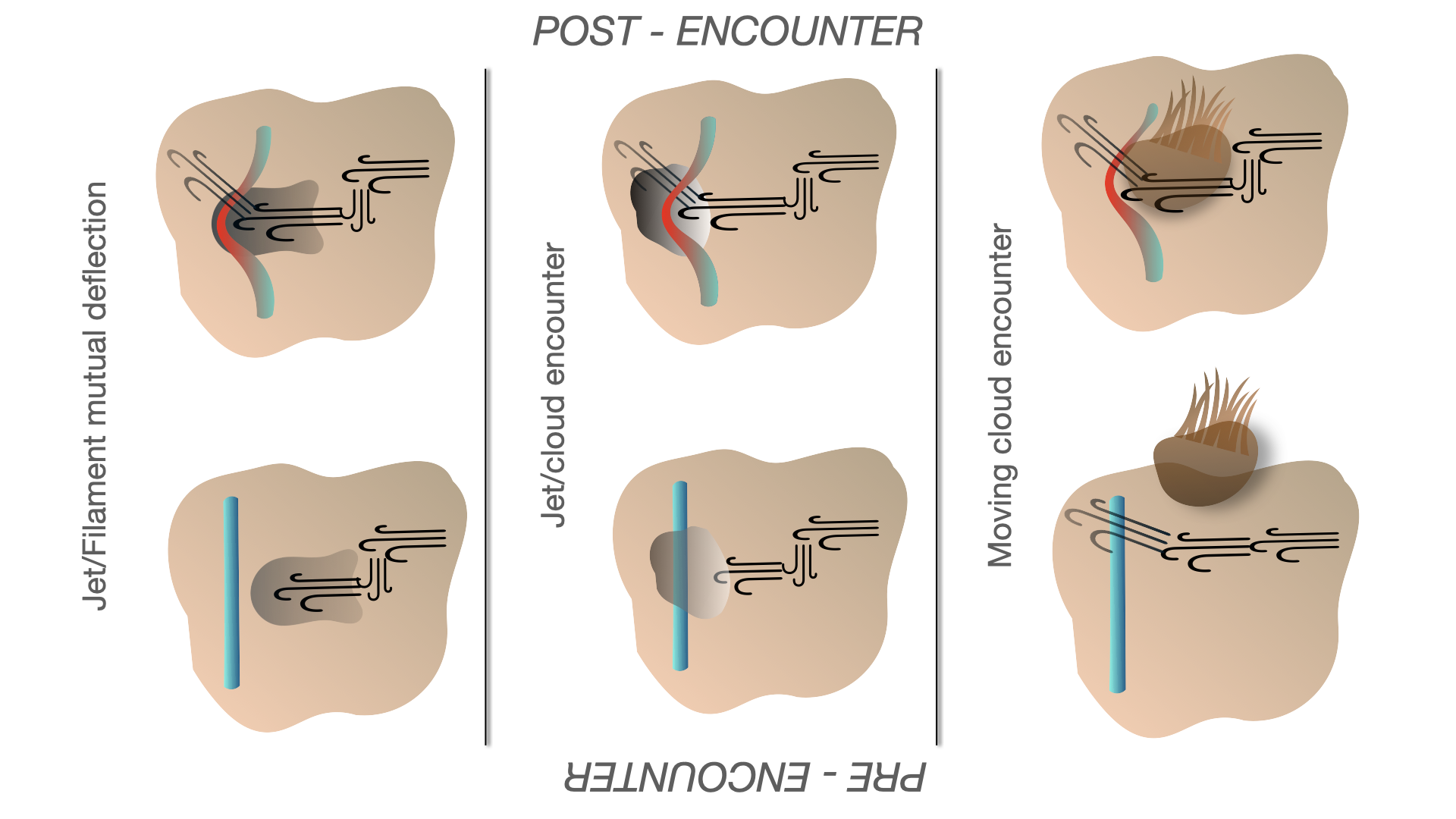}
    \caption{Three possible scenarios for the interaction of 3C40B's jet (represented by the line triplets) with the \efilp In the text, these are referred to as Scenario I (top), II (middle), and III (bottom, preferred). }
\label{fig:cartoon}    
\end{figure}

In addition to stretching the \efil as they encounter the jet and drape over it, we must account for the bending of the jet itself.  As shown above, this likely involves some denser region of the ICM. Fig.~\ref{fig:cartoon} depicts three possible scenarios for this three-way interaction. In Scenario I (top panel), we assume that the filament is broad enough, and has enough magnetic tension, that the  jet is deflected/retarded  during the encounter, and the dense ICM is simply swept up by the jet. In Scenario II (middle), the jet hits a stationary cloud that is intersected by the magnetic filament. Upon impact the jet is deflected by the dense cloud. Momentum is imparted onto the cloud, which is then displaced, dragging the magnetic filament with it. In Scenario III (bottom),  the propagating jet interacts with a moving cloud that crosses its path, generating  a complex interaction.  The \efil are dragged by the subsequent motions in the ICM. For reasons discussed below, we believe that Scenario III is the most likely.

\subsubsection{Bending the jet}\label{sec:bending}

We begin by examining Scenario I. For the magnetic tension in the filament to deflect the incoming jet, the fields will need to be amplified until they become dynamically significant. If the tension force ($\propto B^2/l$, where $l$ is the length over which the filament stretching is concentrated) were comparable to the jet ram pressure, $\rho v_{jet}^2$, then the work to stretch the filament could play a significant role. To put this in context, if the fields near the interaction point were, for example, amplified by a factor of 3, then the tension force could be enhanced by nearly an order of magnitude over analogous tension forces from this filament. However, if the jet is internally supersonic, the momentum flux in the jet will be dominated by the ram pressure $\rho v_j^2$ of the jet material rather than internal pressure in the jet, which is often assumed in pressure equilibrium with the ICM. In this case, even if the filament plasma beta $\beta_p = 8\pi P_{gas}/B^2$. approaches unity, the momentum flux from the kinetic jet will dominate the external tension force. This means the jet material will not be deflected significantly by interaction with the draped filament; this effectively rules out Scenario I.  \rev{In addition, this scenario provides no explanation for the earlier bends in the jet.}

Scenario II implies the presence of a dense cloud which acts to deflect the jet at point C in Figure \ref{fig:RMsliceN}. The Faraday structure, in fact, suggests increased amounts of thermal material in this area (Fig. \ref{fig:moment2}). The impact of the jet onto this cloud imparts momentum to the cloud. The magnetic filaments appear to pass through or near the cloud, so the filaments are then carried along, draped over the cloud. However, such a cloud will not act as a solid body. As jet momentum is deposited in the cloud, the outer portions of the cloud will be stripped away while the core of the cloud may acquire some bulk ``recoil" velocity $v_c$.
Simulations would likely be necessary to fully determine the relationship between an effective recoil velocity, the density ratio between the cloud and the jet material ($\rho_c/\rho_j$), and the jet velocity ($v_j$).

The main difference between Scenarios II and III is the motion of the dense cloud. In Scenario II, the jet must bend sharply by other means prior to the cloud encounter, \rev{such as encounters with additional, otherwise invisible dense clumps,} while in Scenario III, the multiple bends in the jet trajectory are a result of the encounter itself \citep{Nolting22a}. Consideration of the timescales associated with the jet evolution and cloud motion can help establish which of these options is more attractive.  If the cloud's motion is significant during the evolution of the radio jet structure, then we must reject Scenario II. Recent simulations \citep{Nolting22a} show that such interactions strongly alter the character of the jet even during the early phases of an encounter when the jet impacts onto the less dense outer regions of the cloud,
%the character of the jet well before the dense cloud enters fully into the path of the jet, 
so we base our timescale on the half-jet-crossing time. If the cloud is assumed to be in orbit in the cluster, it will likely have a velocity of a few 10\% of the sound speed. If the cluster temperature is 1.4 keV, then the sound speed is close to 600 km/sec. With an estimate of the cloud velocity (taken here to be $v_c \sim 200 \text{km/s}$) and the width of the jet (barely resolved diameter of 2.8~kpc),  we find the time for the cloud to cross half the width of the jet is on the order of 7~Myr. 

Then, we consider the time required for the head of the jet to advance through the ICM. Analytical calculations \citep{Nolting19a} show that the Mach number of the propagation of the head of the jet through the ICM, $M_h$, will be less than the internal Mach number of the jet, $M_j = v_j/c_{s,j}$, where $c_{s,j}$ is the sound speed in the jet. Simulations show $M_h \sim 0.7 M_j$. These radio jets are expected to be supersonic, with internal Mach numbers of a few. Assuming $M_j >$2, then the distance between point A and point C (about $25$~kpc projected) yields an estimate of $\sim$30~Myr for the interaction time. Since the time for the head of the jet to move over the extent of the cloud (30~Myr) is much longer than that for a typical cloud to cross the jet (7~Myr), then the moving cloud scenario is favored.  In addition, the intersection with a moving cloud will cause multiple bends in the jet.  We therefore adopt Scenario III as the most likely option. 

Figure 2 of \citep{Nolting22a} shows a simulation of such a jet-cloud encounter (including an animation in the online journal version), including the multiple resulting bends in the radio jet. The initial glancing deflection could explain emission at point C in 3C40B while the jet propagation prior to the encounter explains the trajectory beyond point D. Since the deflection at point A is approximately 90\degree, there must be a dense core in the cloud near the midpoint of the jet to abruptly reorient the jet. Finally, the deflection of the jet causes it to intersect with the \efilc, whereas it previously passed in front or behind them. Subsequently, ram pressure from the jet and the moving cloud bends and stretches the filaments. In this way, Scenario III naturally explains the multiple sharp bends in the jet given the relative timescales for jet advance and cloud motion.  The presence of the cloud is not \emph{ad hoc}, since it is  supported by the increased Faraday depths in the interaction region.  In this, and any of the scenarios, the presence of the \efil at the interaction point appears fortuitous.  However, if the interaction resulted in the brightening of the \efilc then we would identify them precisely because of this fortuitous interaction.

\subsection{Cosmic Rays}\label{sec:cosmic}
Synchrotron emission in clusters is found in radio galaxies, where the cosmic ray electrons are initially ejected at the AGN, and in diffuse sources such as halos and relics, where more distributed sources of cosmic ray electrons must be invoked.  Filaments provide an interesting intermediate case, where the origin and evolution of their cosmic ray electron populations have not yet been addressed. The \efilc along with the other filaments in Abell~194, present an opportunity to explore the possibilities, as we do in the following.  The discussion is organized as follows:  Sec. \ref{sec:efilCR} explores the possibility that 3C40B is the source of the electrons; in Sec. \ref{sec:CRICM} we provide an overview of the origin of cosmic ray electrons in the ICM, with  Sec. \ref{sec:accfil} then describing a scenario in which cosmic ray electrons from the ICM are accelerated in the \efil. Diffusion of electrons from the filaments is addressed in Sec. \ref{sec:diffusion} and, finally, we discuss in Sec. \ref{sec:distrib} what can be learned from the observed shape of the \efil synchrotron spectrum.

\subsubsection{3C40B as a source of \efil cosmic rays}\label{sec:efilCR}
We now consider whether the radio galaxy 3C40B itself could be the source of the electrons associated with the \efilc  invoking streaming along the magnetic field. 
Modeled via pitch angle scattering, electrons diffusing along field lines in a time, $t$, cover a distance $d \sim 2 \sqrt{D_{||} t}$ where the diffusion coefficient along the field lines is $D_{||} \sim 1/3 (c^2/\nu_s)$. In this expression $\nu_s = \tau_s^{-1}$ is the electron pitch angle scattering rate determined by resonant and non-resonant coupling with kinetic-scale electromagnetic fluctuations in the vicinity of the magnetic filament. 
One major challenge to this possibility is the need for diffusion to be sufficiently fast; that is, electron scattering rates that are slow enough that electrons would propagate at least $\sim$200 kpc along filaments on a time-scale that is shorter than or comparable to their radiative cooling time, ($\tau_{rad}\sim 100$ Myr). This would imply strongly super-Alfvenic streaming along the filaments with a lower limit on the effective scattering time-scale $\tau_s \sim 10^4$ yrs.  % that is, $\tau_s . % $\tau_s \geq 10^{11} / [(\tau_{rad})] \sim ~100$ Myr. 

Streaming electrons from 3C40B could also account for the presence of the diffuse steep spectrum component surrounding the narrow filaments in the \efilp 
The diffuse component could be generated by electron diffusion perpendicular to the local mean field; 
under conditions typical of the ICM, electrons may undergo super-diffusive process up to the Alfven scale \citep[][ and references therein,]{2022ApJ...927...94X} spreading out on $L_{perp} \sim$10-100 kpc while they are diffusing along field lines.

In order for the jet to be the source of electrons for the \efilc the two bright narrow components would need to reconnect with the field of the radio galaxy.  This could happen through the relative motions between the jet and the turbulent ICM.  
In turbulent reconnection theories the reconnection rate (generally defined by the inflow speed of field lines into a reconnection site) is fairly independent of the plasma resistivity and may reach a significant fraction of the local Alfven speed \citep[e.g.,][]{lazarv}. %(e.g., Lazarian \& Vishniac, 1999, ApJ, 517, 700). 
Here the local Alfven speed could be a substantial fraction of the ICM sound speed. Such reconnection modifies the local magnetic field topology and could establish an effective electron transport channel between the radio galaxy lobe/jet and the external ICM.
If magnetic reconnection between radio galaxies and ICM magnetic filaments is happening in Abell~194, this could explain why the filaments are seen so prominently between and near the radio galaxies, and not more broadly distributed throughout the ICM, as traced by the X-ray emission.

Interestingly, this scenario could explain at the same time the increase  in the ratio of brightness between diffuse steep-spectrum emission and the filamentary emission with increasing distance from the radio galaxy. 

\subsubsection{Cosmic rays in the ICM}\label{sec:CRICM}
Relativistic electrons are injected into the ICM by shocks, galactic winds and AGN \citep[e.g.,][ for a review, including the alternative mechanism of producing electrons through cosmic-ray/thermal interactions]{BrunettiJones}. These electrons have a maximum lifetime in the ICM,  $\tau_l \gtrsim$ Gyr, at energies around 100-300 MeV. The lifetime decreases both towards higher energies ($\tau_l \propto E^{-1}$) due to radiative (synchrotron and inverse Compton IC) losses, and towards lower energies ($\tau_l \propto E $) due to Coulomb losses \citep[e.g., ][]{1999ApJ...520..529S}.
The long-lived electrons are advected and distributed on large scales by complex gas motions in the ICM, and constitute a non-thermal component that is effectively mixed with the ICM. 

This non-thermal component can be reaccelerated to higher energies ($\geq$ several GeV) through turbulence and shocks generated within the volume of massive clusters. This is especially true during mergers and through accretion of matter from connecting cosmic filaments.  Turbulence is  expected not only in massive clusters, but also in less massive systems and galaxy groups \citep{2006MNRAS.369L..14V, 2013ApJ...777..137N}%(Vazza, Tormen etal 2006, Nagai etal 2013, ApJ 777, 137).% At some level turbulence is a common ICM feature, although its intensity, distribution and structural characteristics will depend on local drivers and their relations to the host system. After reacceleration, the non-thermal electrons produce observable cluster-scale radio sources in the form of radio halos and relics (e.g., Brunetti \& Jones 2014, van Weeren etal 2019 for reviews).

In relatively relaxed ICM, and in the less massive systems,  low level turbulent electron reacceleration and adiabatic compression resulting from cosmic accretion inflows could balance radiative losses sufficiently to maintain the energy of relativistic electrons at several 100~MeV for times comparable to the cluster lifetime. In this ``minimum maintenance'' situation, synchrotron radiation is expected at ultra-low frequencies, below those accessible to observations. However, even then, localized regions with stronger turbulent energy fluxes, so more effective electron reacceleration and stronger magnetic field amplification, could produce synchrotron emissions at higher frequencies. Filamentary structures with steep synchrotron spectra would be a signature of such localized acceleration/amplification regions, working on the pool of $\sim$100~MeV ``minimum maintenance" electrons.  If radio halos are observed at $>$100~MHz frequencies, then a background source of higher energy electrons already exists, and could feed into more localized filamentary structures.  The very faint diffuse emission south of the \efil (Fig. \ref{fig:diffuse}) suggests the existence of a vast reservoir of seed electrons in Abell~194. Whether or not this is viable could be tested by deeper low frequency observations of the structure and spectra of the diffuse emission. 

\subsubsection{Acceleration in filaments}\label{sec:accfil}
Assuming a background population of $\geq$100~MeV electrons, they will encounter regions where a turbulent magnetic field is stretched and amplified by shear motions and radio galaxy interactions.  %we can look more closely at what would happen in regions of higher turbulence.  Over time, fields in filamentary structures will be amplifed via a turbulent dynamo process (e.g., Tricco etal, 2016, MNRAS,461, 1260). Subsequent stretching by shear motions or radio galaxy interactions, as discussed above, will result in field amplification.  
The electrons spiraling in such an amplifying magnetic fields with $\dot{B}> 0$ are subject to an electromotive force and can consequently be (re)accelerated, e.g.,  via the Betatron mechanism \citep[see ][ for a review]{1980panp.book.....M}.
In the simplest applicable case where the magnetic field grows on a time-scale shorter than the cooling time of electrons, while electron pitch angle isotropy is preserved by scattering, the energy increment of the electrons in this case is $(\Delta pc)^2 \sim (2/3) (pc)^2 \delta B/B = (2/3) (pc)^2 \phi$, with $\phi = \delta B/B$.

The combination of the increase of both magnetic field strength by a factor $\phi>1$ and electron energy leads to a relative increase of the characteristic synchrotron emission frequency, $\nu_s/\nu_{s0} \sim  (1 + 2/3 \phi)(1+\phi) \sim (1+\phi)^2 \sim \phi^2$, where $\nu_{s0}$ is the characteristic frequency prior to the field enhancement. %, where $\phi = \delta B/B$.
This behavior could lead to development of distinctive, steep spectrum synchrotron emissions in the 100-1000 MHz range from the strongest ICM-turbulence-generated filaments as they interact with large scale shear motions and/or radio galaxy jets. This would depend on the presence of magnetic fibers strong enough prior to the amplification that radio emission at ultra-low radio frequencies was already present and sustained at modest levels by reacceleration in the turbulent ICM.

The field stretching that occurs in the \efil is strongest near the encounter with the 3C40B jet, as discussed above. The electron acceleration rate would thus rise strongly closer to the jet, and could at least qualitatively account for the brighter narrow filaments and flatter spectral indices there (Figs. \ref{fig:Broad_Narrow} and \ref{fig:filI}). %More detailed modeling would be required to show whether this is viable, or whether other processes, such as electron diffusion along the filament from higher to lower field regions might be needed to moderate the rate of spectral steepening accompanying radiative energy losses. In the following section, we look at various aspects of electron diffusion.

\subsubsection{Diffusion of cosmic rays from magnetic filaments}\label{sec:diffusion}
In addition to the possibility of diffusion along the \efil with its aligned magnetic fields,  energetic electrons could diffuse perpendicular to the filaments. Such diffusion might be responsible for the broad component of \efilc in which the narrower filaments are embedded. %An intriguing, related issue to understand is whether diffuse emissions seen surrounding one of the filaments 

Electrons propagating along turbulent magnetic field lines on scales $\leq l_A$ ( where $l_A$ is the Alfven scale, i.e. the scale where the turbulent velocity equals the Alfven speed) can effectively exhibit super-linear diffusion in the direction perpendicular to the local mean field due to fast field line wandering. This may sometimes lead to an effective transverse diffusion rate that is comparable to that of pitch-angle-mediated particle diffusion parallel to the mean field \citep[see, e.g.,][]{2021ApJ...923...53L}.%(see, e.g., Lazarian \& Xu (2021) (ApJ, 923, 53)).

In such scenarios, electrons accelerated within the stretching field filaments could diffuse into the surrounding region where the magnetic field is lower.  There,  emissions at a given frequency would come from more energetic components of the electron spectrum that are more impacted by radiation energy losses. The associated electron energy spectrum would then be steeper, consistent with the observed steeper synchrotron component. Such diffusion might even be responsible for the diffuse component filling the space between 3C40A and 3C40B (Fig. \ref{fig:diffuse}), or the faint emission south of the \efilp  %This is the opposite scenario from that presented above, where we suggested these electrons might actually be the seeds for further amplification in the filaments.  Detailed, multifrequency spectral mapping would be required to distinguish between these scenarios. 

\subsubsection{The energy distribution of filament cosmic rays}\label{sec:distrib}
In Fig. \ref{fig:cc} we showed that the observed spectra are significantly broader than, and inconsistent with a simple exponentially cutoff spectrum.  Such broadening would occur if there were substantial inhomogeneities in magnetic field strength along each line of sight, such as expected if the filaments are composed of bundles of fibers )Sec. \ref{sec:shear}).  Such inhomogeneities and broader spectra are also expected in the diffuse lobes of 3C40B (see Sec. \ref{sec:Afilaments}, and Fig. \ref{fig:cc}, respectively). 

The synchrotron spectra would also be broadened if the underlying electron spectra themselves do not cut off exponentially, but fall more gradually at higher energies.  
Such a stretching of the electron spectrum is naturally expected in the case of stochastic re-acceleration, since it induces particle diffusion in momentum space \citep[e.g., ][]{BrunettiJones}.  % Brunetti \& Jones 2014). %Several different models based on turbulent reacceleration are shown in Fig. \ref{fig:ccturb}. 
In particular, Betatron re-acceleration to GeV energies of a population of electrons previously maintained at energies of several 100~MeV by ICM turbulence naturally broadens the spectrum, and thus provides a plausible mechanism for illuminating the \efilp

%\begin{figure}
 %   \centering
%    \includegraphics[width=0.9\columnwidth]{ 40B_CC_TURB_0211.png} \label{fig:ccturb}
%    \caption{Color-color diagram with the same data as Fig. \ref{fig:cc}, highlighting the filament data (black symbols) here. The dot-dashed line corresponds to the models described in Bonafede+20, while the other lines are from TWJ+20 in prep., using a monoenergetic population of 100~MeV fossil electrons.    The dotted line is for  Model 1 (\red{TWJ:explain}), the dashed line is for model i0 (\red{TWJ:explain}), the solid grey line is for Model 30 (\red{TWJ:explain}) and is degenerate with many other models.  }
   
%\end{figure}

\subsection{Filaments and X-ray cavities}\label{sec:cavities}

One of the more striking results from this work is the existence of a channel with extremely low thermal X-ray emitting plasma overlapping with the \efilp  Cavities in the ICM created by radio jets/lobes have been extensively studied \citep[e.g.,][ and references therein]{cavity} and the PdV work needed to create them appears to be consistent with the kinetic luminosity of the radio jets \citep{Birzan}. The southern lobe of 3C40B is another example of a cavity that may have been driven by such jet flows \citep{Bogdan}.

The \efil are perpendicular to the 3C40B jet flow,  so it is unlikely that a bulk flow from the AGN has dumped its energy along the \efil path and created a cavity.  In addition, a cavity produced in such a way  would quickly refill this relatively narrow (35~kpc) low density region.  Shear motions in the ICM   do not themselves create cavities in the medium;  however, by amplifying magnetic fields, accompanied by cosmic ray acceleration, they can build up pressures that could be comparable to those in the thermal ICM, and thus plausibly exclude the thermal plasma. In Abell~194, \citep{Bogdan} suggest that NGC~541 and NGC~545 are falling through the cluster center associated with NGC~547.  The direction of motion is thus W to E, in the same general direction as the \efilc lending support to this picture. % What remains then, to explain the cavities, is that the static pressure of the relativistic plasma in the \efil is comparable to that in the thermal ICM, and thus sufficient to displace that thermal plasma.   The \emph{minimum} pressures calculated from the synchrotron radiation are a substantial fraction of the thermal pressure, so this appears plausible. 

A similar X-ray faint channel was found in Abell~520 by \cite{2016ApJ...833...99W} (see their Fig. 5).  It is observed to be 30~kpc wide and 200~kpc long, very similar to that associated with the \efilp They found, similar to Abell~194, that the observed size of the cavity would not produce a strong enough dip in the X-rays, and assume a much broader evacuated region.(We assumed a flattened thermal medium, instead.)  They also suggest that magnetic fields, enhanced in a minor merger along the direction of the cavity, reached pressures where they could exclude the thermal plasma. % imilar to our arguments about the \efil cavity, \cite{2018ApJ...856..162W} would also require a much larger depth ($>$75~kpc in their case) than the observed width, to explain the dip in X-rays.  They suggested that the cavity could have been formed by a minor merger in the direction, where the magnetic field was enhanced, with its increased pressure expelling the X-ray gas, in the wake of an infalling subcluster. 
In Abell~520, the channel is embedded in extended radio halo-type emission \citep{2019A&A...622A..20H}, but  no radio emission specifically associated with the channel can be recognized.

These observations indicate that shear-amplified magnetic fields can produce X-ray cavities.  It is therefore appropriate to ask whether the buildup of magnetic (and perhaps accompanying cosmic ray) pressure is also important for cavities associated with radio lobes, such as the southern lobe of 3C40B.   Our Faraday results show that the nonthermal synchrotron emissions within the southern lobe are filamentary and intertwined with thermal plasma emissions, since different filaments show different Faraday depths. As discussed by \cite{Bell2019} radio galaxy lobes represent cavities formed by AGN plasma backflowing from the jet head within an ambient thermal plasma. Inside these lobes,  high pressure magnetic flux tubes form intermittently throughout the volume and can accelerate cosmic rays to high energies. Thus, the relative roles of bulk and magnetic pressure in creating X-ray cavities is worth further investigation.

 \color{black}
\section{Concluding Remarks}\label{sec:conclusion}
In Section \ref{global}, we summarized our observational findings on the Abell~194 system. The extensive network of radio filaments, and especially the prominent \efilc illuminate the evolving MHD properties of the ICM and the accompanying acceleration of cosmic rays.  They add to the rapidly rising number of filamentary systems seen in clusters and presented the first opportunity to derive some physical parameters based on their interaction with a jet flow.  

The combination of many factors allowed us to detect and study the \efilp These included observational factors, such as the low redshift of Abell~194, combined with the high resolution of MeerKAT (and LOFAR), combined with excellent sensitivity and dynamic range. Special opportunities were created by Abell~194's low mass and lack of small scale strong Faraday depth fluctuations, combined with the likely large scale motions from the ongoing merger activity.   

The analyses presented here allow us to draw more general conclusions regarding the nature of the ICM and its interactions with the relativistic plasma.  These include:

\begin{itemize}
\item{Magnetized filamentary structures, often or always in bundles, are a natural consequence of high-$\beta$ MHD flows in the lobes of radio galaxies and in the surrounding thermal medium.}
\item{The lengths of filaments are an important indicator of the dominant physical scales in the kinematics of the plasmas;   the widths of the filaments are an important indicator of the physics/microphysics regulating their origin and evolution.}
\item{Magnetic fields in the filaments can be stretched and amplified by interactions by flows in the ICM and radio galaxies to levels where the magnetic pressure reflects the stresses in the ICM flows ($\rho v^2/\ell$, where $\rho$ is the density of the turbulent plasma, with a characteristic velocity $v$ on the driving scale $\ell$). This is comparable to the thermal pressure ($\beta \sim 1$), so magnetic filaments can form X-ray cavities.}
\item{Magnetic filaments embedded in cluster turbulence are an additional site where CRe can be reaccelerated, e.g., out of a ``minimum-maintenance" pool of several~100~MeV electrons.}
\item{Where the foreground fluctuations are small in Faraday depth, the observed Faraday depths to filaments and jets permits  characterization of their 3D structures.}
\item {Inside the lobes of radio galaxies, filamentary synchrotron-emitting structures appear intertwined with and thus have comparable pressures to the surrounding thermal plasma. }
\end{itemize} 

These findings point the way to important opportunities for further study.  For 3D Faraday structure mapping, studies of other radio galaxies in groups or other situations where only minor small-scale Faraday depth fluctuations are present would be important.  Additional examples where filament structures had been perturbed by their interactions with AGN outflows could test the ideas presented here, in particular the roles of magnetic field amplification via stretching.  Multifrequency studies of the spectral structures of filaments, and possible variations as a function of position from potential cosmic ray sources are needed to probe the acceleration histories.   Prospects for much higher resolution studies of filaments, to  characterize their substructures, are challenging, but essential.

\section{Acknowledgements}
The MeerKAT telescope is operated by the South African Radio Astronomy Observatory, which is a facility of the National Research Foundation, an agency of the South Africa Department of Science and Innovation. X-ray data were obtained from the \textit{Chandra} X-ray Observatory Data Archive and the XMM-Newton Science Archive. 
  Funding for the SDSS has been provided by the Alfred P. Sloan Foundation, the Participating Institutions, the National Science Foundation, the U.S. Department of Energy, the National Aeronautics and Space Administration, the Japanese Monbukagakusho, the Max Planck Society, and the Higher Education Funding Council for England. 
  This paper is based in part on results obtained with LOFAR equipment. LOFAR \citep{2013A&A...556A...2V} is the Low Frequency Array designed and constructed by ASTRON in the Netherlands. It has observing, data processing, and data storage facilities in several countries, which are
owned by various parties (each with their own funding sources), and are collectively operated by the ILT foundation under a joint scientific policy. The
ILT resources have benefitted from the following recent major funding sources:
CNRS-INSU, Observatoire de Paris and Universit\'e d’Orl\'eans, France; BMBF,
MIWF-NRW, MPG, Germany; Science Foundation Ireland (SFI), Department
of Business, Enterprise and Innovation (DBEI), Ireland; NWO, The Netherlands; The Science and Technology Facilities Council, UK; Ministry of Science and Higher Education, Poland; Istituto Nazionale di Astrofisica (INAF),
Italy. This research made use of the Dutch national e-infrastructure with support
of the SURF Cooperative (e-infra 180169) and the LOFAR e-infra group, and
of the LOFAR-IT computing infrastructure supported and operated by INAF,
and by the Physics Dept. of Turin University (under the agreement with Consorzio Interuniversitario per la Fisica Spaziale) at the C3S Supercomputing Centre, Italy. The J\"ulich LOFAR Long Term Archive and the German LOFAR
network are both coordinated and operated by the J\"ulich Supercomputing Centre (JSC), and computing resources on the supercomputer JUWELS at JSC were
provided by the Gauss Centre for Supercomputing e.V. (grant CHTB00) through
the John von Neumann Institute for Computing (NIC). This research made use
of the University of Hertfordshire high-performance computing facility and the
LOFAR-UK computing facility located at the University of Hertfordshire and
supported by STFC [ST/P000096/1].

M.B. acknowledges funding by the Deutsche For-schungsgemeinschaft (DFG, German Research Foundation) under Germany’s Excellence Strategy – EXC 2121 ‘Quantum Universe’ – 390833306.  WC acknowledges support from the National Radio Astronomy Observatory, which is a facility of the U.S. National Science Foundation operated under cooperative agreement by Associated Universities, Inc.. Partial support for CN comes from National Science Foundation grant AST19-07850 to the College of Charleston.
Partial support for LR and TWJ comes from National Science Foundation grant AST17-14205 to the University of Minnesota. RJvW acknowledges support from the ERC Starting Grant ClusterWeb 804208.   WF acknowledges support from the Smithsonian Institution, the Chandra High Resolution Camera Project through NASA contract NAS8-03060, and NASA Grants 80NSSC19K0116, GO1-22132X, and GO9-20109X. \rev{GS acknowledges support from NASA through contract  80NSSC19K0116.}

\clearpage
\appendix
\section{Abell 194 Filament System}\label{sec:Afilaments}
The filaments discussed in this paper (\efil) are the brightest of an extensive network of filaments both interior to the lobes of 3C40B and elsewhere in the ICM, especially in
the region between 3C40B and the extended linear twin structures to the north of 3C40A.  Their analysis is beyond the scope of the current paper, but would certainly
contribute to a more complete understanding of filament origins and properties. 

One major question is whether filaments represent minor enhancements in emissivity above the more diffuse surrounding emission, or whether they illuminate a highly ``intermittent" magnetized plasma with large variations in emissivity. To provide some quantitative information on this issue, we measured the rms fluctuations in an $\sim$310\arcsec$\times$240\arcsec~ box in the far northern lobe of 3C40B (in the original, not filtered 7.75\arcsec~ map), and compared that to the mean brightness.  The rms was 70$\mu$Jy/beam and the mean was 460$\mu$Jy/beam, respectively, or $\sim$15\%. If the filaments are cylindrical, then the line of sight through them is $\sim$5-10$\times$ less than that of the more diffuse lobe. \emph{They thus represent changes in emissivity of order unity; they are not minor enhancements.}  The bright southern lobe associated with the X-ray has corresponding values of 1.4~mJy/beam rms and 5.3~mJy/beam mean, or $\sim$25\%, with similar implications.  The enhancements in emissivity are even more pronounced in the filaments in the far southern lobe and the region between 3C40B and 3C40A; the corresponding  rms(mean), \% and [box size] values are 140(280)~$\mu$Jy/beam, $\sim$50\%,[250\arcsec$\times$250\arcsec]  and  90(220)$\mu$Jy/beam, $\sim$40\%,[275\arcsec$\times$100\arcsec], respectively.

\begin{figure}[h]
    \centering
    \includegraphics[width=0.7\columnwidth]{ 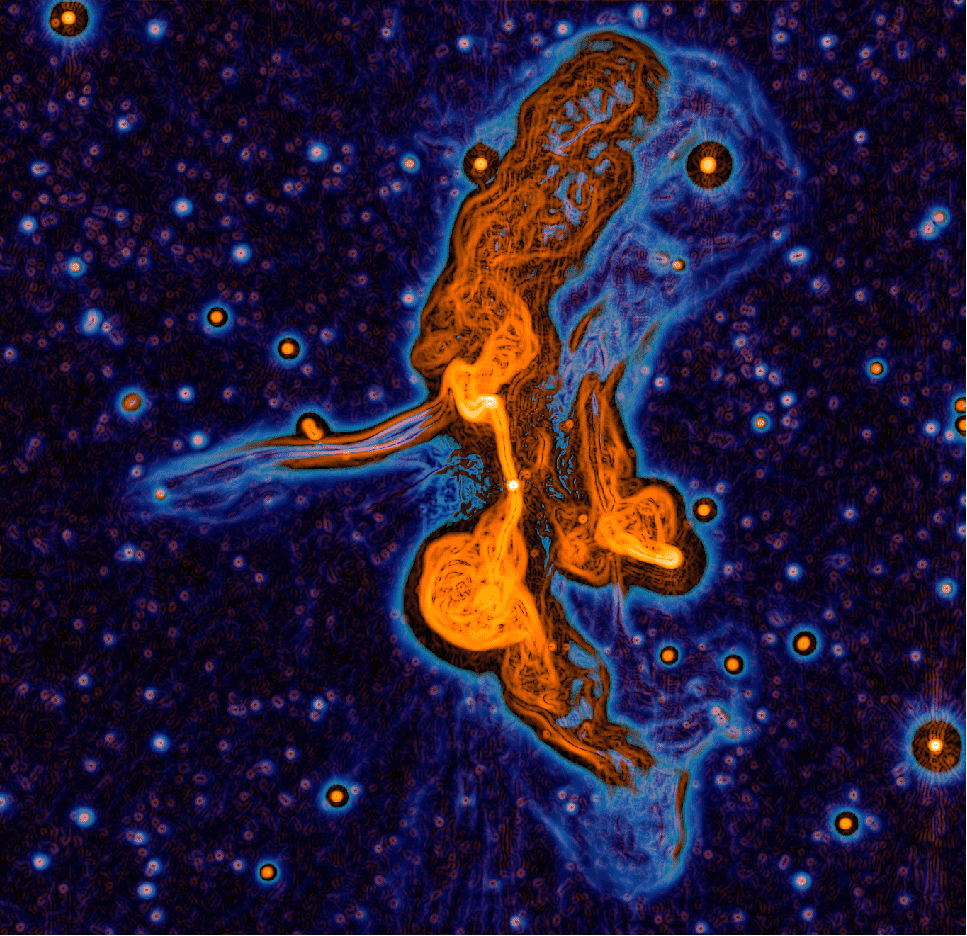}
    \caption{Filtered 7.75\arcsec~ image of Abell~194 to emphasize the filamentary structures inside and outside of the radio galaxy lobes.  In orange, the results from a Sobel edge-detection filter (NINER task in AIPS), and in blue, from Gaussian gradient magnitude (GGM) filtering \citep{2016MNRAS.461..684W}.}
 \label{fig:A194fils}   
\end{figure}
\clearpage
\section{Explorations in 3D}\label{sec:3D}
In this paper, we have inferred the 3D structure of various features using their Faraday depths.  This appears to be a very promising diagnostic tool, under the right conditions.  Here we explore the major assumptions and caveats behind the 3D analysis.

Abell~194 presents an unusual opportunity by virtue of its very small scatter in Faraday depth \citep[][and as observed here, $\sim$10~ \radmm]{Govoni2017}.  In more massive clusters, the typical scatter in RM values is much larger -- e.g.,  $\sim$150~rad/m$^2$ near the center of Abell~2345 \citep{ 2021MNRAS.502.2518S} and 303 and 166~ \radmm~ for the Coma cluster central galaxies \citep{Bonafede10}.  % For example, the rms scatter in RM for the two sources in the central 200~kpc of  is $\sim$150~rad/m$^2$. For the much more massive Coma cluster \citep{Bonafede10}, the scatter for the two central bright radio galaxies were 303 and 166~ \radmm.  None of these values include the contribution from differences in mean RM between the sources which would increase the global scatter further.  By contrast, for the central regions of Abell~194, dominated by emission from 3C40A and B,  the  equivalent global rms is $\sim$10~ \radmm~ an order of magnitude lower than the above examples. This value was calculated from the interquartile ranges of the peak RM and first moment of the Faraday depth, 14.5 and 14~ \radmm, respectively,
The mean RM for Abell~194 is also consistent with the Milky Way foreground (Fig. \ref{fig:RMhist}), implying no detectable \emph{net} contribution from the cluster. We also measured the RMs at the peaks of 21 components from background and possible cluster sources where the polarized intensity was $>$100$\mu$Jy/beam.  The net mean (rms scatter) for these polarized components were 5.7 (8.3~ \radmm).    The low mass Abell~194 system is thus currently unique among clusters with Faraday studies, where the foreground contributions to the RM are small.  %compared to those associated with associated with the ICM in which the sources are embedded.
Random foreground variations in RM, e.g., are very unlikely to produce distinct RM structures right where the northern jet of 3C40B undergoes multiple bends, (Fig. \ref{fig:overview}, bottom right) compared to the adjacent \efil and diffuse lobes. Similarly, in the southern lobe, different filamentary structures have different RMs, which is unlikely to arise with random foreground patches. One  simple interpretation of the RM variations, then, is that they represent the depth to each structure in the embedding thermal medium.

The question remains about how to translate between physical depth and Faraday depth.  The Faraday depth is  $\propto$B~n$_e$~$dl$, where B is the (signed) component of the magnetic field along the line-of-sight, n$_e$ is the local electron density and $dl$ is the incremental path length.  %The observed Faraday depth values will be those associated with the local ICM plus the contribution from the foreground contributions from the Milky Way.
Because of the sign ambiguity, an increasing Faraday depth in a given location could be due to a structure being farther away from us with a local field pointing away from us, \emph{or} a structure being closer to us with a local field pointing towards us.  %In some cases,  this ambiguity can be resolved;  if a large structure has Faraday depths that are all more positive than the Milky Way contribution, then we can conclude that the net field is pointing away from us, and that the more positive values corresponding to structures that are further away along the line-of-sight.  However, 
If the overall structure has Faraday depths that scatter around the Milky Way value, as is true for Abell 194 (see Fig. \ref{fig:RMhist}), then there is no net large scale field to resolve the sign ambiguity. %In this case, the Faraday depth in each beam could, in principle, reflect a different front-back sign.  More likely, however, is that 
The continuity of Faraday depths as a function of position (as seen in multiple figures here) suggests that for large regions the sign stays constant while the line-of-sight depth changes.
For determining 3D structures, changes in the strength of B and the local density n$_e$  will also change the scaling of physical depth with Faraday depth.   %Changes in those parameters will change the Faraday depth from place to place with no corresponding variation in distance along the line-of-sight.  
At present, there is no way to separate out such variations on small scales;  on larger scales e.g., 100s of kpc, some constraints might be available from X-ray modeling of the density, coupled with assumptions about how magnetic field scales with density.

 In Fig. \ref{fig:rmpatfil}, we show the Faraday depth along the the \efilc together with the sky view.  There is a general, but not perfect, correspondence between bends in the jet seen in the plane of the sky and reversals of direction in the magnitude of the Faraday depth along the path.  The simplest explanation is that the jet is underdoing bending in three dimensions.   With a large enough sample of bends and excursions we could make the assumption that the plane of the sky is not a privileged direction and that the bends are of similar magnitude in all 3 dimensions.  With this assumption, we could ask what magnitude of magnetic field was present in the surrounding medium.  We illustrate this exercise using a local density of $\sim$10$^{-3}cm^{-3}$ derived from the X-rays, a distance between bends of $\sim$15~kpc, Faraday depth ($\phi$) and variations between reversals of $\sim$30~rad/m$^2$.  This yields B($\mu$G)=$\phi$~[0.8~n$_e$(cm$^{-3}$)~dl (pc)]$^{-1}\sim2.5\mu$G, similar to the value of $\sim1.4\mu$G derived in Section \ref{sec:closeup}.

As an exploratory tool, we now assume a constant scaling of physical depth with Faraday depth, with no sign changes,  
and present four movies of selected regions \emph{(see the files named in the caption)}. %, assuming that the Faraday depth $\Phi$ can be interpreted as the distance along the line-of-sight .  
For visualization purposes, we adopt a somewhat arbitrary scaling of 0.56~ \radmm~ = 1\arcsec, which yields curvatures in structures along the line-of-sight similar to curvatures in the plane of the sky for the northern jet.    Note that there is a front-back ambiguity, corresponding to the unknown sign of the magnetic field along the line-of-sight. 

In the \efil movie, the filaments curve along the line of sight as they approach the right edge (where they meet 3C40B's northern jet), exactly where they ``wrap" around the jet in the plane of the sky (Fig. \ref{fig:wrapfil}).  In the northern jet movie, the bends along the line-of-sight occur where bends are seen in the plane of the sky.  At the bright spot (point A in Fig. \ref{fig:RMsliceN}), the structure is broadened in Faraday depth space, suggesting a dense thermal region at the bend.  The southern lobe looks flattened, perhaps because the densities there are actually lower than in the assumed scaling. %It is likely that the densities in the southern lobe are lower, e.g., than in the jet, so that it appears artificially flattened.  It could be forced to be closer to circular by increasing the number of arcsec per~ \radmm.  
Similarly, the \efil are likely in lower density regions than the jets, and could be more curved along the line-of-sight than shown here.  We include a movie of 3C40A, which is not otherwise discussed in this paper.  The pair of filaments extended north from the radio galaxy appear to twist around each other, as is also suggested by the apparent braiding of the Faraday depth colored structures in Fig. \ref{fig:overview}.
%\begin{figure*}[h]
%    \centering
%    \includegraphics[width=0.50\textwidth]{ EfilsRM.png} 
%     \includegraphics[width=0.4\textwidth]{ NjetRM.png}
%      \includegraphics[width=0.5\textwidth]{ 3C40SlobeRM.png}
%       \includegraphics[width=0.4\textwidth]{ 3C40ARM.png}

%    \caption{Polarized intensity snapshots of selected regions in Abell~194 at a rotation of 30$^o$ in the RA/Dec/$\Phi$ cube. All three axes are labeled. Each pixel in the image shows the maximum brightness along that line-of-sight in the projected/rotated cube. Animated versions of each of these figures are available in the HTML version \emph{(for review, see file names below)}.  The animations are each 12~seconds long, and rotate around the Declination axis, starting with an orientation in the plane of the sky (0$^o$), rotating 180$^o$ to view the source from ``behind", and then returning to 0$^o$. Midway through the rotation, at 90$^o$, the view is in the Declination (vertical) vs. $\phi$ (horizontal) plane, effectively a ``side view" of the structure.  The animations do not have axis labels. The selected regions (and corresponding file names) are:
%    Top left: {\efil} (Efil-s.mpeg); Top right: {3C40B northern jet} (3C40Net-s.mpeg); Bottom left: {3C40B southern lobe} (3C40Slobe-s.mpeg); and Bottom right: {3C40A} (3C40A-s.mpeg).
 %       Top left: \href{https://drive.google.com/file/d/1jJ6UalY6j3EZ7L8qEg23uVBumgbFYLpv/view?usp=sharing}{\efil}; Top right: \href{https://drive.google.com/file/d/1jE1EuH-BP18C-Qv2tKEwX5EyiUKqA3_X/view?usp=sharing}{3C40B northern jet}; Bottom left: \href{https://drive.google.com/file/d/1jI6pSALawFG2JDsjBifL5cqefPKxeXzq/view?usp=sharing}{3C40B southern lobe}; and Bottom right: \href{https://drive.google.com/file/d/1jCoownC0HeuEGxYPlkTYyrs6wBT_dKKX/view?usp=sharing}{3C40A}.
%    }
%    \label{fig:RMcubes}
%\end{figure*}

\begin{figure*}[h]
    \centering
    \includegraphics[width=0.75\textwidth]{ 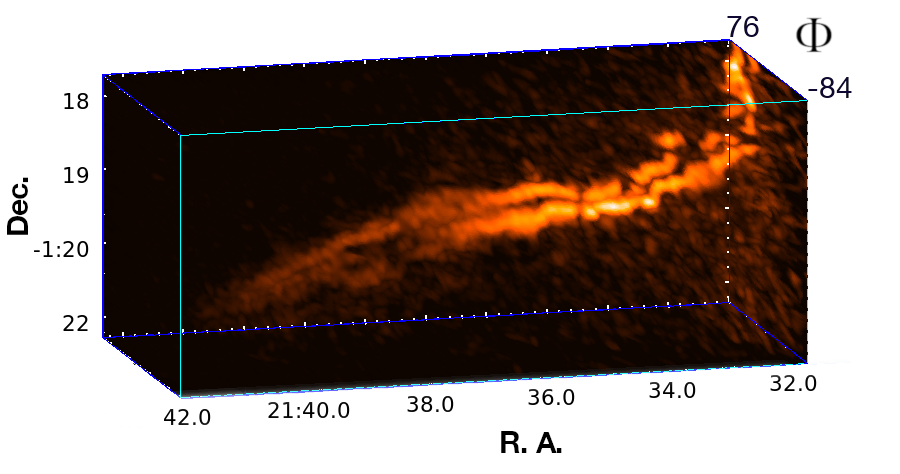} 

    \caption{Polarized intensity snapshot \rev{of the \efil} at a rotation of 30$^o$ in the RA/Dec/$\Phi$ cube. All three axes are labeled. Each pixel in the image shows the maximum brightness along that line-of-sight in the projected/rotated cube.The animation is available as Efil-s.mpeg, under Ancillary Files on the arXiv download page for this preprint  The animation is 12~seconds long, and rotates around the Declination axis, starting with an orientation in the plane of the sky (0$^o$), then rotating 180$^o$ to view the source from ``behind", and then returning to 0$^o$. Midway through the rotation, at 90$^o$, the view is in the Declination (vertical) vs. $\phi$ (horizontal) plane, effectively a ``side view" of the structure.  The animation does not have axis labels. 
   }
    \label{fig:RMcubesA}
\end{figure*}

\begin{figure*}[h]
    \centering
  
     \includegraphics[width=0.65\textwidth]{ 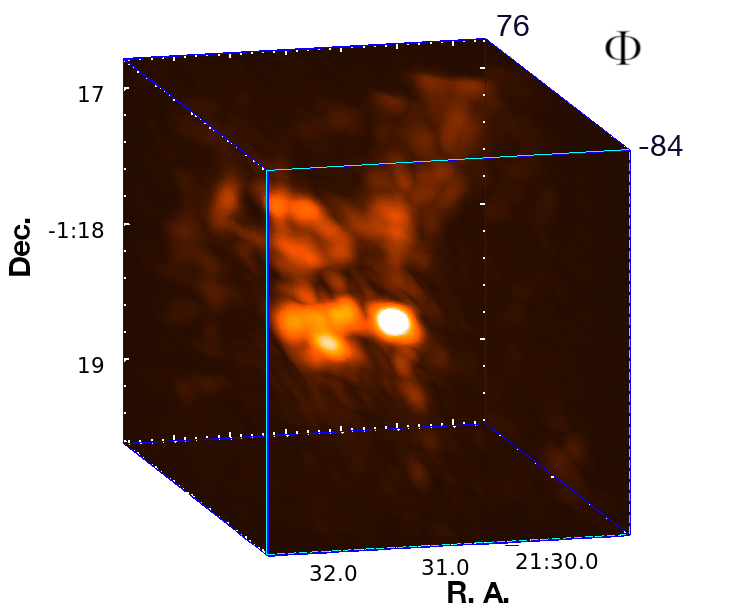}

    \caption{\rev{Polarized intensity snapshot of the region around the bends in the north jet of 3C40B.  See the detailed description of the accompanying animation in the caption of Fig. \ref{fig:RMcubesA}. The animation is available as 3C40Net-s.mpeg, under Ancillary Files on the arXiv download page for this preprint. }
   }
    \label{fig:RMcubesB}
\end{figure*}

\begin{figure*}[h]
    \centering
 
      \includegraphics[width=0.6\textwidth]{ 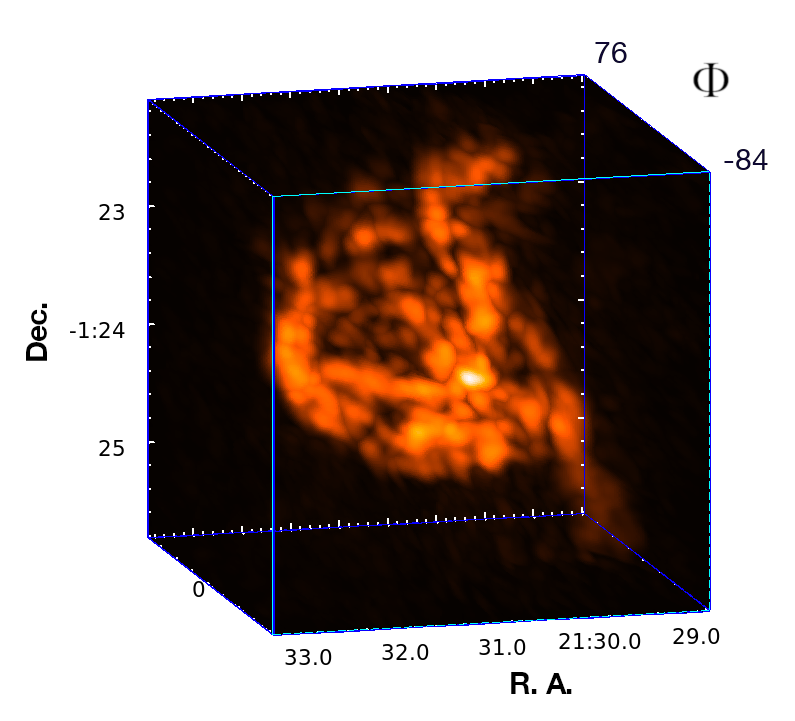}

    \caption{\rev{Polarized intensity snapshot of the southern lobe of 3C40B.  See the detailed description of the accompanying animation in the caption of Fig. \ref{fig:RMcubesA}. The animation is available as 3C40Slobe-s.mpeg, under Ancillary Files on the arXiv download page for this preprint. }
   }
    \label{fig:RMcubesC}
\end{figure*}
\begin{figure*}[h]
    \centering
       \includegraphics[width=0.55\textwidth]{ 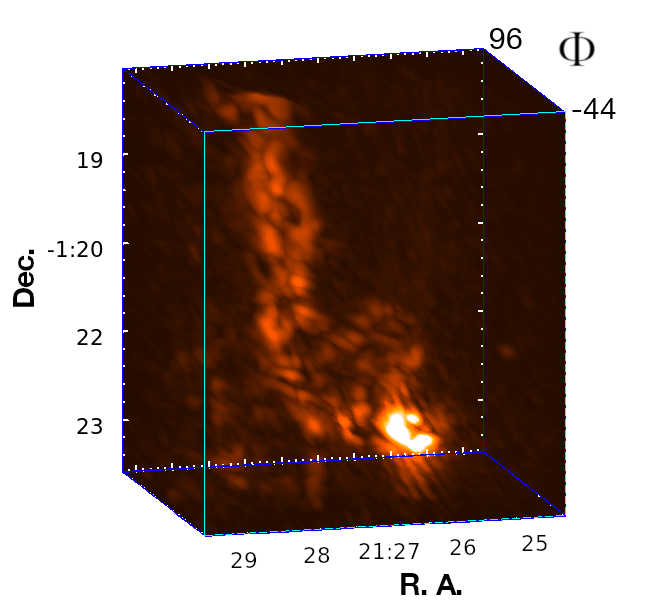}

     \caption{\rev{Polarized intensity snapshot of 3C40A.  See the detailed description of the accompanying animation in the caption of Fig. \ref{fig:RMcubesA}. The animation is available as 3C40A-s.mpeg, under Ancillary Files on the arXiv download page for this preprint.}
   }
    \label{fig:RMcubesD}
\end{figure*}


\clearpage
\begin{thebibliography}{}
%%%%%%%%%%%%%%%%%%%%%%%%%%%%%
%%%%%%%%%%%%%%%%%%%%%%%%%%%%%
\bibitem[Abell et al.(1989)]{Abell89} Abell, G.~O., Corwin, H.~G., \& Olowin, R.~P.\ 1989, \apjs, 70, 1. doi:10.1086/191333
\bibitem[Alam et al.(2015)]{2015ApJS..219...12A} Alam, S., Albareti, F.~D., Allende Prieto, C., et al.\ 2015, \apjs, 219, 12. doi:10.1088/0067-0049/219/1/12
\bibitem[Bell et al. (2019)]{Bell2019} https://doi.org/10.1093/mnras/stz1604
\bibitem[B{\^\i}rzan et al.(2004)]{Birzan} B{\^\i}rzan, L., Rafferty, D.~A., McNamara, B.~R., et al.\ 2004, \apj, 607, 800. doi:10.1086/383519
\bibitem[Bogdan et al. (2011)]{Bogdan} Bogdan, A, Kraft, R.P., Forman, W. R. et al., 2011, ApJ 743:59
\bibitem[Botteon et al.(2020)]{messA2255} Botteon, A., Brunetti, G., van Weeren, R.~J., et al.\ 2020, \apj, 897, 93. doi:10.3847/1538-4357/ab9a2f
%\bibitem[Brodie et al.(1985)]{1985ApJ...293L..59B} Brodie, J.~P., Bowyer, S., \& McCarthy, P.\ 1985, \apjl, 293, L59. doi:10.1086/184491
\bibitem[Bonafede et al.(2021)]{2021ApJ...907...32B} Bonafede, A., Brunetti, G., Vazza, F., et al.\ 2021, \apj, 907, 32. doi:10.3847/1538-4357/abcb8f %SW bridge particle accel
\bibitem[Bonafede et al.(2010)]{Bonafede10} Bonafede, A., Feretti, L., Murgia, M., et al.\ 2010, \aap, 513, A30. doi:10.1051/0004-6361/200913696
\bibitem[Botteon et al.(2020)]{2020ApJ...897...93B} Botteon, A., Brunetti, G., van Weeren, R.~J., et al.\ 2020, \apj, 897, 93. doi:10.3847/1538-4357/ab9a2f
\bibitem[Brodie et al.(1985)]{1985ApJ...293L..59B} Brodie, J.~P., Bowyer, S., \& McCarthy, P.\ 1985, \apjl, 293, L59. doi:10.1086/184491
\bibitem[Brienza et al.(2022)]{Brienza} Brienza, M., Lovisari, L., Rajpurohit, K., et al.\ 2022, arXiv:2201.04591
\bibitem[Brentjens \& de Bruyn(2005)]{Brentjens} Brentjens, M.~A. \& de Bruyn, A.~G.\ 2005, \aap, 441, 1217. doi:10.1051/0004-6361:20052990
\bibitem[Brienza et al.(2021)]{2021NatAs...5.1261B} Brienza, M., Shimwell, T.~W., de Gasperin, F., et al.\ 2021, Nature Astronomy, 5, 1261. doi:10.1038/s41550-021-01491-0
\bibitem[Brienza et al.(2022)]{2022arXiv220104591B} Brienza, M., Lovisari, L., Rajpurohit, K., et al.\ 2022, arXiv:2201.04591
\bibitem[Brunetti et al.(1997)]{1997A&A...325..898B} Brunetti, G., Setti, G., \& Comastri, A.\ 1997, \aap, 325, 898
\bibitem[Brunetti \& Jones(2014)]{BrunettiJones} Brunetti, G. \& Jones, T.~W.\ 2014, International Journal of Modern Physics D, 23, 1430007-98. doi:10.1142/S0218271814300079
\bibitem[Camilo et al.(2018)]{Camilo2018} Camilo, F., Scholz, P., Serylak, M., et al.\ 2018, \apj, 856, 180. doi:10.3847/1538-4357/aab35a
\bibitem[Churazov et al.(2000)]{2000A&A...356..788C} Churazov, E., Forman, W., Jones, C., et al.\ 2000, \aap, 356, 788
\bibitem[Churazov et al.(2012)]{2012MNRAS.421.1123C} Churazov, E., Vikhlinin, A., Zhuravleva, I., et al.\ 2012, \mnras, 421, 1123. doi:10.1111/j.1365-2966.2011.20372.x
\bibitem[Condon et al.(2021)]{Condon21} Condon, J.~J., Cotton, W.~D., White, S.~V., et al.\ 2021, \apj, 917, 18. doi:10.3847/1538-4357/ac0880
\bibitem[Cotton (2008)]{Cotton08}Cotton, W. D. 2008, \pasp, 120, 439
\bibitem[Cotton et al. (2018)]{SourceSize} Cotton, W., Condon, J., Kellermann, K. et al. 2018 ApJ 856, 67
\bibitem[Cotton \& Rudnick (2022)]{Complex} Cotton, W. \& Rudnick, L. 2022, in preparation.
\bibitem[Croft et al.(2006)]{2006ApJ...647.1040C} Croft, S., van Breugel, W., de Vries, W., et al.\ 2006, \apj, 647, 1040. doi:10.1086/505526
\bibitem[de Gasperin et al.(2017)]{2017SciA....3E1634D} de Gasperin, F., Intema, H.~T., Shimwell, T.~W., et al.\ 2017, Science Advances, 3, e1701634. doi:10.1126/sciadv.1701634
\bibitem[de Gasperin et al.(2019)]{2019A&A...622A...5D} de Gasperin, F., Dijkema, T.~J., Drabent, A., et al.\ 2019, \aap, 622, A5. doi:10.1051/0004-6361/201833867
\bibitem[de Gasperin et al.(2022)]{2022A&A...659A.146D} de Gasperin, F., Rudnick, L., Finoguenov, A., et al.\ 2022, \aap, 659, A146. doi:10.1051/0004-6361/202142658
\bibitem[Ekers et al.(1978)]{precess} Ekers, R.~D., Fanti, R., Lari, C., et al.\ 1978, \nat, 276, 588. doi:10.1038/276588a0
\bibitem[Fabian(2012)]{2012ARA&A..50..455F} Fabian, A.~C.\ 2012, \araa, 50, 455. doi:10.1146/annurev-astro-081811-125521
\bibitem[Galishnikova et al.(2022)]{2022arXiv220107757G} Galishnikova, A.~K., Kunz, M.~W., \& Schekochihin, A.~A.\ 2022, arXiv:2201.07757
\bibitem[Gendron-Marsolais et al.(2021)]{2021ApJ...911...56G} Gendron-Marsolais, M.-L., Hull, C.~L.~H., Perley, R., et al.\ 2021, \apj, 911, 56. doi:10.3847/1538-4357/abddbb
\bibitem[Govoni et al.(2005)]{2005A&A...430L...5G} Govoni, F., Murgia, M., Feretti, L., et al.\ 2005, \aap, 430, L5. doi:10.1051/0004-6361:200400113
\bibitem[Govoni et al.(2017)]{Govoni2017} Govoni, F., Murgia, M., Vacca, V., et al.\ 2017, \aap, 603, A122. doi:10.1051/0004-6361/201630349
\bibitem[Guidetti et al.(2011)]{2011MNRAS.413.2525G} Guidetti, D., Laing, R.~A., Bridle, A.~H., et al.\ 2011, \mnras, 413, 2525. doi:10.1111/j.1365-2966.2011.18321.x
\bibitem[Guidetti et al.(2012)]{2012MNRAS.423.1335G} Guidetti, D., Laing, R.~A., Croston, J.~H., et al.\ 2012, \mnras, 423, 1335. doi:10.1111/j.1365-2966.2012.20961.x
\bibitem[Jetha et al.(2006)]{Jetha2006} Jetha, N.~N., Hardcastle, M.~J., \& Sakelliou, I.\ 2006, \mnras, 368, 609. doi:10.1111/j.1365-2966.2006.10155.x
\bibitem[Hines et al.(1989)]{1989ApJ...347..713H} Hines, D.~C., Owen, F.~N., \& Eilek, J.~A.\ 1989, \apj, 347, 713. doi:10.1086/168163
\bibitem[Hlavacek-Larrondo et al.(2015)]{cavity} Hlavacek-Larrondo, J., McDonald, M., Benson, B.~A., et al.\ 2015, \apj, 805, 35. doi:10.1088/0004-637X/805/1/35
es, D.~C., Owen, F.~N., \& Eilek, J.~A.\ 1989, \apj, 347, 713. doi:10.1086/168163
\bibitem[Hoang et al.(2019)]{2019A&A...622A..20H} Hoang, D.~N., Shimwell, T.~W., van Weeren, R.~J., et al.\ 2019, \aap, 622, A20. doi:10.1051/0004-6361/201833900
\bibitem[Hu et al.(2022)]{2022arXiv220304977H} Hu, H., Qiu, Y., Gendron-Marsolais, M.-L., et al.\ 2022, arXiv:2203.04977
\bibitem[Hudaverdi et al.(2006)]{Hudaverdi06} Hudaverdi, M., Kunieda, H., Tanaka, T., et al.\ 2006, \pasj, 58, 931. doi:10.1093/pasj/58.6.931

\bibitem[Jaffe \& Perola(1973)]{1973A&A....26..423J} Jaffe, W.~J. \& Perola, G.~C.\ 1973, \aap, 26, 423
\bibitem[Johnson et al.(2020)]{JohnsonRM} Johnson, A.~R., Rudnick, L., Jones, T.~W., et al.\ 2020, \apj, 888, 101. doi:10.3847/1538-4357/ab5d30
\bibitem[Jonas \& MeerKAT Team(2016)]{Jonas2016} Jonas, J. \& MeerKAT Team\ 2016, MeerKAT Science: On the Pathway to the SKA, 1

\bibitem[Katz-Stone et al.(1993)]{color-color} Katz-Stone, D.~M., Rudnick, L., \& Anderson, M.~C.\ 1993, \apj, 407, 549. doi:10.1086/172536
\bibitem[Katz-Stone \& Rudnick(1997)]{tomog} Katz-Stone, D.~M. \& Rudnick, L.\ 1997, \apj, 479, 258. doi:10.1086/303882
\bibitem[Knowles et al.(2022)]{MGCLS} Knowles, K., Cotton, W.~D., Rudnick, L., et al.\ 2022, \aap, 657, A56. doi:10.1051/0004-6361/202141488
\bibitem[Lame'e(2017)]{Lamee} Lame'e, M.~M.\ 2017, Ph.D. Thesis
\bibitem[Lazarian \& Vishniac(1999)]{lazarv} Lazarian, A. \& Vishniac, E.~T.\ 1999, \apj, 517, 700. doi:10.1086/307233
\bibitem[Lazarian \& Xu(2021)]{2021ApJ...923...53L} Lazarian, A. \& Xu, S.\ 2021, \apj, 923, 53. doi:10.3847/1538-4357/ac2de9
\bibitem[Lovisari et al.(2015)]{Lovisari15} Lovisari, L., Reiprich, T.~H., \& Schellenberger, G.\ 2015, \aap, 573, A118. doi:10.1051/0004-6361/201423954
\bibitem[Lupton \& Gott(1982)]{orbit} Lupton, R.~H. \& Gott, J.~R.\ 1982, \apj, 255, 408. doi:10.1086/159841
\bibitem[Mandal et al.(2020)]{2020A&A...634A...4M} Mandal, S., Intema, H.~T., van Weeren, R.~J., et al.\ 2020, \aap, 634, A4. doi:10.1051/0004-6361/201936560
\bibitem[Mahdavi et al. (2005)]{Mahdavi} Mahdavi, A., Finoguenov, A., B{\"o}hringer, H., et al.\ 2005, \apj, 622, 187. doi:10.1086/427916  %Background clsuter in X-ray
\bibitem[Markevitch \& Vikhlinin(2007)]{2007PhR...443....1M} Markevitch, M. \& Vikhlinin, A.\ 2007, \physrep, 443, 1. doi:10.1016/j.physrep.2007.01.001
\bibitem[McNamara et al.(2000)]{2000ApJ...534L.135M} McNamara, B.~R., Wise, M., Nulsen, P.~E.~J., et al.\ 2000, \apjl, 534, L135. doi:10.1086/312662
\bibitem[Melrose (1980)]{1980panp.book.....M} Melrose, D.~B.\ 1980, New York: Gordon and Breach, 1980
\bibitem[Milne(1995)]{1995MNRAS.277.1435M} Milne, D.~K.\ 1995, \mnras, 277, 1435. doi:10.1093/mnras/277.4.1435
\bibitem[Miniati \& Beresnyak(2015)]{2015Natur.523...59M} Miniati, F. \& Beresnyak, A.\ 2015, \nat, 523, 59. doi:10.1038/nature14552
\bibitem[M{\"u}ller et al.(2021a)]{2021MNRAS.508.5326M} M{\"u}ller, A., Pfrommer, C., Ignesti, A., et al.\ 2021, \mnras, 508, 5326. doi:10.1093/mnras/stab2928
\bibitem[M{\"u}ller et al.(2021b)]{2021NatAs...5..159M} M{\"u}ller, A., Poggianti, B.~M., Pfrommer, C., et al.\ 2021, Nature Astronomy, 5, 159. doi:10.1038/s41550-020-01234-7
\bibitem[Murgia et al.(2004)]{2004A&A...424..429M} Murgia, M., Govoni, F., Feretti, L., et al.\ 2004, \aap, 424, 429. doi:10.1051/0004-6361:20040191
\bibitem[Nagai et al.(2013)]{2013ApJ...777..137N} Nagai, D., Lau, E.~T., Avestruz, C., et al.\ 2013, \apj, 777, 137. doi:10.1088/0004-637X/777/2/137
\bibitem[Nolting et al. (2019)]{Nolting19a} {Nolting}, Chris and {Jones}, T.~W., {O'Neill}, Brian J. and {Mendygral}, P.~J., 2019, \apj, 876, 154.
\bibitem[Nolting et al. (2022)]{Nolting22a} {Nolting}, C. and {Lacy}, M. and {Croft}, S. and {Fragile}, P. C. and {Linden}, S. T. and {Nyland}, K.  and {Patil}, P., 2022, submited to \apj .
\bibitem[O'Dea \& Owen(1985)]{Odea85} O'Dea, C.~P. \& Owen, F.~N.\ 1985, \aj, 90, 927. doi:10.1086/113801
\bibitem[Offringa et al.(2014)]{2014MNRAS.444..606O} Offringa, A.~R., McKinley, B., Hurley-Walker, N., et al.\ 2014, \mnras, 444, 606. doi:10.1093/mnras/stu1368
\bibitem[Offringa \& Smirnov(2017)]{2017MNRAS.471..301O} Offringa, A.~R. \& Smirnov, O.\ 2017, \mnras, 471, 301. doi:10.1093/mnras/stx1547
\bibitem[Offringa et al.(2012)]{2012A&A...539A..95O} Offringa, A.~R., van de Gronde, J.~J., \& Roerdink, J.~B.~T.~M.\ 2012, \aap, 539, A95. doi:10.1051/0004-6361/201118497
\bibitem[Owen et al.(2014)]{2014ApJ...794...24O} Owen, F.~N., Rudnick, L., Eilek, J., et al.\ 2014, \apj, 794, 24. doi:10.1088/0004-637X/794/1/24
\bibitem[Perley \& Butler(2013)]{2013ApJS..206...16P} Perley, R.~A. \& Butler, B.~J.\ 2013, \apjs, 206, 16. doi:10.1088/0067-0049/206/2/16
\bibitem[Pizzo et al.(2011)]{2011A&A...525A.104P} Pizzo, R.~F., de Bruyn, A.~G., Bernardi, G., et al.\ 2011, \aap, 525, A104. doi:10.1051/0004-6361/201014158
\bibitem[Porter et al.(2015)]{porter} Porter, D.~H., Jones, T.~W., \& Ryu, D.\ 2015, \apj, 810, 93. doi:10.1088/0004-637X/810/2/93
\bibitem[Rajpurohit et al.(2022a)]{2022A&A...657A...2R} Rajpurohit, K., Hoeft, M., Wittor, D., et al.\ 2022, \aap, 657, A2. doi:10.1051/0004-6361/202142340
\bibitem[Rajpurohit et al.(2022b)]{2022ApJ...927...80R} Rajpurohit, K., van Weeren, R.~J., Hoeft, M., et al.\ 2022, \apj, 927, 80. doi:10.3847/1538-4357/ac4708
\bibitem[Ramatsoku et al.(2020)]{2020A&A...636L...1R} Ramatsoku, M., Murgia, M., Vacca, V., et al.\ 2020, \aap, 636, L1. doi:10.1051/0004-6361/202037800
%\bibitem[Ramatsoku et al.(2020)]{threads} Ramatsoku, M., Murgia, M., Vacca, V., et al.\ 2020, \aap, 636, L1. doi:10.1051/0004-6361/202037800
\bibitem[Rines et al.(2013)]{Rines2003} Rines, K., Geller, M.~J., Diaferio, A., et al.\ 2013, \apj, 767, 15. doi:10.1088/0004-637X/767/1/15
\bibitem[Rudnick(2002)]{Rudnick02} Rudnick, L.\ 2002, \pasp, 114, 427. doi:10.1086/342499
\bibitem[Rudnick(2019)]{stormy} Rudnick, L.\ 2019, arXiv:1901.09448
\bibitem[Rudnick \& Blundell(2003)]{RudMag} Rudnick, L. \& Blundell, K.~M.\ 2003, \apj, 588, 143. doi:10.1086/373891
\bibitem[Sakelliou et al.(2008)]{Sakelliou2008} Sakelliou, I., Hardcastle, M.~J., \& Jetha, N.~N.\ 2008, \mnras, 384, 87. doi:10.1111/j.1365-2966.2007.12465.x
\bibitem[Sanders \& Fabian(2007)]{Perseus} Sanders, J.~S. \& Fabian, A.~C.\ 2007, \mnras, 381, 1381. doi:10.1111/j.1365-2966.2007.12347.x
\bibitem[Sarazin(1999)]{1999ApJ...520..529S} Sarazin, C.~L.\ 1999, \apj, 520, 529. doi:10.1086/307501
\bibitem[Schekochihin \& Cowley(2007)]{2007mhet.book...85S} Schekochihin, A.~A. \& Cowley, S.~C.\ 2007, Magnetohydrodynamics: Historical Evolution and Trends, 85
\bibitem[Stroe et al.(2014)]{sausagespec} Stroe, A., Harwood, J.~J., Hardcastle, M.~J., et al.\ 2014, \mnras, 445, 1213. doi:10.1093/mnras/stu1839
\bibitem[Shimwell et al.(2017)]{2017A&A...598A.104S} Shimwell, T.~W., R{\"o}ttgering, H.~J.~A., Best, P.~N., et al.\ 2017, \aap, 598, A104. doi:10.1051/0004-6361/201629313
\bibitem[Shimwell et al.(2019)]{2019A&A...622A...1S} Shimwell, T.~W., Tasse, C., Hardcastle, M.~J., et al.\ 2019, \aap, 622, A1. doi:10.1051/0004-6361/201833559
\bibitem[Shimwell et al.(2022)]{2022A&A...659A...1S} Shimwell, T.~W., Hardcastle, M.~J., Tasse, C., et al.\ 2022, \aap, 659, A1. doi:10.1051/0004-6361/202142484
\bibitem[Snowden et al.(2008)]{2008A&A...478..615S} Snowden, S.~L., Mushotzky, R.~F., Kuntz, K.~D., et al.\ 2008, \aap, 478, 615. doi:10.1051/0004-6361:20077930
\bibitem[Stuardi et al.(2021)]{2021MNRAS.502.2518S} Stuardi, C., Bonafede, A., Lovisari, L., et al.\ 2021, \mnras, 502, 2518. doi:10.1093/mnras/stab218

\bibitem[{\v{S}}uhada et al.(2011)]{Suhada11} {\v{S}}uhada, R., Fassbender, R., Nastasi, A., et al.\ 2011, \aap, 530, A110. doi:10.1051/0004-6361/201116876
\bibitem[Struble \& Rood(1999)]{1999ApJS..125...35S} Struble, M.~F. \& Rood, H.~J.\ 1999, \apjs, 125, 35. doi:10.1086/313274
\bibitem[Tasse et al.(2021)]{2021A&A...648A...1T} Tasse, C., Shimwell, T., Hardcastle, M.~J., et al.\ 2021, \aap, 648, A1. doi:10.1051/0004-6361/202038804
%\bibitem[xxx ]{2016ApJS..223...2V}

\bibitem[Tasse(2014)]{2014A&A...566A.127T} Tasse, C.\ 2014, \aap, 566, A127. doi:10.1051/0004-6361/201423503
%\bibitem[Tasse(2014)]{2014arXiv1410.8706T} Tasse, C.\ 2014, arXiv:1410.8706
\bibitem[Tasse et al.(2018)]{2018A&A...611A..87T} Tasse, C., Hugo, B., Mirmont, M., et al.\ 2018, \aap, 611, A87. doi:10.1051/0004-6361/201731474
\bibitem[van Breugel et al.(1985)]{1985ApJ...293...83V} van Breugel, W., Filippenko, A.~V., Heckman, T., et al.\ 1985, \apj, 293, 83. doi:10.1086/163216
\bibitem[van Haarlem et al.(2013)]{2013A&A...556A...2V} van Haarlem, M.~P., Wise, M.~W., Gunst, A.~W., et al.\ 2013, \aap, 556, A2. doi:10.1051/0004-6361/201220873
\bibitem[van Weeren et al.(2019)]{vWreview} van Weeren, R.~J., de Gasperin, F., Akamatsu, H., et al.\ 2019, \ssr, 215, 16. doi:10.1007/s11214-019-0584-z
\bibitem[van Weeren et al.(2021)]{2021A&A...651A.115V}  van Weeren, R.~J., Shimwell, T.~W., Botteon, A., et al.\ 2021, \aap, 651, A115. doi:10.1051/0004-6361/202039826
\bibitem[Vazza et al.(2006)]{2006MNRAS.369L..14V} Vazza, F., Tormen, G., Cassano, R., et al.\ 2006, \mnras, 369, L14. doi:10.1111/j.1745-3933.2006.00164.x
\bibitem[Vazza et al.(2018a)]{vazza18} Vazza, F., Angelinelli, M., Jones, T.~W., et al.\ 2018, \mnras, 481, L120. doi:10.1093/mnrasl/sly172
\bibitem[Vazza et al.(2018b)]{2018MNRAS.474.1672V} Vazza, F., Brunetti, G., Br{\"u}ggen, M., et al.\ 2018, \mnras, 474, 1672. doi:10.1093/mnras/stx2830

\bibitem[Venturi et al.(2022)]{Shapley} Venturi, T., Giacintucci, S., Merluzzi, P., et al.\ 2022, arXiv:2201.04887
\bibitem[Walker, Sanders \& Fabian (2016)]{2016MNRAS.461..684W} Walker, S.~A., {Sanders}, J.~S. and {Fabian}, A.~C., 2016, MNRAS 461, 684
\bibitem[Wang et al.(2016)]{2016ApJ...833...99W} Wang, Q.~H.~S., Markevitch, M., \& Giacintucci, S.\ 2016, \apj, 833, 99. doi:10.3847/1538-4357/833/1/99
%\bibitem[Wang et al.(2018)]{2018ApJ...856..162W} Wang, Q.~H.~S., Giacintucci, S., \& Markevitch, M.\ 2018, \apj, 856, 162. doi:10.3847/1538-4357/aab2aa
\bibitem[Wilber et al.(2018)]{2018MNRAS.476.3415W} Wilber, A., Br{\"u}ggen, M., Bonafede, A., et al.\ 2018, \mnras, 476, 3415. doi:10.1093/mnras/sty414
\bibitem[Williams et al.(2016)]{2016MNRAS.460.2385W} Williams, W.~L., van Weeren, R.~J., R{\"o}ttgering, H.~J.~A., et al.\ 2016, \mnras, 460, 2385. doi:10.1093/mnras/stw1056
\bibitem[Wykes et al.(2014)]{2014MNRAS.442.2867W} Wykes, S., Intema, H.~T., Hardcastle, M.~J., et al.\ 2014, \mnras, 442, 2867. doi:10.1093/mnras/stu1033
\bibitem[Xu \& Lazarian(2022)]{2022ApJ...927...94X} Xu, S. \& Lazarian, A.\ 2022, \apj, 927, 94. doi:10.3847/1538-4357/ac4dfd
\bibitem[Yusef-Zadeh et al.(2021)]{2021MNRAS.500.3142Y} Yusef-Zadeh, F., Wardle, M., Heinke, C., et al.\ 2021, \mnras, 500, 3142. doi:10.1093/mnras/staa3257
\bibitem[Zhang et al.(2019)]{2019MNRAS.488.5259Z} Zhang, C., Churazov, E., Forman, W.~R., et al.\ 2019, \mnras, 488, 5259. doi:10.1093/mnras/stz2135
\bibitem[ZuHone et al.(2021a)]{Zuhone21} ZuHone, J.~A., Markevitch, M., Weinberger, R., et al.\ 2021, \apj, 914, 73. doi:10.3847/1538-4357/abf7bc
\bibitem[ZuHone et al.(2021b)]{Zuhone21b} ZuHone, J., Ehlert, K., Weinberger, R., et al.\ 2021, Galaxies, 9, 91. doi:10.3390/galaxies9040091
\bibitem[Zhuravleva et al.(2016)]{2016MNRAS.458.2902Z} Zhuravleva, I., Churazov, E., Ar{\'e}valo, P., et al.\ 2016, \mnras, 458, 2902. doi:10.1093/mnras/stw520
\bibitem[Zhuravleva et al.(2014)]{2014Natur.515...85Z} Zhuravleva, I., Churazov, E., Schekochihin, A.~A., et al.\ 2014, \nat, 515, 85. doi:10.1038/nature13830
\bibitem[Zhuravleva et al.(2019)]{Zhura} Zhuravleva, I., Churazov, E., Schekochihin, A.~A., et al.\ 2019, Nature Astronomy, 3, 832. doi:10.1038/s41550-019-0794-z



%%%%%%%%%%%%%%%%%%%%%%%%%%%%
%%%%%%%%%%%%%%%%%%%%%%%%%%%
\end{thebibliography}
\end{document}